\documentclass[11pt,a4paper]{article}
\usepackage{verbatim}
\usepackage[UKenglish]{babel}
\usepackage[UKenglish]{isodate}
\usepackage[T1]{fontenc}
\usepackage{lmodern}
\usepackage[top=0.75in,left=0.75in,right=0.75in,bottom=1.0in]{geometry}
\cleanlookdateon
\usepackage{epsfig}
\usepackage{graphicx}
\usepackage[margin=15pt,font={footnotesize},labelfont=bf,format=hang]{caption}
\usepackage[margin=15pt,font={footnotesize},labelfont=bf,format=hang]{subcaption}
\usepackage{bm}
\usepackage{amsbsy}
\usepackage{amssymb}
\usepackage{amsmath}
\usepackage{mathtools}
\usepackage{isomath}
\usepackage{enumitem}
\usepackage{natbib}
\usepackage[]{hyperref}
\usepackage[dvipsnames]{xcolor}
\usepackage[ruled,vlined,linesnumbered]{algorithm2e}
\usepackage[capitalise]{cleveref}
\usepackage{xpatch}
\xpretocmd{\algorithm}{\hsize=\linewidth}{}{}
\usepackage[normalem]{ulem}
\usepackage{booktabs}
\usepackage{soul}
\usepackage{siunitx}
\usepackage{color}
\usepackage{tikz}
\usepackage{authblk}

\renewcommand{\v}[1]{\mathbfit{#1}}      % Generic vector
\newcommand{\uv}[1]{\widehat{\mathbfit{#1}}}      % Generic unit vector
\newcommand{\q}[1]{\mathsfbfit{#1}}        % Generic quaternion
\renewcommand{\t}[1]{\mathsfbfit{#1}}	 % Generic tensor
\newcommand{\vu}[1]{\mathbf{#1}}         % Upright vector (for zero vector)
\newcommand{\tu}[1]{\bm{\mathsf{#1}}}	 % Upright tensor (for zero tensor)

\newcommand{\qprod}{\mathbin{\bullet}} % So we can easily change the symbol later if we want to. Mathbin command makes it behave like a binary operator, e.g. line breaks work correctly.

\renewcommand{\d}{{\mathrm d}}
\def \p {\partial}

\newcommand{\vnabla}{\bm{\nabla}}

\newcommand{\nx}{t}
\newcommand{\ny}{\mu}
\newcommand{\nz}{\nu}

\renewcommand{\hat}{\widehat}

\renewcommand{\bar}{\overline}

\newcommand{\fd}[2]{\mathchoice{\frac{\d #1}{\d #2}}{\d #1/\d #2}{\d #1/\d #2}{\d #1/\d #2}}

\newcommand{\pd}[2]{\mathchoice{\frac{\p #1}{\p #2}}{\p #1/\p #2}{\p #1/\p #2}{\p #1/\p #2}}

\DeclareMathSymbol{\Delta}{\mathalpha}{operators}{1} % Keeps Delta upright

\newcommand{\diag}{\operatorname{diag}}
\newcommand{\order}{\mathcal{O}}
\newcommand{\BCH}{\mathcal{B}}
\newcommand{\Sp}{\text{\rm \textit{Sp}}}
\DeclareMathOperator{\dexp}{dexp}
\newcommand{\dexpinv}{\dexp^{-1}}
\newcommand{\Reals}{\mathbb{R}}
\newcommand{\twopartdef}[4]{
	\left\{
		\begin{array}{ll}
			#1 & \mbox{} #2 \\[6pt]
			#3 & \mbox{} #4
		\end{array}
	\right.
}

\DeclareRobustCommand{\legendline}[1]{%
\raisebox{2pt}{\tikz{\draw[-,#1,solid,line width = 0.9pt](0,0) -- (5mm,0);}}%
}

\begin{document}

\newcommand\blfootnote[1]{%
  \begingroup
  \renewcommand\thefootnote{}\footnote{#1}%
  \addtocounter{footnote}{-1}%
  \endgroup
}

\title{\bf Methods for suspensions of passive and active filaments}
\author{Simon F Schoeller\thanks{Contributed equally.},\,
Adam K Townsend,$\hspace{-3pt}^*$
Timothy A Westwood,$\hspace{-3pt}^*$
Eric E Keaveny\thanks{Email addresses: \url{simon.schoeller14@imperial.ac.uk}, \url{adam.townsend@imperial.ac.uk}, \\ \url{t.westwood16@imperial.ac.uk}, \url{e.keaveny@imperial.ac.uk}}}
\affil{\it\normalsize Department of Mathematics, Imperial College London, London SW7 2AZ, UK}
\maketitle

\begin{abstract}

    Flexible filaments and fibres are essential components of important complex fluids that appear in many biological and industrial settings.  Direct simulations of these systems that capture the motion and deformation of many immersed filaments in suspension remain a formidable computational challenge due to the complex, coupled fluid--structure interactions of all filaments, the numerical stiffness associated with filament bending, and the various constraints that must be maintained as the filaments deform.  In this paper, we address these challenges by describing filament kinematics using quaternions to resolve both bending and twisting, applying implicit time-integration to alleviate numerical stiffness, and using quasi-Newton methods to obtain solutions to the resulting system of nonlinear equations.  In particular, we employ geometric time integration to ensure that the quaternions remain unit as the filaments move.  We also show that our framework can be used with a variety of models and methods, including matrix-free fast methods, that resolve low Reynolds number hydrodynamic interactions.  We provide a series of tests and example simulations to demonstrate the performance and possible applications of our method.  Finally, we provide a link to a MATLAB/Octave implementation of our framework that can be used to learn more about our approach and as a tool for filament simulation.

\end{abstract}
\tableofcontents
\section{Introduction}
Microscopic flexible filaments and rigid fibres suspended in fluid arise in many industrial and biological applications, ranging from the wood pulp fibres found in paper production \citep{derakhshandeh_rheology_2011, stockie_simulating_1998,ross_dynamic_1997, pettersson_measurement_2017,du_roure_dynamics_2017} to the flagella and cilia used by cells to swim, pump, or mix fluid \citep{lauga_hydrodynamics_2009,elgeti_physics_2015, brennen_fluid_1977,supatto_cilia_2011,smith_symmetry-breaking_2019,faubel_cilia-based_2016,elgeti_emergence_2013}. In suspension, such as in polymeric fluids, their presence alters the bulk rheological properties of the medium leading to non-Newtonian responses such as shear thinning or viscoelasticity \citep{larson_structure_1999}.  When entangled or connected in networks, filaments and fibres form gels and disordered solids as is the case in important biological materials such as mucus \citep{hwang_rheological_1969,sheehan_hydrodynamic_1984,quraishi_rheology_1998,lai_micro-_2009} and the extra-cellular matrix \citep{baker_extracellular_2009, heck_modeling_2017}.  Filaments can also be active biological elements that facilitate transport through self-deformation \citep{lauga_hydrodynamics_2009, brennen_fluid_1977}, or through growth and the motion of motor proteins along their lengths \citep{shelley_dynamics_2016}.

Simulation of filament or fibre motion requires solving a complex low Reynolds number fluid--structure interaction problem where the motion and deformation of all filaments are coupled through the surrounding fluid.  As a result, computations involving filaments use many existing models and methods developed over the years to resolve the hydrodynamic aspects of the problem at various levels of approximation, ranging from drag-based resistive force theory \citep{moreau_asymptotic_2018,yamamoto_dynamic_1995} and point and regularised singularity methods \citep{cortez_method_2001, cortez_method_2005,cosentino_lagomarsino_hydrodynamic_2005, delmotte_general_2015, olson_modeling_2013, smith_boundary_2009, smith_discrete_2007} for single filament problems, to the immersed boundary method \citep{peskin_immersed_2002,fauci_computational_1988, stockie_simulating_1998, lim_dynamics_2008,lim_dynamics_2010,wiens_simulating_2015} and discretisations of the integral equations given by slender-body theory coupled with fast-summation techniques \citep{saintillan_smooth_2005,tornberg_simulating_2004, gustavsson_gravity_2009, nazockdast_fast_2017, tornberg_numerical_2006} for larger collections of flexible filaments or rigid fibres.

While there are a variety of models and methods available to treat filament hydrodynamics, a similar catalogue of methods and algorithms for filament elasticity is less developed.  Due to the higher-order derivatives associated with bending and twisting, the equations of filament motion are numerically stiff.  In addition, conditions such as inextensibility give rise to constraints that must be satisfied as filaments move and deform.  Despite this, explicit time integration is often used \citep{peskin_immersed_2002,li_sedimentation_2013,olson_modeling_2013,simons_fully_2015} to advance the filaments in time, while constraints of inextensibility are mimicked through stiff springs \citep{chelakkot_migration_2010,stockie_simulating_1998}, both of which tend to limit stable timestep sizes.  There have been recent efforts to use implicit time integration schemes \citep{moreau_asymptotic_2018,hall-mcnair_efficient_2019}, as well as treat the constraints directly using Lagrange multipliers \citep{delmotte_general_2015} or through a suitable choice of the filament degrees of freedom \citep{moreau_asymptotic_2018,hall-mcnair_efficient_2019}.  These studies, however, are limited in the number of filaments that they can treat as they rely on explicit knowledge of the filament mobility matrix and require dense linear algebra to solve for the unknown Lagrange multipliers, or to implicitly integrate the equations of motion.  For specific hydrodynamic approaches intended for larger-scale simulation of many interacting filaments, similar techniques have been used to reduce computational times. For example, predictor--corrector schemes have been successfully used with the immersed boundary method \citep{wiens_simulating_2015}, while recent work \citep{nazockdast_fast_2017}, building from \citet{tornberg_simulating_2004}, presents a complete approach utilising Lagrange multipliers and implicit time integration along with matrix-free hydrodynamics using the fast multipole method.

In this study, we provide a comprehensive, computationally-scalable methodology for simulating filament dynamics that is not dependent on the specific method or model used to describe the hydrodynamic forces experienced by the filaments.  Our approach only assumes that hydrodynamics forces and torques experienced by the discretised beam segments are linearly related to their translational and rotational velocities, consistent with low Reynolds number hydrodynamics.  The methods that we present in this paper expand on those we have already used for planar simulations of thousands of interacting sperm cells \citep{schoeller_flagellar_2018} and the motion of undulatory swimmers through structured \citep{majmudar_experiments_2012} and unstructured \citep{kamal_enhanced_2018} complex environments.  This paper presents the fully three-dimensional version of the methodology that takes advantage of unit quaternions to describe filament kinematics and geometric time integration schemes \citep{iserles_lie-group_2000,faltinsen_multistep_2001,park_geometric_2005} to advance the unit quaternions in time.

Beginning from Kirchhoff rod theory \citep{kirchhoff_uber_1859,lim_dynamics_2008}, we provide a description of the filament model and highlight the differences between deformations restricted to a plane and those that are fully three-dimensional.  In particular, for fully three-dimensional deformations, we show how both filament bending and twisting can be captured through the use of unit quaternions to describe how the filament's material frame varies with arclength.  We employ Lagrange multipliers to obtain the constraint forces and moments necessary to link the material frame with the positions of points along the filament, and implicit time integration to handle numerical stiffness.  By comparing Jacobian-free Newton--Krylov \citep{knoll_jacobian-free_2004} and Broyden's methods \citep{broyden_class_1965,kvaalen_faster_1991}, we show how the resulting system of nonlinear equations, whose solution provides the updated generalised positions and Lagrange multipliers, can be solved most effectively using Broyden's method with a suitable choice of approximate Jacobian based on a diagonal mobility matrix.  Also novel to our approach is our application of geometric time integration to ensure the quaternions remain unit.  We find these schemes help to guarantee robustness of the iterative solver for the nonlinear system of equations.  We provide a number of tests of our method and explore several examples using both a hydrodynamic approach based on direct summation of the Rotne--Prager--Yamakawa (RPY) tensor \citep{wajnryb_generalization_2013} and matrix-free computations based on the force-coupling method (FCM) \citep{maxey_localized_2001,lomholt_force-coupling_2003}.  In doing so, we show that our method provides an effective tool for studying a variety of applications involving filament dynamics including tethered filaments and cilia arrays, interacting undulatory swimmers, and sedimenting suspensions and clouds of filaments.  We provide a basic MATLAB/Octave implementation \citep{schoeller_github_2019} of the method that can be used to reproduce our numerical results, applied to other problems of interest, or altered to couple with the user's preferred hydrodynamic solver.

\section{Filament model}\label{sec:filament-model}
We begin by describing the filament model for a single filament.  In this work, a filament refers to an inextensible, slender (i.e. high-aspect ratio) body, whose shape is given by a curve in space and whose deformations are related to the rotations of a right-handed frame defined along the length of the curve.  The filament has circular cross-section with radius $a$ and length $L$.  Its physical size is assumed to be sufficiently small such that its inertia, and that of the surrounding fluid, can be ignored.  It can be subject to external forces and moments along its length, such as those due to the surrounding fluid, in addition to the internal moments due to bending and twist, and the internal force that arises due to constraints on filament motion.  In addition, we consider the important case where, in the absence of external forces and torques, the filament may relax to a curve with nonzero curvature and twist.

\subsection{Kinematics}
The positions of the points along the filament's centreline at time $t$ are given by $\v{Y}(s,t)$, where $s \in [0, L]$ is the filament arclength.  Along with the position, there is a right-handed, orthonormal material frame $\{\uv{t}(s,t),\uv{\ny}(s,t),\uv{\nz}(s,t)\}$ at each $s$ and $t$, see \cref{pic:arclength}.  At this stage, the unit vector $\uv{t}$ and position, $\v{Y}$, are independent of each other, but later we will introduce the kinematic constraint, $\pd{\v{Y}}{s} = \uv{t}$, that links these quantities and will lead to forces within the filament.

Filament deformation is determined by how the material frame varies with $s$.  Following \citet[\S 18]{landau_theory_1986}; \citet{powers_dynamics_2010}, this deformation can be described by the vector $\v{\Phi}(s,t)$ that describes the spatial-rate of rotation of the local frame with $s$ such that $\pd{\v{v}}{s} = \v{\Phi} \times \v{v}$ for any vector in the material frame, $\v{v}$.  Using $\uv{t}$ and $\pd{\uv{t}}{s}$, $\v{\Phi}$ can be expressed as
\begin{align}
    	\v{\Phi} &= \uv{t} \times\pd{\uv{t}}{s} + \uv{t} \left(\v{\Phi}\cdot \uv{t} \right),
\end{align}
where the first term is related to bending based on a curve with tangent $\uv{t}$, while the second term describes twisting about $\uv{t}$.  By combining the derivatives of the frame vectors with the fact that $\pd{(\uv{t}\cdot\uv{\mu})}{s} = \pd{(\uv{t}\cdot\uv{\nu})}{s} = \pd{(\uv{\nu}\cdot\uv{\mu})}{s} = 0$, we can write the bending contribution as
  \begin{align}
  	\uv{t} \times \pd{\uv{t}}{s} &= \uv{\ny}\left(\uv{t}\cdot\pd{\uv{\nz}}{s}\right) + \uv{\nz}\left(\uv{\ny}\cdot\pd{\uv{t}}{s}\right),
	\label{eqn:bendingkin}
  \end{align}
  while the twist contribution is given by
  \begin{align}
  	\v{\Phi}\cdot \uv{t} &= \uv{\nz}\cdot \pd{\uv{\ny}}{s}.
	\label{eqn:twistingkin}
  \end{align}
\subsection{Constitutive law and internal moments}
The internal moments, $\v{M}$, are linearly related to filament bending, \cref{eqn:bendingkin}, and twisting, \cref{eqn:twistingkin}, through the bending, $K_B$, and twist, $K_T$,  moduli. These moduli are related to the shear modulus, $G$, and the Young's modulus, $E$, of the underlying filament material via $K_B = EI$, $K_T = 2GI$, where $I = \pi a^4 /4$ is the moment of inertia about any radial axis lying in the circular filament cross-section \citep[\S 16, 17]{landau_theory_1986}. Specifically, we substitute \cref{eqn:bendingkin,eqn:twistingkin} into (19.7) of \citet[\S 19]{landau_theory_1986} to yield
  \begin{align} \begin{split} \label{equation:3d_moment_frame_no_preferred_curvature}
  	\v{M}(s,t)=& K_B \left(\left(\uv{t}\cdot\pd{\uv{\nz}}{s}  \right)\uv{\ny} + \left(\uv{\ny}\cdot\pd{\uv{t}}{s} \right)\uv{\nz} \right) \\ &+ K_T  \left(\uv{\nz}\cdot \pd{\uv{\ny}}{s}\right)\uv{t}.
  \end{split} \end{align}
With this form of $\v{M}$, the filament equilibrium configuration is straight and untwisted.  Following \citet{lim_dynamics_2008, lim_dynamics_2010, olson_modeling_2013}, to incorporate a nontrivial equilibrium shape into the model, we can introduce the preferred curvatures, $\kappa_{\ny}(s,t)$ and $\kappa_{\nz}(s,t)$, and preferred twist, $\gamma_0(s,t)$, such that
\begin{align}
\begin{split} \label{equation:3d_moment_frame}
  	\v{M}(s,t)=& K_B \left(\uv{\ny}\left(\uv{t}\cdot\pd{\uv{\nz}}{s} - \kappa_\ny(s,t) \right) + \uv{\nz}\biggl(\uv{\ny}\cdot\pd{\uv{t}}{s}- \kappa_\nz(s,t)\biggr)\right) \\ &+ K_T \uv{t} \left(\uv{\nz}\cdot \pd{\uv{\ny}}{s} -\gamma_0(s,t) \right).
\end{split}
\end{align}
In the absence of applied forces and torques acting on the filament, its shape is determined completely by $\kappa_\ny$, $\kappa_\nz$ and $\gamma_0$.  Additionally, by allowing $\kappa_\ny$, $\kappa_\nz$ and $\gamma_0$ to be time dependent, we can introduce shape changes that allow the filament to propel itself in a surrounding fluid without experiencing a net force or net torque, in accordance with the constraints on swimming at low Reynolds number \citep{purcell_life_1977,lauga_hydrodynamics_2009}.

\subsection{Kinematic constraint and internal stress}
Along with the internal moments, filament deformation will give rise to an internal force, $\v{\Lambda}(s,t)$, or the force on the filament cross-section.  In \citet{lim_dynamics_2008, lim_dynamics_2010, olson_modeling_2013}, this force is captured through a constitutive law that relates $\pd{\v{Y}}{s}$ and the local frame such that in equilibrium,
\begin{align}
	\pd{\v{Y}}{s} = \uv{t}.
	\label{eqn:constraint}
\end{align}
We instead impose \cref{eqn:constraint} as a kinematic constraint, requiring that $\uv{t}$ be the unit tangent to the curve defined by $\v{Y}$. This constraint gives rise to the internal force, $\v{\Lambda}(s,t)$ and as such, it can be viewed as the Lagrange multiplier\footnote{Rather than considering force and moment balances along the filament, the filament model can alternatively be derived through the introduction of a Lagrangian \citep[appendix]{lim_dynamics_2008}.  The internal force arises from the constraint term,
\begin{equation}
    \int_{0}^L \v{\Lambda}\cdot\left(\pd{\v{Y}}{s} - \uv{t}\right) \d s,
\end{equation}
in the Lagrangian, which will also contain the bending and twist energies, as well as any external potential.} that enforces \eqref{eqn:constraint}.  The internal force provides both the tension in the direction $\uv{t}$ that keeps the filament inextensible, but also the necessary normal force that couples the force and moment balances as we describe below.

\subsection{Force and moment balances}
In general, the internal force and moment described above will balance external forces, $\v{f}$, and torques, $\v{\tau}$, per unit length that act along the filament, see \cref{pic:beam-forces-cts}.  The external forces and torques can arise due to, for example, an external field such as gravity or short-ranged filament--filament interactions. In particular, we insist that they include the force and torque per unit length exerted on the filament by the surrounding fluid.  We assume that the interaction with the fluid is governed by low Reynolds number hydrodynamics, and hence these forces and torques are linearly related to the filament's motion.  As described in \citet[\S 19]{landau_theory_1986}; \citet{powers_dynamics_2010}, the resulting force and moment balances are
 \begin{align}
		\pd{\v{\Lambda}}{s} + \v{f} &= \vu{0}, \label{eqn:KirchhoffForce}\\
    	\pd{\v{M}}{s} + \uv{t} \times \v{\Lambda} + \v{\tau} &= \vu{0}.\label{eqn:KirchhoffMoment}
\end{align}
In addition to these equations, we will have conditions at the filament ends.  For example, for filaments with free ends, we have\footnote{The boundary conditions $\v{\Lambda}(0,t)=\v{\Lambda}(L,t)=\vu{0}$ require that $\int_0^L \v{f} \,\d s = \vu{0}$.  In our case, this will be satisfied since $\v{f}$ includes forces due to the surrounding fluid which will automatically balance any net force applied to the filament by other means (e.g.\ gravity, filament--filament repulsion).} $\v{\Lambda}(0,t)=\v{\Lambda}(L,t)=\vu{0}$ and $\v{M}(0,t)=\v{M}(L,t)=\vu{0}$.  Filaments that are tethered will need to satisfy conditions imposed on $\v{Y}$ and $\{\uv{t},\uv{\ny},\uv{\nz}\}$ at the tethered end.  At this stage, it is important to note that \cref{eqn:KirchhoffForce,eqn:KirchhoffMoment} establish the low Reynolds number mobility problem that couples the motion of all filaments through the fluid flows that they generate.  This mobility problem, which we describe and solve in the discrete setting (see \cref{sec:hydrodynamics}), yields equations for the translational and angular velocities of the discrete filament segments.  The values of these velocities, however, will depend on $\v{\Lambda}$ which remains unknown.  The values of $\v{\Lambda}$ are determined later by imposing the kinematic constraint, \cref{eqn:constraint}, when integrating the differential equations for the filament segment positions and orientations.
\subsection{Discrete force and moment balances}
\begin{figure}[t!]
    \centering

    \begin{subfigure}[t]{0.45\textwidth}
    \centering
    \small
    \textbf{continuous}
    \end{subfigure}%
    \begin{subfigure}[t]{0.45\textwidth}
    \centering
    \small
    \textbf{discrete}
    \end{subfigure}
    \vspace{2ex}

    \begin{subfigure}[t]{0.45\textwidth}
    \centering
    \includegraphics[scale=1]{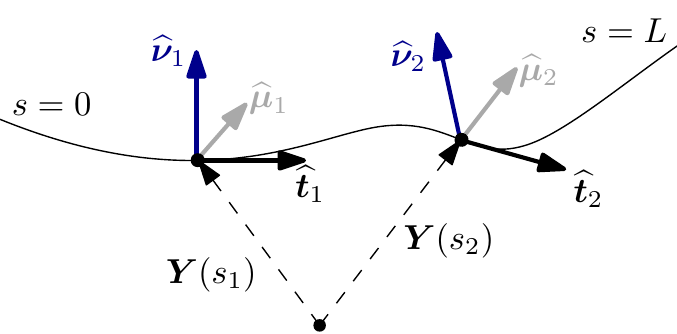}
    \caption{Position $\v{Y}(s)$ and local frame $\{\uv{\nx}(s),\uv{\ny}(s),\uv{\nz}(s)\}$ at arclengths $s_1$ and $s_2$.}
    \label{pic:arclength}
    \end{subfigure}%
    \begin{subfigure}[t]{0.45\textwidth}
    \centering
    \includegraphics[scale=1]{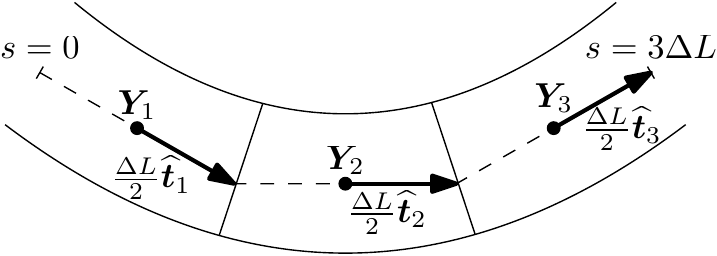}
    \caption{Spatial discretisation of the filament with positions $\v{Y}_n$ and vectors $\uv{t}_n$ satisfying the constraint \cref{equation:inex_constraints}}
    \label{pic:beam}
    \end{subfigure}

    \begin{subfigure}[t]{0.45\textwidth}
    \centering
    \includegraphics[scale=1]{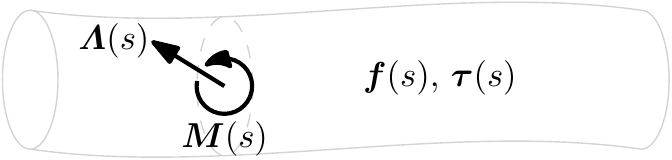}
    \caption{The internal forces $\v{\Lambda}(s)$ and moments $\v{M}(s)$. External force and torque densities along the filament are $\v{f}(s)$ and $\v{\tau}(s)$, respectively.}
    \label{pic:beam-forces-cts}
    \end{subfigure}%
    \begin{subfigure}[t]{0.45\textwidth}
    \centering
    \includegraphics[scale=1]{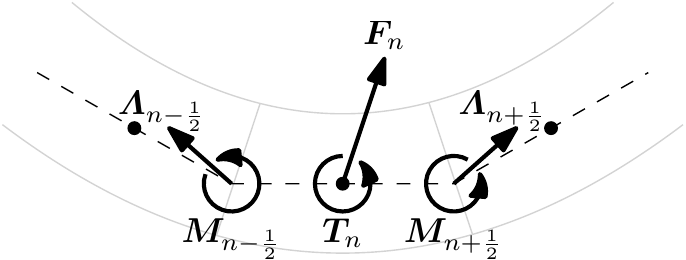}
    \caption{Forces and torques in the discrete system with $\v{F}_n = \Delta L \v{f}_n$ and $\v{T}_n = \Delta L \v{\tau}_n$}
    \label{pic:beam-forces}
    \end{subfigure}
\caption{Continuous and discrete representations of the filament model}
\label{pic:continuous-and-discrete-filament-models}
\end{figure}

To solve the equations numerically, the filament is first discretised into $N$ segments of length $\Delta L$ as shown in \cref{pic:beam}.  The segments have positions, $\v{Y}_n$, and orientations, $\uv{t}_n$, for $n = 1,\dots,N$.  The segment orientations are the discrete representation of the frame vector, $\uv{t}$, at the segment position.  Applying central differencing to \cref{eqn:KirchhoffForce,eqn:KirchhoffMoment}, the force and moment balances become
\begin{align}
	\frac{\v{\Lambda}_{n+1/2} - \v{\Lambda}_{n-1/2}}{\Delta L} + \v{f}_n &= \vu{0}, \label{eqn:DifferencingForce} \\
	\frac{\v{M}_{n+1/2} - \v{M}_{n-1/2}}{\Delta L} + \frac{1}{2} \uv{t}_n \times \left(\v{\Lambda}_{n+1/2} + \v{\Lambda}_{n-1/2}\right) + \v{\tau}_n &= \vu{0}, \label{eqn:DifferencingMoments}
\end{align}
where $\v{M}_{n+1/2}$ and $\v{\Lambda}_{n+1/2}$ are the internal moment and stress, respectively, between segments $n$ and $n+1$; $\v{f}_n$ is the external force per unit length on segment $n$; and $\v{\tau}_n$ is the external torque per unit length on $n$; see \cref{pic:beam-forces}.  Free end conditions are established by requiring that $\v{M}_{1/2} = \v{M}_{N+1/2} = \vu{0}$ and $\v{\Lambda}_{1/2} = \v{\Lambda}_{N+1/2} = \vu{0}$.  Similarly, for the kinematic constraint, \cref{eqn:constraint}, we have
\begin{equation}
\v{g}_{n+1/2} = \v{Y}_{n+1} - \v{Y}_n - \frac{\Delta L}{2}\left( \uv{t}_n + \uv{t}_{n+1} \right) = \vu{0}. \label{equation:inex_constraints}
\end{equation}
Multiplying \cref{eqn:DifferencingForce,eqn:DifferencingMoments} by $\Delta L$, we convert the force and moment per unit length balances to the force and torque balances such that
        \begin{align}
        	\v{F}^C_n + \v{F}_n &= \vu{0}, \label{eqn:DiscreteForce} \\
           \v{T}^E_n + \v{T}^C_n + \v{T}_n &= \vu{0}, \label{eqn:DiscreteTorque}
        \end{align}
for segment $n$. The elastic torque on $n$ is given by
\begin{equation}
    \v{T}^E_n = \v{M}_{n+1/2} - \v{M}_{n-1/2}.
    \label{eqn:elastic-torque}
\end{equation}
Since the internal moment is directly linked to how the local frame changes with arclength, the computation of $\v{M}_{n+1/2}$ will depend on whether filament deformation is strictly planar, or fully three-dimensional.  We present both cases below in \cref{sec:internal-moments-2d,sec:internal-moments-3d}.

The constraint force and torque, respectively, on segment $n$, are given by
\begin{align}
\v{F}^C_n &=\v{\Lambda}_{n+1/2} - \v{\Lambda}_{n-1/2}, \label{eqn:constraint-force}\\
\v{T}^C_n &= \frac{\Delta L}{2} \uv{t}_n\times\left(\v{\Lambda}_{n+1/2} + \v{\Lambda}_{n-1/2}\right). \label{eqn:constraint-torque} %
% Sorry - need to refer to these later as well, hence the displaystyle -- Adam
\end{align}
The values of $\v{\Lambda}_{n+1/2}$ are the Lagrange multipliers associated with the discrete version of the kinematic constraint, \cref{equation:inex_constraints}, that relates the translational and rotational motions of segments $n$ and $n+1$.  The Lagrange multipliers are unknown and must be computed as the filament moves and deforms.

Finally, the external forces and torques acting on a segment are defined as $\v{F}_n = \Delta L \v{f}_n$ and $\v{T}_n = \Delta L \v{\tau}_n$, respectively.  The external forces and torques can arise due to an external field, such as gravity or an external flow, but also through interactions with neighbouring filaments as mediated by the fluid, or through direct collisions and steric interactions.  We describe here how we capture these interactions and, in particular, how hydrodynamic interactions lead to segment motion.  In the case where the external forces and torques are due only to steric interactions and the surrounding fluid, we have
\begin{align}
    \v{F}_n &= \v{F}^B_n - \v{F}^H_n,\\
    \v{T}_n &= -\v{T}^H_n,
\end{align}
where $\v{F}^B_n$ are the steric forces experienced by segment $n$ and $-\v{F}^H_n$ and $-\v{T}^H_n$ are the hydrodynamic force and torque  on segment $n$.  We have chosen to introduce the negative sign here so that $\v{F}^H_n$ and $\v{T}^H_n$ denote, respectively, the force and torque segment $n$ exerts on the fluid.

\subsection{Steric interactions}\label{sec:steric-interactions}
In our simulations, segments from different filaments and non-neighbouring segments from the same filament repel each other via a short-ranged, pairwise barrier force.  For simplicity, we take this force to be a function of the centre-to-centre separation of the segments, though a more detailed representation of the segment shape could also be taken.

Specifically, in our simulations, the total barrier force on segment $n$ is
\begin{align}
    \v{F}^B_n = \sum_m \v{F}^B_{nm},
\end{align}
where the force on segment $n$ due to $m$ is \citep{dance_collision_2004}
\begin{align}
    \v{F}^B_{nm} = F^S \left(\frac{4a^2\chi^2 - r_{nm}^2}{4a^2(\chi^2 - 1)}\right)^4 \frac{\v{r}_{nm}}{2a} \quad  (\text{if } r_{nm} < 2\chi a),
    \label{eqn:barrier-force}
\end{align}
and $\v{F}^B_{nm} = \vu{0}$ if $r_{nm} \geq 2\chi a$.  We also have that $\v{F}^B_{nm} = \vu{0}$ for any separation distance if $n$ and $m$ are neighbouring segments from the same filament.  In \cref{eqn:barrier-force}, the displacement between segments is $\v{r}_{nm} = \v{Y}_n - \v{Y}_m$, while $r_{nm} = \|\v{r}_{nm}\|$.  The parameter $F^S$ controls the strength of repulsion at contact, while $\chi$ controls the range over which the barrier force acts.  Recall that $2a$ is the filament thickness.  In practice, to prevent the segments from overlapping, we choose the reference force $F^S$ to be large compared to the typical bending force and set $\chi=1.1$ to ensure that the force only acts for particle separations that are 10\% of their contact distance.

\subsection{Hydrodynamics interactions and segment motion}\label{sec:hydrodynamics}
The hydrodynamic forces and moments on the segments balance the internal stresses and moments and the inter-segment repulsive forces.  In the absence of fluid inertia, this leads to a low Reynolds number mobility problem which provides a linear relationship between the forces and torques on the segments and their velocities and angular velocities,
\begin{equation}\label{eq:mobility-problem}
	\begin{pmatrix}
	\v{V}\\\v{\Omega}
	\end{pmatrix}
    = \mathcal{M}\cdot
    \begin{pmatrix}
	\v{F}^H\\\v{T}^H
	\end{pmatrix},
\end{equation}
where $\v{V}^\top = (\v{V}_1^\top,\dots,\v{V}_N^\top)$ is the vector of all segment velocity components.  Similarly $\v{\Omega}$ contains the angular velocity components for all segments, while $\v{F}^H$ and $\v{T}^H$ are the vectors that hold the forces and torques, respectively, that all the filament segments exert on the surrounding fluid.
The methods presented in this paper can be coupled with a variety of hydrodynamic models and associated computational methodologies, including those that are mobility matrix-free, provided that they retain this linear relationship between the forces that the segments exert on the fluid and segment motion.  Such approaches include Stokesian Dynamics \citep{brady_stokesian_1988,sierou_accelerated_2001}, boundary integral methods \citep{pozrikidis_boundary_1992}, the method of regularised Stokeslets \citep{cortez_method_2001} and the immersed boundary method applied to the Stokes equations \citep{peskin_immersed_2002}.

In this work, we employ two approaches to perform the mobility matrix multiplication and resolve segment motion: a direct computation based on the Rotne--Prager--Yamakawa (RPY) tensors \citep{wajnryb_generalization_2013}, and a matrix-free approach based on the force-coupling method (FCM) \citep{maxey_localized_2001,lomholt_force-coupling_2003}.  Both approaches provide a similar resolution of segment hydrodynamic interactions based on the Stokes flows generated by singular and regularised force and torque distributions.  In addition, RPY and FCM are closely related to the methods mentioned above and are chosen to demonstrate how our filament model can be used in conjunction with a wide range of methods, models, and solvers for filament motion in viscous fluids. We note that both RPY and FCM link the radius of the filament cross-section to the segment length.  Thus, increasing the number of segments changes the aspect ratio of the filament, and hence its hydrodynamic response.  While we will see in \cref{sec:accuracy-of-hydrodynamic-model} that this hydrodynamic model provides accurate values for the hydrodynamic force on the filament over a range of aspect ratios, it could be important to be able to increase (or decrease) the spatial resolution of filament deformation without affecting hydrodynamic response.  This decoupling is achieved, for example, with slender body theory \citep{tornberg_simulating_2004}, as well as through a modified version of the method of regularised Stokeslets \citep{cortez2018regularized}.  With this drawback in mind, we designed the overarching computational framework to be flexible and readily accommodate changes to the hydrodynamic model provided it satisfies \cref{eq:mobility-problem}.

\subsubsection{RPY mobility matrix}\label{sec:RPY}
In one approach, we perform a direct pairwise evaluation of the RPY tensor \citep{wajnryb_generalization_2013} to compute the mobility matrix--vector multiplication.  The operation count for this computation scales like the number of segments squared, though more sophisticated, faster techniques \citep{liang_fast_2013} could also be used.  With RPY, and given a collection of $N$ non-overlapping particles of radius $a$ and separations of $\v{r}_{nm} = \v{Y}_n - \v{Y}_m,$ $r_{nm} = \|\v{r}_{nm}\| > 2a$, in a fluid with viscosity $\eta$, the velocities of particle $n$ are given by
\begin{align}
    \begin{pmatrix}
	\v{V}_n\\\v{\Omega}_n
	\end{pmatrix} = \sum_{m=1}^N \begin{pmatrix} \t{M}^{tt}_{nm} & \t{M}^{tr}_{nm} \\ \t{M}^{rt}_{nm} & \t{M}^{rr}_{nm} \end{pmatrix} \begin{pmatrix}
	\v{F}_m\\\v{T}_m
	\end{pmatrix}.
\end{align}
For an unbounded fluid, the sub-tensors are defined as
\begin{align}
    \t{M}^{tt}_{nm} &= \twopartdef{\frac{1}{6\pi\eta a}\t{I}}{n = m,}{\frac{1}{8\pi\eta r_{nm}}\left(\left(1 + \frac{2a^2}{3r_{nm}^2}\right)\t{I} + \left(1 - \frac{2a^2}{r_{nm}^2}\right)\uv{r}_{nm}\uv{r}_{nm}\right)}{n\neq m,} \\
    \t{M}^{rr}_{nm} &= \twopartdef{\frac{1}{8\pi\eta a^3}\t{I}}{n = m,}{\frac{1}{16\pi\eta r_{nm}^3}\left(3\uv{r}_{nm}\uv{r}_{nm} - \t{I}\right)}{n\neq m,} \\
    \t{M}^{tr}_{nm} = \t{M}^{rt}_{nm} &= \twopartdef{\tu{0}}{n = m,}{\frac{1}{8\pi\eta r_{nm}^2}\t{\varepsilon}\cdot\uv{r}_{nm}}{n\neq m,}
\end{align}
where $\uv{r}_{nm} = \v{r}_{nm}/r_{nm}$, $\t{I}$ is the identity matrix, and $\t{\varepsilon}$ is the three-dimensional Levi-Civita symbol.  We note that there are extensions and variants of these tensors which allow for particles of different sizes, overlapping particles, the inclusion of background shear flows, and the presence of no-slip boundaries \citep{wajnryb_generalization_2013,zuk_rotneprageryamakawa_2014,swan_simulation_2007}.

\subsubsection{Force-coupling method}\label{sec:fcm}
We also present computations where the mobility matrix--vector multiplication is performed using the matrix-free FCM \citep{maxey_localized_2001, lomholt_force-coupling_2003, liu_force-coupling_2009}.  With FCM, the forces and torques the segments exert on the fluid are transferred to the fluid through a truncated and regularised force multipole expansion in the Stokes equations,
\begin{align}
 -\vnabla p + \eta \nabla^2\v{u} + \sum_{n} \v{F}^H_n \mathit{\Delta}_n(\v{x}) %\\
  - \frac{1}{2}\sum_{n} \v{T}^H_n \times \vnabla \Theta_n(\v{x}) &= \vu{0}, \label{eqn:stokes} \\
\vnabla\cdot \v{u} &= 0,
\end{align}
where the sums are over all segments, $\eta$ is the fluid viscosity, $\v{u}$ is the fluid velocity and $p$ is the pressure.  The Dirac delta functions in the multipole expansions are replaced by the Gaussians,
\begin{align}
\mathit{\Delta}_n(\v{x}) &=  \left(2\pi \sigma_{\mathit{\Delta}}^2\right)^{-3/2}\exp \left( -\frac{\left\|\v{x} - \v{Y}_n\right\|^2}{2\sigma_{\mathit{\Delta}}^2}\right),\\
\Theta_n(\v{x}) &= \left(2\pi \sigma_{\Theta}^2\right)^{-3/2}\exp \left( -\frac{\left\|\v{x} - \v{Y}_n\right\|^2}{2\sigma_{\Theta}^2}\right).
\end{align}
Taking advantage of the ratios established for spherical particles \citep{maxey_localized_2001,lomholt_force-coupling_2003}, we take the Gaussian envelope sizes to be related to the filament thickness through $ \sigma_{\mathit{\Delta}} = a/\sqrt{\pi}$ and $\sigma_{\Theta} = a/(6\sqrt{\pi})^{1/3}$.  In \cref{eqn:stokes}, we have ignored the contribution of the stresslets that would enforce a vanishing rate of strain within the segments \citep{lomholt_force-coupling_2003}.  The stresslets can also be included at the cost of several conjugate gradient iterations \citep{yeo_simulation_2010} each time the segment velocities and angular velocities are determined.

After the fluid velocity is found, it is spatially averaged using the same Gaussian functions to obtain the velocity $\v{V}_n$ and angular velocity $\v{\Omega}_n$ of each segment $n$.  Specifically, we have that
 \begin{align}
\v{V}_n &= \int \v{u}\,\mathit{\Delta}_n(\v{x}) \mathop{\d^3\v{x}}, \label{eq:segvel} \\
\v{\Omega}_n &= \frac{1}{2}\int (\vnabla \times \v{u})\, \Theta_n(\v{x}) \mathop{\d^3\v{x}}. \label{eq:segangvel}
\end{align}
Following \citet{yeo_simulation_2010,keaveny_fluctuating_2014}, in our computations, the regularised forcing is first evaluated on a regular grid.  The Stokes equations subject to periodic boundary conditions are then solved using a Fourier spectral method.  Finally, the trapezoidal rule is used to integrate numerically \cref{eq:segvel,eq:segangvel} to obtain the translational and angular velocity for each segment.

\subsection{Planar filament deformation and motion}

In order to keep track of filament deformation and motion, as well as compute the internal moments and forces, we need to describe how the local frame, $\{\uv{t}(s,t),\uv{\ny}(s,t),\uv{\nz}(s,t)\}$, at each point along the filament rotates with time.

In many important problems, such as sperm locomotion through planar flagellar beats \citep{schoeller_flagellar_2018,yang_cooperation_2008}, filament deformation and motion are restricted to a single plane and only bending occurs.  Due to the absence of both elastic and preferred twist, the local frame at each point along the filament will rotate about a single, fixed direction.  Without loss of generality, we can take this direction to be aligned with the fixed vector $\uv{e}_z$, as well as $\uv{\nz} = \uv{e}_z$ for all $s$ and $t$.  This direction is always normal to the plane of motion.

\subsubsection{Computing the internal moments from the angles}\label{sec:internal-moments-2d}
As a result of taking the out-of-plane direction as $\uv{\nz} = \uv{e}_z$, we have that $\kappa_\ny = 0$ in \cref{equation:3d_moment_frame}.  The internal moment is then purely the result of bending and is given by
\begin{align} \label{eqn:continuous2Dmoments}
  	\v{M}  = K_B \uv{t} \times \pd{\uv{t}}{s} - K_B \kappa_\nz \uv{e}_z,
  \end{align}
with the understanding that since $\uv{t}$ lies in the same plane for all $s$, $\uv{t} \times \pd{\uv{t}}{s}$ will be in the direction $\uv{e}_z$ for all $s$ and $t$. In the discretised model, the elastic moments are then given by
\begin{subequations}
\begin{align}
        	\v{M}_{n+1/2} &= K_B \left(\uv{t}_n \times \left( \frac{\uv{t}_{n+1} - \uv{t}_{n}}{\Delta L}\right) - \kappa_\nz(s_{n+1},t)\uv{e}_z  \right) \\
            &= K_B \left(\frac{1}{\Delta L} \left(\uv{t}_n \times \uv{t}_{n+1}\right) - \kappa_\nz(s_{n+1},t) \uv{e}_z \right). \label{eqn:discrete2Dmoments}
\end{align}
\end{subequations}
where again it is understood that $\uv{t}_n \times \uv{t}_{n+1}$ is in the direction $\uv{e}_z$ for all $n$ and $t$.  The restriction to a single plane also allows for $\uv{t}_n$ to be conveniently expressed in terms of an angle $\theta_n(t)$ such that $\uv{t}_n = (\cos\theta_n, \sin\theta_n,0)$.

With this expression for $\v{M}_{n+1/2}$ found, we now have expressions for all the forces and torques on the segments appearing in the discrete force and torque balances, \cref{eqn:DiscreteForce,eqn:DiscreteTorque}.

\subsubsection{Differential-algebraic system and time integration in two dimensions}\label{2d-system-of-equations}
With expressions for forces $\v{F}^C_n$ and $\v{F}^B_n$, and torques $\v{T}^C_n$ and $\v{T}^E_n$, we now move on to integrating the system in time. The mobility problem in \cref{eq:mobility-problem} expresses the velocities, $\v{V}_n$, and angular velocities, $\v{\Omega}_n$, for each segment $n$, in terms of these aforementioned forces and torques through
\begin{equation}\label{eq:mobility-problem-timeint}
	\begin{pmatrix}
	\v{V}\\\v{\Omega}
	\end{pmatrix}
    = \mathcal{M}\cdot
    \begin{pmatrix}
	\v{F}^C + \v{F}^B\\ \v{T}^C + \v{T}^E
	\end{pmatrix}.
\end{equation}
Due to the motion being restricted to the plane, we have $\v{V}_n \cdot \uv{e}_z = 0$ and $\v{\Omega}_n = (0,0,\Omega_n)$ for each $n$.  We emphasise that we cannot simply evaluate \cref{eq:mobility-problem-timeint} to obtain the segment motion since the values of $\v{\Lambda}_{n+1/2}$ appearing in $\v{F}^C_n$ (see \cref{eqn:constraint-force}) are unknown and are obtained by applying the kinematic constraint, \cref{equation:inex_constraints}, for each $n$.  Thus, the segment positions and angles can be updated by integrating numerically in time the differential-algebraic system of equations,
    \begin{align}
    	\fd{\v{Y}_n}{t} &= \v{V}_n, \\
        \fd{\theta_n}{t} &= \Omega_n, \\
        \v{Y}_{n+1} - \v{Y}_n - \frac{\Delta L}{2}\left( \uv{t}_n + \uv{t}_{n+1}\right) &= \vu{0}.
    \end{align}
Due to the numerical stiffness associated with the overdamped dynamics of elastic beams \citep{powers_dynamics_2010,nazockdast_fast_2017}, as well as the need to ensure that the constraints are satisfied after advancing in time, we discretise in time using the implicit second-order backwards differential formula (BDF) \citep{ascher_computer_1998} and impose the constraint at the updated positions and orientations (time $j+1$), such that
    \begin{subequations}\label{eqn:2d-system}
    \begin{align}
    	\v{Y}_n^{j+1} -\frac{4}{3} \v{Y}_n^{j} + \frac{1}{3} \v{Y}_n^{j-1} -\frac{2}{3} \Delta t \, \v{V}_n^{j+1} &= \vu{0}, \label{eqn:2d-system-1}\\
        \theta_n^{j+1} -\frac{4}{3} \theta_n^{j} + \frac{1}{3} \theta_n^{j-1} -\frac{2}{3} \Delta t \, \Omega_n^{j+1} &= 0, \label{eqn:2d-system-2}\\
        \v{Y}_{n+1}^{j+1} - \v{Y}_n^{j+1} - \frac{\Delta L}{2}\left( \uv{t}_n^{j+1} + \uv{t}_{n+1}^{j+1} \right) &= \vu{0},\label{eqn:2d-system-3}
    \end{align}%
    \end{subequations}
where $\Delta t$ is the timestep size and $\v{V}_n^{j+1}$ and $\Omega_n^{j+1}$ are given by the mobility problem, \cref{eq:mobility-problem-timeint}, with the right-hand side evaluated at $j+1$ using $\v{Y}_n^{j+1}$, $\theta_n^{j+1}$, and $\v{\Lambda}^{j+1}_{n+1/2}$. This yields a nonlinear system of equations for the positions, $\v{Y}_{n+1}^{j+1}$, and angles, $\theta_n^{j+1}$, as well as the internal force Lagrange multipliers, $\v{\Lambda}^{j+1}_{n+1/2}$, associated with each constraint.  If $M$ is the number of filaments and $N$ is the number of segments per filament, the position and orientation updates, \cref{eqn:2d-system-1,eqn:2d-system-2}, and the constraints, \cref{eqn:2d-system-3}, constitute a system of $M(5N - 2)$ nonlinear equations.

Before solving this system numerically, we first reduce the system size by substituting the constraint, \cref{eqn:2d-system-3}, into the position updates, \cref{eqn:2d-system-1}, as to reduce the number of degrees of freedom to those of a so-called robot arm whose motion is completely described by a single position and the orientations of each link comprising the arm.  Specifically, in the case of a single filament, we replace $\v{Y}^{j+1}_n$ for $n>1$ by
\begin{align} \label{eqn:positionSubstitution}
	\v{Y}_n^{j+1} &= \v{Y}_1^{j+1} + \frac{\Delta L}{2}\sum_{m = 2}^n \left(\uv{t}_{m-1}^{j+1} + \uv{t}_m^{j+1}\right)
\end{align}
to obtain the new, reduced system of equations,
\begin{subequations}\label{eqn:robotArm}\label{eqn:RA-2d-system}
\begin{align}
    \v{Y}_1^{j+1} -\frac{4}{3} \v{Y}_1^{j} + \frac{1}{3} \v{Y}_1^{j-1} -\frac{2}{3} \Delta t \, \v{V}_1^{j+1} &= \vu{0}, \label{eqn:RA-2d-system-1}\\
    \theta_n^{j+1} -\frac{4}{3} \theta_n^{j} + \frac{1}{3} \theta_n^{j-1} -\frac{2}{3} \Delta t \, \Omega_n^{j+1} &= 0, \label{eqn:RA-2d-system-2} \\
	 \v{Y}_1^{j+1} + \frac{\Delta L}{2}\sum_{m = 2}^n \left(\uv{t}_{m-1}^{j+1} + \uv{t}_m^{j+1}\right) - \frac{4}{3}\v{Y}_n^{j} + \frac{1}{3} \v{Y}_n^{j-1}   -\frac{2}{3} \Delta t \, \v{V}^{j+1}_n &= \vu{0}. \label{eqn:RA-2d-system-3}
\end{align}
\end{subequations}
This system may be expressed as $\v{f}(\v{X}^*) = \vu{0}$, where the solution $\v{X}^*$ contains the updated position for the first segment, all orientation angles and the Lagrange multipliers.  For $M$ filaments each with $N$ segments, as a result of this substitution, the dimension of the nonlinear system is reduced from $M(5N-2)$ to $3MN$.

\subsection{Three-dimensional filament deformation and motion}
In general, there is no restriction about how the filament can bend or twist, and the local frame along the filament can rotate about any axis.  Rather than considering the local frame vectors explicitly, we keep track of these rotations using the unit quaternions that map the standard basis to the local frame at each point along the filament.  Unit quaternions allow for successive rotations to be computed easily while requiring less storage and fewer floating point operations than rotation matrices.  Quaternions also avoid gimbal lock that is typically experienced with Euler angles \citep{allen_computer_2017}.  Before describing how we employ quaternions in our methodology, we provide a brief overview of representing rotations using quaternions.

\subsubsection{Representing rotations as quaternions}

Quaternions \citep{allen_computer_2017,dunn_3d_2011,vince_quaternions_2011} can be viewed as an extension of the complex numbers in that they have one real part and three imaginary components.  As such, the quaternions inherit the notion of conjugation from the complex numbers, $\q{q}^* = (q_0,-\v{q})$, and the norm of a quaternion is the Euclidean norm $\left\|\q{q}\right\|^2 = q_0^2 + \left\|\v{q}\right\|^2 = q_0^2 + q_1^2 + q_2^2 + q_3^2$.  Quaternions are also frequently identified with elements of $\Reals^4$ as
\begin{align}
    \q{q} = \left(q_0,q_1,q_2,q_3\right) = \left(q_0,\v{q}\right).
\end{align}
We use the operation $[(q_0,\v{q})]_{\Reals^3} = \v{q}$ to extract the vector in $\Reals^3$ constructed from the last three entries of a quaternion.

Quaternions are subject to standard element-wise addition, real multiplication by a scalar, and the associative, non-commutative product,
\begin{subequations}
\begin{align}
    \q{p}\qprod\q{q} &= \left(p_0,\v{p}\right)\qprod\left(q_0,\v{q}\right)\\
    &= \left(p_0q_0 - \v{p} \cdot \v{q},\; p_0\v{q} + q_0\v{p} + \v{p} \times \v{q}\right).
\end{align}
\end{subequations}
Under this product, the unit quaternions, i.e.\ quaternions with $\left\|\q{q}\right\|^2 = 1$, form a group with identity $\q{I}_q = (1,\vu{0})$ and inverse $\q{q}^{-1} = \q{q}^*$.

As unit quaternions can be written as
\begin{align} \label{equation:quaternion_polar_form}
    \q{q} = \left(\cos\left(\frac{\theta}{2}\right), \sin\left(\frac{\theta}{2}\right)\uv{v}\right),
\end{align}
they can be identified with spatial rotations of an angle, $\theta$, anticlockwise about the unit vector, $\uv{v}$.  Specifically, the rotation applied to a vector $\v{w}$ is given by
\begin{align}
	\left(0,\v{w}'\right) = \q{q} \qprod \left(0,\v{w}\right) \qprod \q{q}^*,
\end{align}
where $\v{w}'$ is the image of $\v{w}$ after its rotation.  This rotation can also be written as
\begin{align}
	\v{w}' = \t{R}\left(\q{q}\right)\v{w},
\end{align}
where $\t{R}(\q{q})$ is the rotation matrix whose entries are related to those of the quaternions through
\begin{align} \label{equation:quaternion_rotation_matrix}
    \t{R}\left(\q{q}\right) = \begin{pmatrix} 1 - 2q^{2}_{2} - 2q^{2}_{3} & 2\left(q_{1}q_{2} - q_{3}q_{0}\right) & 2\left(q_{1}q_{3} + q_{2}q_{0}\right) \\
2\left(q_{1}q_{2} + q_{3}q_{0}\right) & 1 - 2q^{2}_{1} - 2q^{2}_{3} & 2\left(q_{3}q_{2} - q_{1}q_{0}\right) \\
2\left(q_{1}q_{3} - q_{2}q_{0}\right) & 2\left(q_{3}q_{2} + q_{1}q_{0}\right) & 1 - 2q^{2}_{2} - 2q^{2}_{1}
\end{pmatrix}.
\end{align}
For successive rotations first by $\q{q}$, say, followed by $\q{p}$, we have that $\t{R}(\q{p}\qprod \q{q}) = \t{R}(\q{p})\t{R}(\q{q})$.

%\end{enumerate}

\subsubsection{Computing the internal moments from the quaternions}\label{sec:internal-moments-3d}
As we use quaternions to describe how the local frame rotates as the filament bends and twists, it is convenient to have an expression for the internal moments in terms of the quaternions themselves.

Specifically, the unit quaternions provide the rotation of the standard basis to the local frame vectors, and we have that at each point along the filament and at each time,
\begin{align}\label{eq:RotToFrame}
    \t{R}\left(\q{q}\left(s,t\right)\right) = \left(\uv{\nx}\left(s,t\right)\;\; \uv{\ny}\left(s,t\right)\;\; \uv{\nz}\left(s,t\right)\right).
\end{align}
In \cref{appendix:initial_quaternion}, we describe how to obtain the quaternion that satisfies this condition at $t=0$.

Using \cref{eq:RotToFrame} and the expression for the internal moments in terms of the frame vectors, \cref{equation:3d_moment_frame}, the internal moments can be expressed as
\begin{align}
\v{M} = \t{R} \left( \q{q} \right) \t{D} \left( \left[ 2\q{q}^*\qprod\pd{\q{q}}{s}\right]_{\Reals^3} - \left(\begin{matrix}
\gamma_0 \\ \kappa_\ny \\ \kappa_\nz
\end{matrix}\right) \right),
\end{align}
where $\t{D} = \diag(K_T,K_B,K_B)$.  The details of this derivation are presented in \cref{appendix:quaternion_elasticity}.  This expression is a special case of more general constitutive laws discussed in \citet{zupan_quaternion-based_2009}.

For the discretised system, the internal moments are then given by
\begin{align}\label{eq:quaternionM}
\v{M}_{n+1/2} = \t{R} \left( \q{q}_{n+1/2} \right) \t{D} \left( 2\left[ \q{q}_{n+1/2}^*\qprod\left(\frac{\q{q}_{n+1} - \q{q}_n}{\Delta L}\right)\right]_{\Reals^3} - \left(\begin{matrix}
\gamma_0 \\ \kappa_\ny \\ \kappa_\nz
\end{matrix}\right) \right),
\end{align}
where the interpolated quaternion, $\q{q}_{n+1/2}$, is constructed by performing half of the rotation from $\q{q}_n$ to $\q{q}_{n+1}$.  Specifically, we have $\q{q}_{n+1/2} = (\q{q}_{n+1}\qprod\q{q}_{n}^*)^{1/2}\qprod\q{q}_n$ where the square root is defined in \cref{appendix:initial_quaternion}.

\subsubsection{Differential-algebraic system and time integration in three dimensions}\label{3d-system-of-equations}

With the translational and angular velocities of the segments given by the mobility problem, \cref{eq:mobility-problem-timeint}, but with \cref{eq:quaternionM} used to compute $\v{T}^E$, the positions and quaternions are updated by integrating the differential-algebraic system,
\begin{align}
\fd{\v{Y}_n}{t} &= \v{V}_n, \label{equation:position_ode} \\
\fd{\q{q}_n}{t} &= \frac{1}{2}\left(0,\v{\Omega}_n\right)\qprod\q{q}_n,\label{equation:quaternion_ode}\\
\v{Y}_{n+1} - \v{Y}_n - \frac{\Delta L}{2}\left(\uv{t}_n + \uv{t}_{n+1}\right) &= \vu{0}.
\end{align}
The differential equation \cref{equation:quaternion_ode} for the quaternions describes the time evolution of the entire frame of segment $n$ such that taking $\uv{v}(t) = \t{R}(\q{q}_n(t))\uv{v}(0)$ for a vector $\uv{v}$ is equivalent to (see \cref{appendix:quaternion_ode}) integrating in time the perhaps more familiar expression
\begin{equation}
    \fd{ \uv{v} }{t} = \v{\Omega}_n\times \uv{v}.%
    % I need to refer to this later - sorry for making it displaystyle -- Adam
    \label{equation:angular-velocity}
\end{equation}

As when filament motion and deformation were restricted to a plane, we can update the positions using the second-order BDF scheme.  While we would like to update the quaternions using a similar scheme, we must also ensure that after each update they continue to have unit norm as to continue to represent rotations.  This could be accomplished by simply applying the second-order BDF scheme to \cref{equation:quaternion_ode} and subsequently normalising the result.  We have found in practice, however, that this approach interfaces poorly with the quasi-Newton methods we use to solve the nonlinear system for the segment positions, orientations and Lagrange multipliers as it introduces timestep restrictions for numerical stability.  Instead, we apply a geometric multi-step method \citep{faltinsen_multistep_2001} to perform the multiplicative update
\begin{align} \label{equation:general_multiplicative_q_update}
    \q{q}^{j+1}_n = \q{p}^{j+1}_n\qprod\q{q}^{j}_n,
\end{align}
for some appropriate unit quaternion $\q{p}^{j+1}_n$.  This approach preserves the norm of the quaternion to machine precision and, in practice, leads to better numerical stability properties of the larger numerical method.

Applying the general framework set out in \citet{faltinsen_multistep_2001} to our specific case, we have that for sufficiently small $\Delta t$,  \cref{equation:quaternion_ode} has solution
\begin{align}
\q{q}\left(t_0 + \Delta t\right) = \exp\left(\v{u}\left(t_0 + \Delta t, \q{q}\left(t_0 + \Delta t\right)\right)\right)\qprod\q{q}\left(t_0\right),
\end{align}
where the vector $\v{u}$ is an element of the Lie algebra (in this case $\mathfrak{so}(3)$) and the exponential map \citep{iserles_lie-group_2000} is given by
\begin{align}
\exp\left(\v{u}\right) = \left(\cos\left(\frac{\|\v{u}\|}{2}\right), \sin\left(\frac{\|\v{u}\|}{2}\right)\frac{\v{u}}{\|\v{u}\|}\right).
\end{align}
The Lie algebra element itself satisfies the differential equation
\begin{align} \label{equation:lie_algebra_ode}
\fd{\v{u}}{t} = \dexpinv_{\v{u}}\left(\v{\Omega}\right),
\end{align}
with $\v{u}(t_0) = \vu{0}$.  The function $\dexpinv_{\v{u}}$ is the inverse of the differential of the exponential mapping and is given by \citep{iserles_lie-group_2000}
\begin{align} \label{eq:dexpinv_definition}
\dexpinv_{\v{u}}\left(\v{\Omega}\right) = \v{\Omega} - \frac{1}{2}\v{u}\times\v{\Omega} - \frac{1}{2\|\v{u}\|^2}\left(\|\v{u}\|\cot\left(\frac{\|\v{u}\|}{2}\right) - 2\right)\v{u}\times\left(\v{u}\times\v{\Omega}\right).
\end{align}
In a general sense, the Lie algebra element can be interpreted as the integral of the angular velocity over time.  This holds true if the rotation is about a single axis.  For example, if $\v{\Omega} \equiv \alpha \uv{e}_z$ for a constant $\alpha$, then \cref{equation:lie_algebra_ode} with initial condition $\q{q}(0) = \q{I}_q$ gives $\v{u}(t) = \alpha t \uv{e}_z$ and we obtain $\q{q}(t) = (\cos(\alpha t/2),0,0,\sin(\alpha t/2))$.

By introducing \cref{equation:lie_algebra_ode}, the problem of updating the unit quaternions is transferred to one of updating the Lie algebra elements.  This can be done using a standard, additive scheme with the only caveat being that since the previously updated Lie algebra elements are in the tangent spaces of the previous quaternions, we must also re-centre our coordinate system at the current quaternion when using a multi-step method \citep{faltinsen_multistep_2001}.  We show in \cref{appendix:lie_algebra_bdf2} that re-centring this yields the simpler second-order BDF scheme
\begin{align} \label{equation:lie-algebra-update}
\v{u}^{j+1}_{n} = \frac{1}{3}\v{u}^{j}_{n} + \frac{2}{3} \Delta t \dexpinv_{\v{u}^{j+1}_{n}}\left(\v{\Omega}_{n}^{j+1}\right),
\end{align}
for the Lie algebra element of segment $n$.

Combining the Lie algebra element update with that for the positions, as well as the constraints linking the segment orientations and positions, we obtain the system of equations
\begin{subequations} \label{equation:3d-nonlinear-system}
\begin{align}
\v{Y}_n^{j+1} - \frac{4}{3}\v{Y}_n^j + \frac{1}{3}\v{Y}_n^{j-1} - \frac{2 \Delta t}{3}\v{V}_n^{j+1} &= \vu{0},\label{equation:3d-nonlinear-system-1}
\\
\v{u}^{j+1}_{n} - \frac{1}{3}\v{u}^{j}_{n} - \frac{2}{3} \Delta t \dexpinv_{\v{u}^{j+1}_{n}}\left(\v{\Omega}_{n}^{j+1}\right) &= \vu{0},\label{equation:3d-nonlinear-system-2}
\\
\v{Y}_{n+1}^{j+1} - \v{Y}_n^{j+1} - \frac{\Delta L}{2}\left( \uv{t}_n^{j+1} + \uv{t}_{n+1}^{j+1} \right) &= \vu{0},\label{equation:3d-nonlinear-system-3}
\end{align}
\end{subequations}
for the updated segment positions, Lie algebra elements, and Lagrange multipliers. As in 2D, $\v{V}_n^{j+1}$ and $\v{\Omega}_{n}^{j+1}$ are given by the mobility problem, \cref{eq:mobility-problem-timeint}, with the right-hand side evaluated at $j+1$. For $M$ filaments, each discretised into $N$ segments, this is an $M(9N-3)$ system of nonlinear equations.

As done in the case of planar filament deformations, we reduce the system size before seeking a numerical solution by substituting
\begin{align} \label{eqn:positionSubstitution-3d}
	\v{Y}_n^{j+1} &= \v{Y}_1^{j+1} + \frac{\Delta L}{2}\sum_{m = 2}^n \left(\uv{t}_{m-1}^{j+1} + \uv{t}_m^{j+1}\right)
\end{align}%
into the position updates.  This yields the new system of equations
\begin{subequations}\label{eqn:RA-3d-system}
\begin{align}
    \v{Y}_1^{j+1} -\frac{4}{3} \v{Y}_1^{j} + \frac{1}{3} \v{Y}_1^{j-1} -\frac{2}{3} \Delta t \,\v{V}_1^{j+1} &= \vu{0}, \label{eqn:RA-3d-system-1}\\
    \v{u}^{j+1}_{n} - \frac{1}{3}\v{u}^{j}_{n} - \frac{2}{3} \Delta t \dexpinv_{\v{u}^{j+1}_{n}}\left(\v{\Omega}_{n}^{j+1}\right) &= \vu{0}, \label{eqn:RA-3d-system-2} \\
	 \v{Y}_1^{j+1} + \frac{\Delta L}{2}\sum_{m = 2}^n \left(\uv{t}_{m-1}^{j+1} + \uv{t}_m^{j+1}\right) - \frac{4}{3}\v{Y}_n^{j} + \frac{1}{3} \v{Y}_n^{j-1}   -\frac{2}{3} \Delta t \, \v{V}^{j+1}_n &= \vu{0}. \label{eqn:RA-3d-system-3}
\end{align}
\end{subequations}
As done in the planar case, we may again express this system as $\v{f}(\v{X}^*) = \vu{0}$, where the solution $\v{X}^*$ contains the updated position for the first segment, all the Lie algebra elements and the Lagrange multipliers.  This substitution reduces the dimension of the nonlinear system from $M(9N-3)$ to $6MN$.

\section{Numerical solvers for the nonlinear systems}\label{sec:jfb}

As described in the previous section, updating the filament positions, orientations, and the Lagrange multipliers requires finding a solution $\v{X}^*$ of the nonlinear system of equations of the form $\v{f}(\v{X}^*)=\vu{0}$.  This system is given by \cref{eqn:RA-2d-system} for planar filament deformations and \cref{eqn:RA-3d-system} in the fully three-dimensional case.

To solve the system, one would typically apply Newton's method.  Starting with $\v{X}_0$ as an initial guess of the solution, the solution is computed iteratively from
\begin{equation}
    \v{X}_{k+1} = \v{X}_k - \t{J}^{-1}(\v{X}_k) \v{f}(\v{X}_k),
    \label{equation:newton-iteration}
\end{equation}
where $\t{J}$ is the Jacobian of the system, $\t{J} = \vnabla_{\v{X}}\v{f}$ (i.e. $J_{ij}(\v{X}) = \p f_i(\v{X})/\p X_j$).  The process is continued until $\|\v{f}(\v{X})\|$ is smaller than a given tolerance.

For the systems given by \cref{eqn:RA-2d-system,eqn:RA-3d-system}, the Jacobian $\t{J}$ will contain the mobility matrix, $\mathcal{M}$ (see \cref{eq:mobility-problem}), as well as its derivative with respect to segment positions. In general, when one uses fast, matrix-free methods, the mobility matrix is not known explicitly.  Even if the mobility matrix was known, the Jacobian would still be very complicated and challenging to compute, especially given the dependence of the mobility matrix on particle positions.  Additionally, storing the Jacobian would incur significant memory costs when the number of filament segments is large.

To avoid the costs and complications associated with standard Newton's method, we explore how solutions can be obtained using Jacobian-free Newton--Krylov and Broyden's methods.  The remainder of this section provides a description of these methods and, in particular, highlights the advantages and drawbacks associated with these approaches when applied to our specific computation.  We also provide results from a series of tests comparing these methods and show that Broyden's method with a suitable approximate Jacobian provides the most effective approach to finding the solution.

\subsection{Jacobian-free Newton Krylov (JFNK)}

The Jacobian-free Newton--Krylov (JFNK) method \citep{knoll_jacobian-free_2004} uses Newton's method to find the solution of the nonlinear system, but avoids an explicit computation of the Jacobian through judicious evaluations of the function, $\v{f}$.  With this method, the linear system of the form, $\t{J}(\v{X}_k) \v{x} = \v{f}(\v{X}_k)$ for unknown $\v{x}$, that arises during each Newton iteration is solved using a Krylov subspace method, typically GMRES \citep{saad_gmres_1986}.  Within GMRES, the Jacobian matrix--vector multiplication required at each iteration is approximated by the finite difference formula,
 \begin{align}
    	\t{J}\v{v} = \frac{\v{f}(\v{X}+\delta \v{v}) - \v{f}(\v{X})}{\delta} + \order(\delta).
        \label{equation:JFD}
    \end{align}
Since in JFNK multiplication by the Jacobian is replaced by \cref{equation:JFD}, each GMRES iteration requires evaluating $\v{f}(\v{X})$.

In our case, this involves evaluating \cref{eqn:RA-2d-system} or \cref{eqn:RA-3d-system} and computing the translational and angular velocities of each segment, which itself requires performing a multiplication by the mobility matrix, the most computational costly aspect of the simulation.  Thus, although JFNK does not require explicit knowledge of the Jacobian, it could still incur high computational costs if GMRES is slow to converge and many GMRES iterations are required for each Newton iteration.  As a result, and as is often the case with GMRES, preconditioning may be required to limit iteration counts and to ensure good convergence rates.

\subsection{Broyden's method}

Another approach, and the one that we will ultimately adopt, to obtain the solution of the nonlinear system is Broyden's quasi-Newton method \citep{broyden_class_1965}.  Like the traditional Newton's method, Broyden's method finds the solution iteratively, but instead uses an initial approximate Jacobian, $\t{J}_0$, that is improved at each iteration through rank-one updates based on the secant equation.  The choice of update, however, is not unique and the so-called ``good'' Broyden's method updates the Jacobian itself (see \cref{alglin:update-good} in \cref{alg:good-broydens-method}), while the so-called ``bad'' Broyden's method updates its inverse (\cref{alglin:update-bad} in \cref{alg:bad-broydens-method}).  In both cases, we can employ a limited-memory version \citep{van_de_rotten_limited_2003} of Broyden's method where the updates are stored as vectors rather than altering and writing to memory the full Jacobian at each iteration.  Thus, the only matrix that may require storage is $\t{J}_0$, which can be chosen with this memory requirement in mind.

\begin{algorithm}
    Take initial guess of solution $\v{X}_0$, convergence tolerance $\varepsilon>0$, estimate $\t{J}_0$ for the Jacobian at $\v{X}_0$\\
    $k=0$\\
    \While{$\|\v{f}(\v{X}_k)\| > \varepsilon$}{
    $\Delta\v{X}_k = - \Big[\t{J}_0 + \sum_{i=1}^{k}\v{c}_i\v{d}_i^\top\Big]^{-1} \v{f}(\v{X}_k)$\label{alglin:update-good}\\
    $\v{X}_{k+1} = \v{X}_k + \Delta\v{X}_k$\\
    $\v{c}_{k+1} = \v{f}(\v{X}_{k+1})/\|\Delta \v{X}_k\|$, $\v{d}_{k+1} = \Delta \v{X}_k /\|\Delta \v{X}_k\|$\\
    $k \coloneqq k+1$\\
    }
    \Return $\v{X}_k$
    \caption{Limited-memory, good Broyden's method, to solve $\v{f}(\v{X}) = \vu{0}$}
    \label{alg:good-broydens-method}
    \end{algorithm}
\begin{algorithm}
    Take initial guess of solution $\v{X}_0$, convergence tolerance $\varepsilon>0$, estimate $\t{J}_0$ for the Jacobian at $\v{X}_0$\\
    $k=0$\\
    Evaluate $\t{J}^{-1}_0$\\
    \While{$\|\v{f}(\v{X}_k)\| > \varepsilon$}{
    $\Delta\v{X}_k = -\Big[ \t{J}^{-1}_0 + \sum_{i=1}^{k}\v{c}_i\v{d}_i^\top \Big] \v{f}(\v{X}_k)$ \label{alglin:update-bad}\\
    $\v{X}_{k+1} = \v{X}_k + \Delta\v{X}_k$\\
    $\v{c}_{k+1} = -\Big[ \t{J}^{-1}_0 + \sum_{i=1}^{k}\v{c}_i\v{d}_i^\top \Big] \v{f}(\v{X}_{k+1})      /\|\Delta \v{f}_k\|$, $\v{d}_{k+1} = \Delta \v{f}_k/\|\Delta \v{f}_k\|$, where $\Delta \v{f}_k = \v{f}(\v{X}_{k+1}) - \v{f}(\v{X}_k)$ \label{alglin:bad-broydens-method-find-f} \\
    $k \coloneqq k+1$\\
    }
    \Return $\v{X}_k$
    \caption{Limited-memory, bad Broyden's method, to solve $\v{f}(\v{X}) = \vu{0}$}
    \label{alg:bad-broydens-method}
    \end{algorithm}

For Broyden's method to converge, not only must the convergence conditions for Newton's method be satisfied, but also the initial guess for the Jacobian, $\t{J}_0$, must be sufficiently close to $\t{J}$.  The process of choosing $\t{J}_0$ is similar to that of finding a suitable preconditioner in that one seeks $\t{J}_0 \approx \t{J}$ with the condition that systems involving $\t{J}_0$ should be easily solved.  Below we present two approaches that we have explored to obtain approximate Jacobians based on simplified, yet related, physical systems.

\subsubsection{Jacobian-free Broyden's method (JFB)}
Our first approach involves applying the JFNK-style finite differencing, \cref{equation:JFD}, to a function, $\v{f}_0(\v{X}) \approx \v{f}(\v{X})$, to perform the matrix--vector multiplications by $\t{J}_0 = \vnabla_{\v{X}}\v{f}_0$ and solve the systems appearing in \cref{alg:good-broydens-method} or \cref{alg:bad-broydens-method} using GMRES.  As this combines a Jacobian-free approach with Broyden's method, we refer to this approach as Jacobian-free Broyden's (JFB) method.

In the tests presented below, the function $\v{f}_0$ involves all the steps of the model presented in \cref{sec:filament-model}, however, we omit both the hydrodynamic interactions between filament segments by replacing the mobility matrix $\mathcal{M}$ by $\mathcal{M}_0 = \diag(\mathcal{M})$ and remove the steric interactions between filament segments.  In this way, the computationally expensive mobility matrix multiplication is only performed once per Broyden iteration.

\subsubsection{A block diagonal, explicit \texorpdfstring{$\t{J}_0$}{J\_0} (EJB)}\label{sec:block-diagonal-j0}
Along with JFB, we also construct explicitly the approximate Jacobian based on $\v{f}_0(\v{X})$ with a diagonal mobility matrix and no steric interactions between filaments.  By differentiating $\v{f}_0(\v{X})$, we obtain expressions for the entries of $\t{J}_0$ which we evaluate at each timestep.  These expressions are provided in \cref{sec:form-of-jacobian}.

In general, for simulations involving $M$ filaments, each with $N$ segments, $\t{J}_0$ will be $3NM \times 3NM$ for filament motion confined to a plane, and $6NM \times 6NM$ for the fully 3D case.  When interactions between the filaments are ignored, $\t{J}_0$ will be block diagonal and consist of $M$ blocks of size $3 N\times 3 N$ for 2D motion and $6 N\times 6 N$ for fully 3D simulations.  As a result, solutions to systems involving $\t{J}_{0}$ can be found by solving the $M$ smaller, independent linear systems corresponding to each block.  Additionally, only the $M$ blocks, rather than the entire matrix, need to be stored in memory.

\subsection{A comparison of the nonlinear system solvers}\label{sec:numerical-tests-and-method-choice}
Here, we test and compare the different approaches for solving $\v{f}(\v{X}) = \vu{0}$.  Based on the tests presented below, we find that bad Broyden's method using the explicit expressions for $\t{J}_{0}$ (see \cref{sec:form-of-jacobian}) provides the most effective approach for solving the system arising from the discretisation of the differential-algebraic system, \cref{eqn:RA-2d-system,eqn:RA-3d-system}.

In our tests, we consider a monolayer of $M$ filaments, each formed of $N$ segments and subject to a constant force per unit length of magnitude $W$ in the plane of the monolayer.  The simulations are performed using the MATLAB with FCM implementation described in the next section.  The filaments are initially straight and are distributed uniformly and isotropically in the centre-plane of a periodic domain of size $141 \Delta L \times141 \Delta L \times8.8 \Delta L$.  Varying $N$ between $5$ and $60$, we explore a range of filament lengths $L=N\Delta L$. FCM is used to resolve the hydrodynamic interactions between the filaments, and the Stokes equations are solved on a grid of $1024\times1024\times64$ points.  With the domain size fixed, the computational cost of FCM scales linearly with number of filament segments.  The filaments are allowed to settle for two characteristic sedimentation times defined as $T=\eta L/W$. The parameter $W$ is set such that the dimensionless elasto-gravitational number \citep{cosentino_lagomarsino_hydrodynamic_2005} is $B = L^3 W/K_B = 1000$ and filament deformation is significant.  The timestep size is set to be $\Delta t = T/300$.

In the following tests, the GMRES residual tolerance for both JFNK and JFB is $10^{-4}$ and the JFNK finite difference parameter $\delta$ is set to $10^{-7}$. The Newton and Broyden iteration tolerances for the infinity norm of the residual relative to $a$ are both $\varepsilon = 10^{-4}$.   These parameters have been selected to minimise wall times while still ensuring sufficient accuracy.  To quantify solver performance, we measure average wall times, the average number of mobility matrix multiplications per timestep, and the average iteration count per timestep.  These averages of are computed between simulation times $t=T$ and $t=2T$ once the suspension has evolved away from its initial configuration.  All simulations were performed on a 16-core 2.5GHz AMD Opteron 6380 processor.

\subsubsection{JFNK versus Broyden's method (JFNK v EJB-b)}\label{sec:timestep-preconditioner}
We first compare wall times per timestep for the simulation run with JFNK with those using ``bad'' Broyden's method with the explicit approximate Jacobian (EJB-b), \cref{alg:bad-broydens-method}.  A key aspect of using JFNK effectively is the choosing the correct preconditioner to accelerate GMRES convergence.  Here, we explore two preconditioners.  In the first instance, employ a diagonal right preconditioner that mitigates ill-conditioning due to the factor of $\Delta t$ appearing in entries of the Jacobian associated with the system \cref{eqn:RA-2d-system,eqn:RA-3d-system}.  Since other entries in the Jacobian are $\order(1)$, small timesteps can increase the Jacobian condition number, resulting in slow GMRES convergence.  As demonstrated explicitly in \cref{sec:form-of-jacobian} for a drag-only mobility matrix, this factor of $\Delta t$ appears in the Jacobian as a result of variations with respect to the Lagrange multipliers.

Accordingly, the relevant diagonal entries of the right preconditioner are set to $\Delta t/(6\pi \eta a)$.  Specifically, for a single, $N$ segment filament in 2D, the right preconditioner is
\begin{equation}
\t{P}_{\Delta t} = \begin{pmatrix}
\t{I}_{N+2} & \tu{0}\\
\tu{0} & \frac{\Delta t}{6\pi\eta a} \t{I}_{2N-2}
\end{pmatrix},
\label{eqn:timestep-preconditioner}
\end{equation}
with the system ordered as in \cref{eqn:robotArm}.

To provide a balanced comparison with Broyden's method, we also use as a left preconditioner the block-diagonal, explicit Jacobian described in \cref{sec:block-diagonal-j0} and derived in \cref{sec:form-of-jacobian}.  Based on the filament model with a diagonal mobility matrix and no steric iteractions, this preconditioner captures the local physics of the problem and its block structure allows for rapid system solves and reduced storage costs.

\begin{figure}
    \centering
    \includegraphics[width=0.4\textwidth]{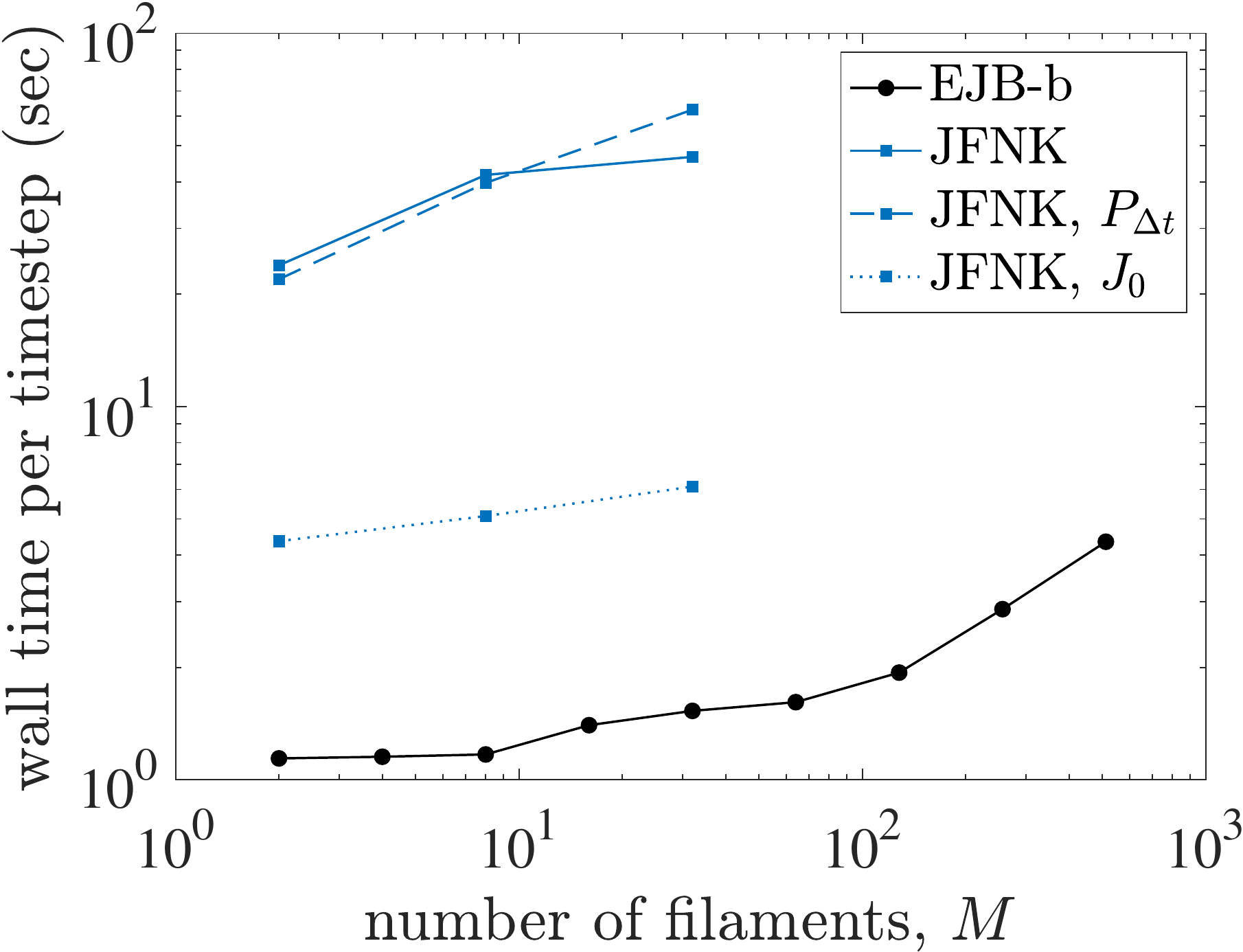}
    \includegraphics[width=0.4\textwidth]{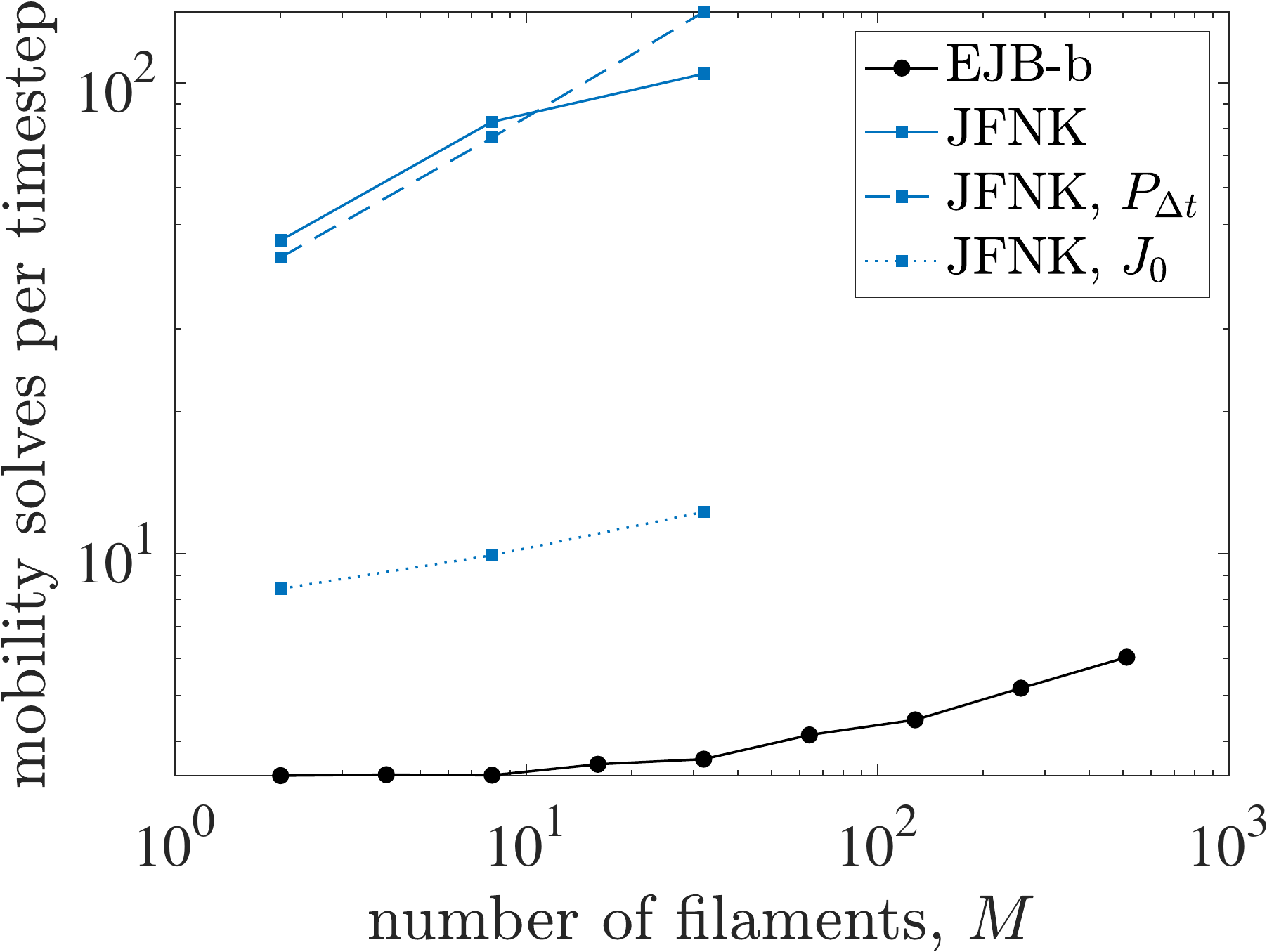}
    \caption{Comparison between bad Broyden's method with $\t{J}_0$ (EJB-b) and JFNK with three different preconditioner options: (i) unpreconditioned, (ii) $\t{P}_{\Delta t}$, and (iii) $\t{J}_0$. \textbf{Left:} Average wall time per timestep for $M$ filaments formed of $N=15$ segments. \textbf{Right:} Average number of mobility solves required per timestep.}
    \label{fig:walltime-ejb-v-jfnk}
\end{figure}

In \cref{fig:walltime-ejb-v-jfnk}, we compare the performance of JFNK with these preconditioners against that of EJB-b. The left panel of \cref{fig:walltime-ejb-v-jfnk} shows the average wall time per timestep, while the right panel indicates the average number of mobility solves required per timestep.  By ``mobility solve'', we mean multiplying by the mobility matrix using the force-coupling method (FCM) as described in \cref{sec:fcm}. We see that left preconditioning with the approximate Jacobian improves the performance of JFNK much more so than $\t{P}_{\Delta t}$.  Based on our wall time measurements, however, EJB-b is still approximately five times faster than JFNK.  The additional costs per timestep associated with JFNK are due to the higher number of mobility solves that it requires.  While EJB-b needs only one mobility solve per Broyden iteration, each Newton iteration in JFNK requires multiple GMRES iterations that each cost one mobility solve.  Due to these additional costs, JFNK is not competitive for solving the nonlinear system.

\subsubsection{Choice of Broyden's method (EJB-b v JFB-g)}

In this section, we compare the performance of a JFB implementation of good Broyden's method (JFB-g) with EJB-b.  Implementing \emph{good} Broyden's method with the explicit approximate Jacobian is computationally expensive since it would require solving a dense linear system that changes with each Broyden iteration (see \cref{alglin:update-good} in \cref{alg:good-broydens-method}).  In contrast, for bad Broyden's method, we avoid the JFB implementation since we update directly the inverse of the explicit approximate Jacobian.  As a result, the system we need to solve at each Broyden iteration remains the same.

As discussed in \cref{sec:timestep-preconditioner} for JFNK, preconditioning is also required to accelerate GMRES convergence in JFB-g.  In the tests presented below, we again trial the right preconditioner $\t{P}_{\Delta t}$, as well as the approximate Jacobian, $\t{J}_0$, as a left preconditioner.

\begin{figure}
    \centering
    \begin{subfigure}[t]{0.4\textwidth}
        \centering
        \includegraphics[width=\textwidth]{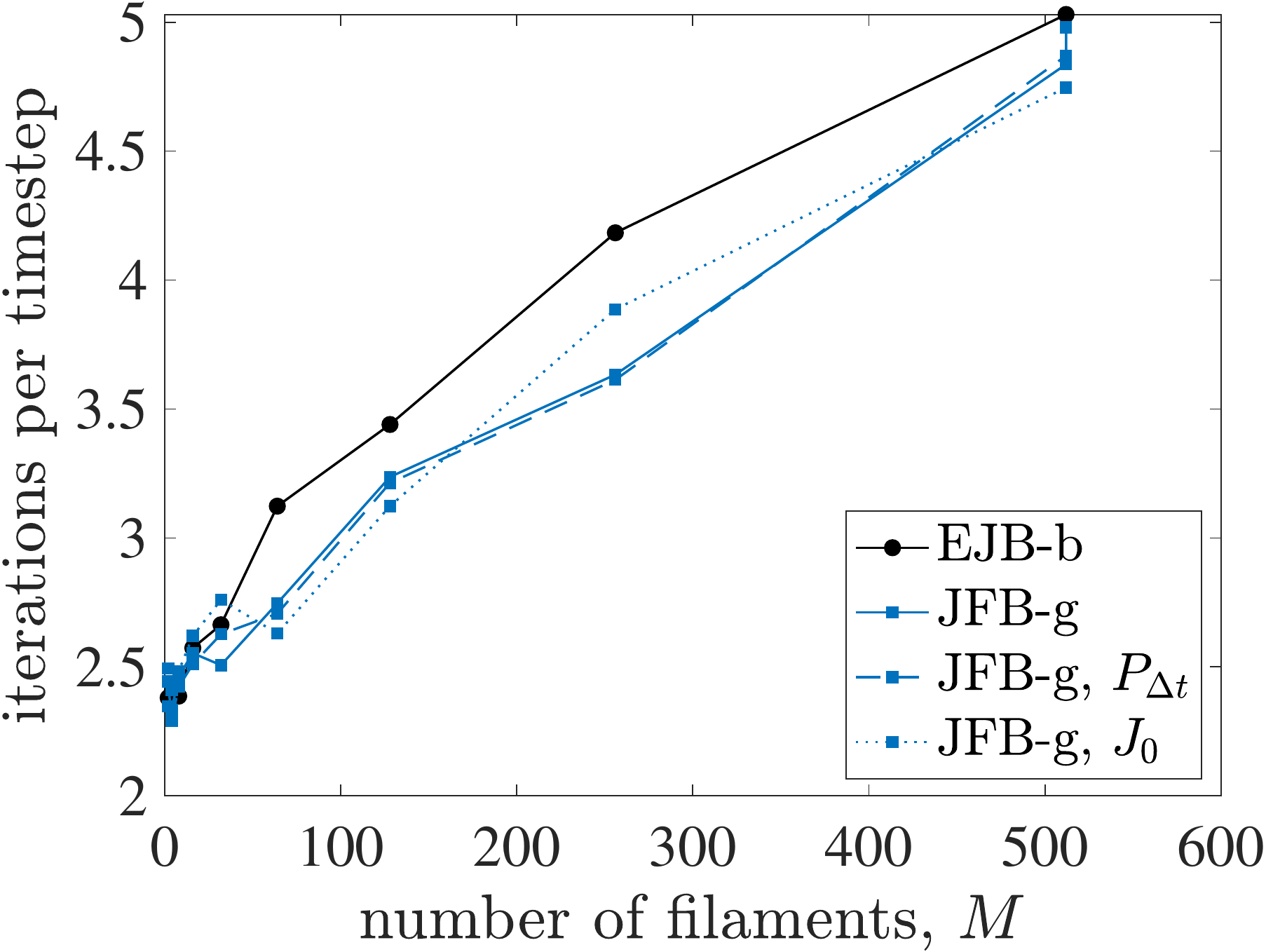}
        \caption{Average number of Broyden iterations per timestep for $M$ filaments, formed of $N=15$ segments. Filaments are initially straight and randomly oriented.}
    \end{subfigure}
    \begin{subfigure}[t]{0.4\textwidth}
        \centering
        \includegraphics[width=\textwidth]{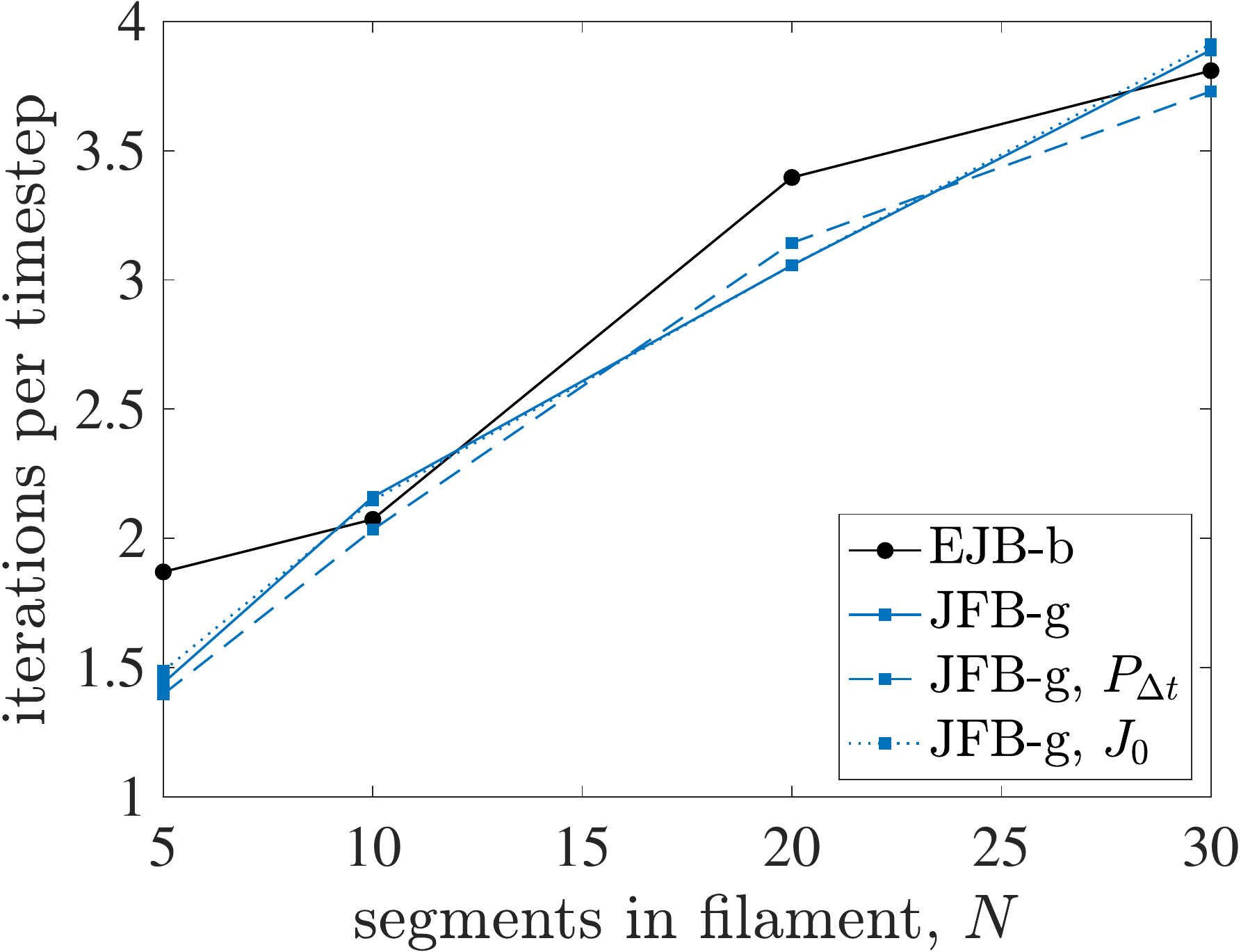}
        \caption{Average number of Broyden iterations per timestep for $M=128$ filaments, formed of $N$ segments. Filaments are initially straight and aligned horizontally.}
    \end{subfigure}
    \begin{subfigure}[t]{0.4\textwidth}
        \centering
        \includegraphics[width=\textwidth]{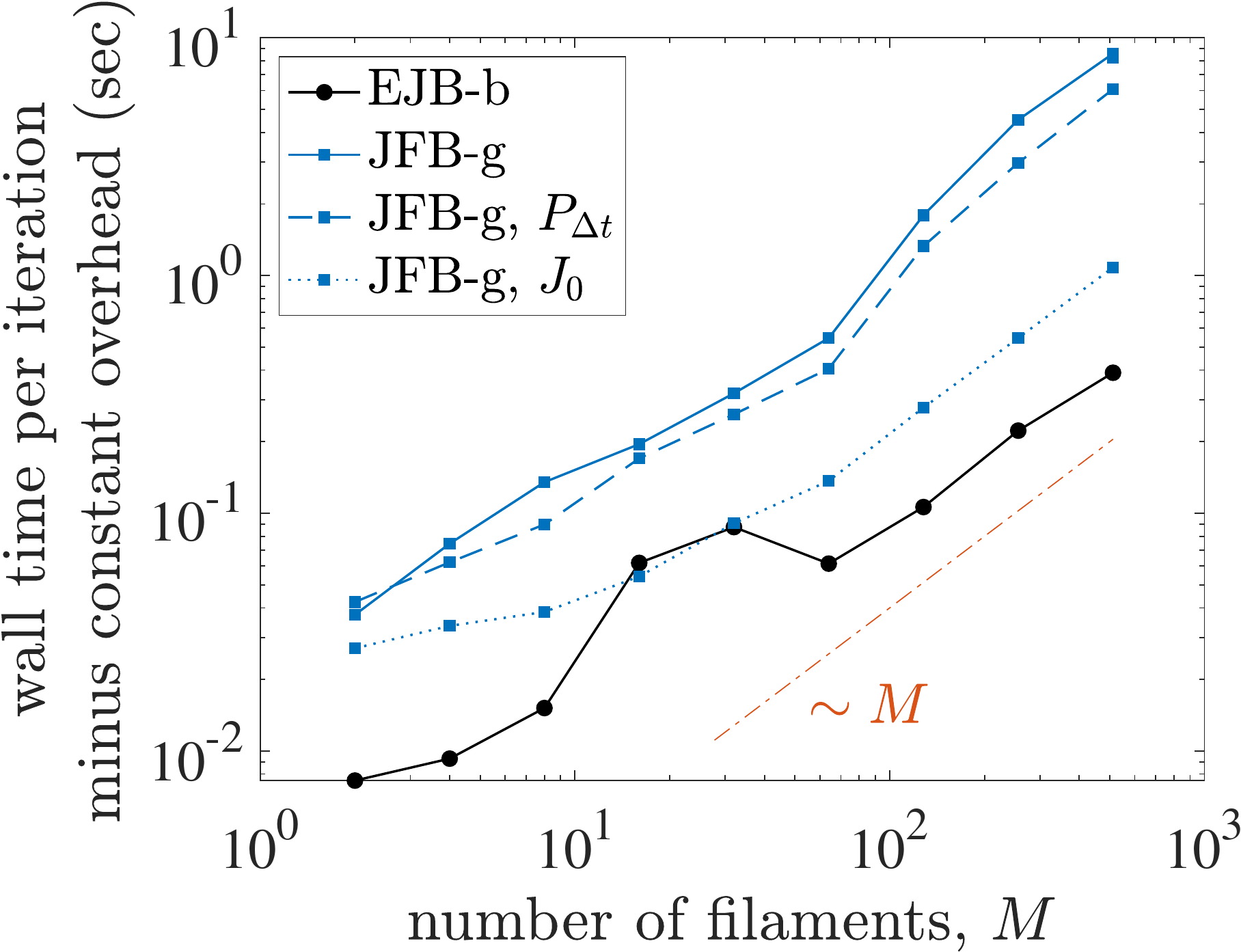}
        \caption{Average wall time per iteration for $M$ filaments, formed of $N=15$ segments.  Filaments are initially straight and randomly oriented.}
    \end{subfigure}
      \begin{subfigure}[t]{0.4\textwidth}
        \centering
        \includegraphics[width=\textwidth]{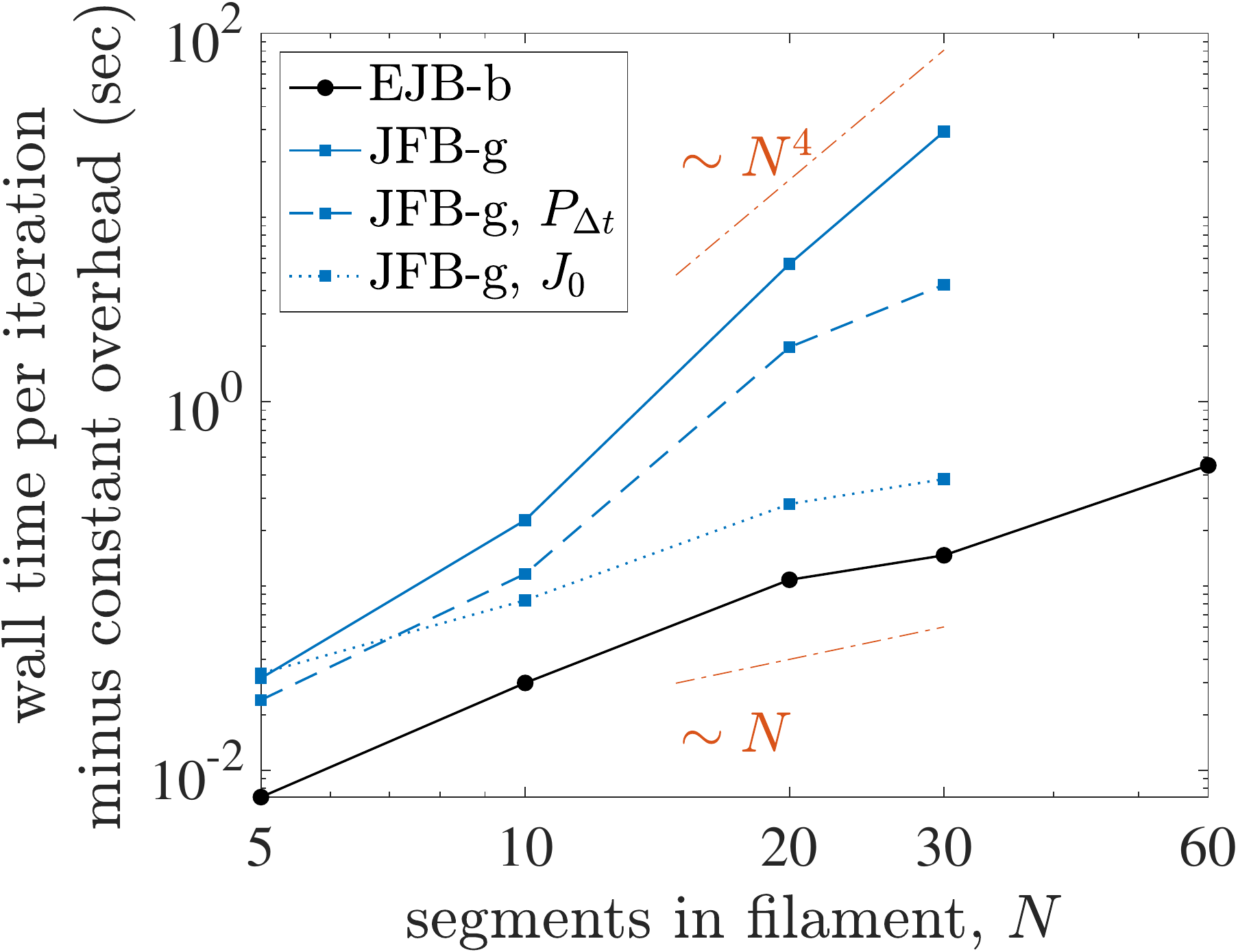}
        \caption{Average wall time per iteration for $M=128$ filaments, formed of $N$ segments. Filaments are initially straight and aligned horizontally.}
    \end{subfigure}
    \caption{Comparison between bad Broyden's method with $\t{J}_0$ (EJB-b) and JFB with three different preconditioner options: (i) unpreconditioned, (ii) $\t{P}_{\Delta t}$, and (iii) $\t{J}_0$.  For the wall time measurements, we have subtracted the initial wall time per iteration of approximately $\SI{0.3}{s}$ that is independent of $M$ and $N$ and is associated with setting up the fluid solver.
    }
    \label{fig:walltime-ejb-v-jfbg-1}
\end{figure}

The performance measurements for JFB-g with and without preconditioners and EJB-b simulations are presented in \cref{fig:walltime-ejb-v-jfbg-1}. For all approaches, the number of Broyden iterations per timestep increases similarly with both the number of segments per filament (\cref{fig:walltime-ejb-v-jfbg-1}(b)) and the number of filaments (\cref{fig:walltime-ejb-v-jfbg-1}(a)).  This increase is expected, as increasing the number of filaments or segments increases the concentration of the suspension. This in turn increases filament interactions and the number of iterations required.

When we examine the wall time per Broyden iteration, however, differences between the approaches begin to emerge.
\cref{fig:walltime-ejb-v-jfbg-1}(c) shows the wall time per iteration as a function of the number of filaments.  We find that for all methods, the wall time per iteration scales linearly with the number of filaments $M$.  This is a result of the FCM mobility solve being the most expensive part of the iteration and its cost scaling linearly with the total number of segments.

The wall time per iteration as the number of segments is varied is indicated in \cref{fig:walltime-ejb-v-jfbg-1}(d).  For JFB-g with $\t{P}_{\Delta t}$ as the preconditioner, or no preconditioner at all, the computational time increases at a more rapid rate of approximately $N^4$.  This rapid increase in cost is a result of the approximate Jacobian system size growing with $N$ and poor GMRES convergence.  For example, we find that for a JFB-g simulation of $M=512$ filaments each with $N=15$ segments, without preconditioning, 94\% of the wall time per iteration is spent in the GMRES routine to solve the Jacobian system.  We do see, however, using an effective preconditioner, in this case $\t{J}_0$, can bring down this cost.  For EJB-b and JFB-g with preconditioner $\t{J}_0$, we observe a slightly faster than linear growth in computational time with the number of segments.  Despite the reduction due to preconditioning, EJB-b still outperforms JFB by approximately a factor of 2.

In addition to being the more computationally efficient method, EJB-b is demonstrably more robust.  In practice, for JFB-g, we found that the convergence of Broyden's method relied heavily on the quality of the solution given by GMRES.  Consequently, a low residual tolerance was typically required.  We also observed that GMRES could be prone to stagnation and found that GMRES convergence is sensitive to the choice of $\delta$ in \cref{equation:JFD}.
\subsubsection{Effects of tolerance and timestep size on performance}
The Broyden iteration tolerance and the timestep size not only affect the accuracy of the simulation but also the computation time.  In particular, while increasing the timestep size reduces the number of timesteps, larger timestep sizes require more Broyden iterations in order to reach the prescribed tolerance as initial guesses are extrapolated from solutions at past timesteps.

\begin{figure}
    \centering
    \includegraphics[width=0.4\textwidth]{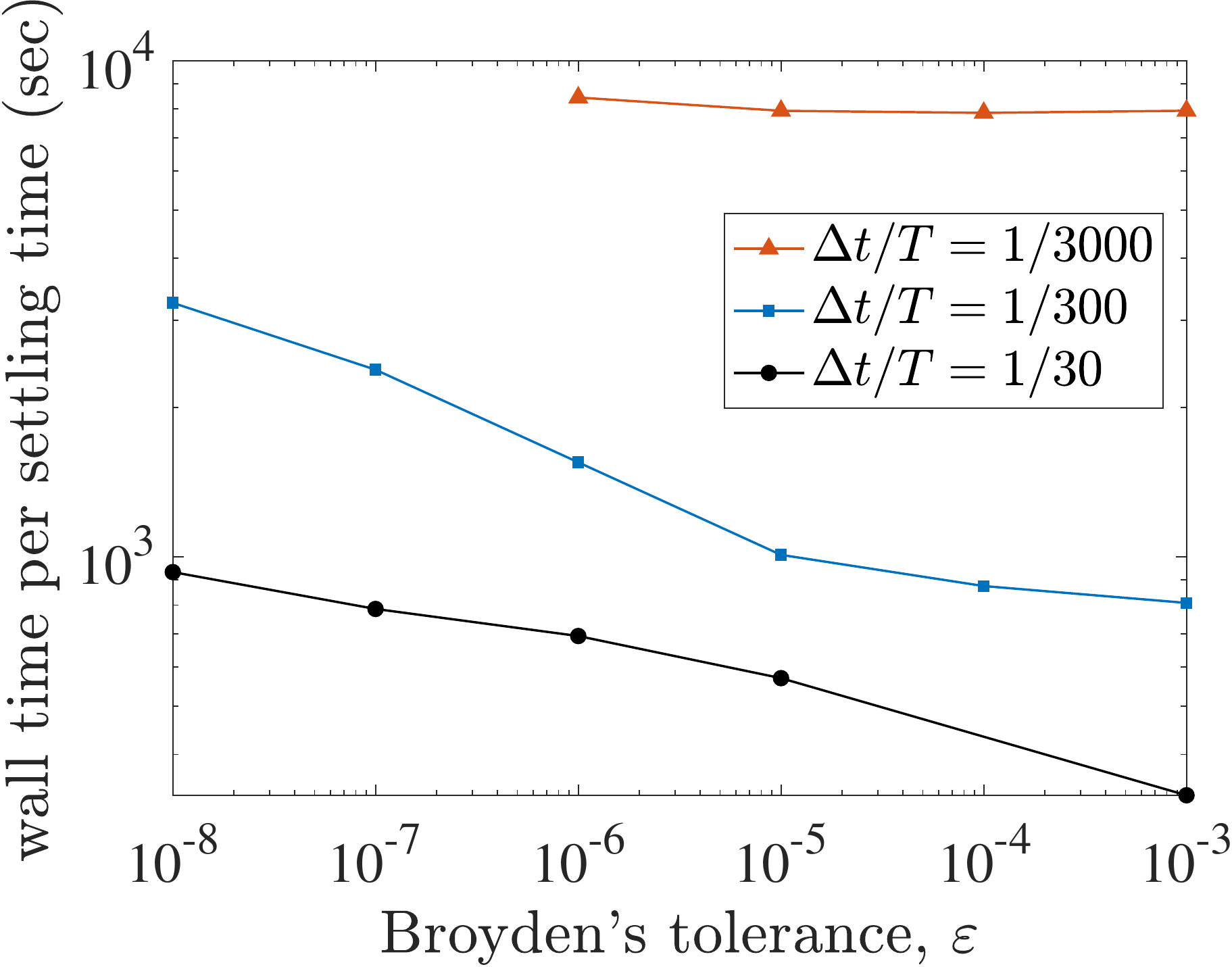}
    \caption{Average wall time per sedimentation time for EJB-b as the Broyden tolerance and timestep size are changed.  The simulations are performed with $M=256$ filaments, formed of $N=15$ segments and $B=1000$.  The filaments are initially straight and randomly oriented. The simulation with $\varepsilon=10^{-4}$ and $\Delta t/T=1/30$ failed to converge.}
    \label{fig:walltime-ejb-tol-and-dt}
\end{figure}

Measurements of the wall time for a full settling time, $T$, at different timestep sizes and different Broyden's tolerances are presented in \cref{fig:walltime-ejb-tol-and-dt} for EJB-b simulations of a concentrated suspension (area fraction 12.5\%).  In general, increasing the Broyden's tolerance decreases the wall time, though with reduced effect at the smallest timestep sizes.  Due to differences in iteration numbers, we find that decreasing the timestep from $\Delta t/T=1/30$ to $1/300$ increases the computation time by only a factor of $\sim 3$.  We also note that for $\Delta t/T=1/30$, Broyden's method fails to converge in the case where the tolerance was $\varepsilon=10^{-4}$.  Reducing the timestep further to $\Delta t/T=1/3000$, however, does not lead to additional reductions in the number of Broyden iterations, and consequently, the computation times for these simulations are approximately $10$ times greater than those with $\Delta t/T=1/300$.

\subsubsection{Effects of segment and filament number on performance}
In general, the wall time impact of changing the number of segments, $N$, or filaments, $M$, can be complicated.  In practice, we found that increasing local concentrations of filaments can lead to higher numbers of Boyden iterations. For each iteration, the predominant computations under EJB-b are the hydrodynamic mobility solve and the inversion of the approximate Jacobian, $\t{J}_0$. Given that $\t{J}_0$ is block diagonal, the complexity of this operation is at worst $\order(N^3)$. For filaments with $N \lesssim 50$, we have found that the hydrodynamic mobility solve is the more costly aspect of the computation.

The total number of segments impacts the computational work required to apply the mobility matrix.  In \cref{sec:hydrodynamics}, we discussed the two hydrodynamic models used in this work: the force-coupling method (FCM) for periodic domains and RPY for unbounded domains.  As we use a direct pairwise summation, the computational work with our implementation of RPY scales like $\order((NM)^2)$.  For FCM, the computational work scales as $\order(NM)$, and is linked to the cost of constructing the body-force in the Stokes equations used to represent the forces the segments exert on the fluid.  This scaling is seen in \cref{fig:walltime-ejb-v-jfbg-1}(c, d).  It is important to note that for FCM this cost is in addition to the fixed cost of the FFT-based Stokes solver that scales like $\order(N_g \log N_g)$, where $N_g$ is the total number of grid points.  Thus, the computational work associated with FCM will scale linearly with the number of segments only once the filament density is sufficiently high.

\section{Accuracy and convergence}
In this section, we present results from numerical experiments that demonstrate the finite difference scheme's second-order convergence in space, as well as the second-order convergence in time for the 2D and 3D implementations of the BDF schemes.  In addition, we provide a comparison between the RPY version of the hydrodynamic model and numerical solutions to the boundary integral equations for the hydrodynamic forces experienced by a straight filament.

\subsection{Convergence in space}
\begin{figure}
    \centering
    \includegraphics[width=0.5\textwidth]{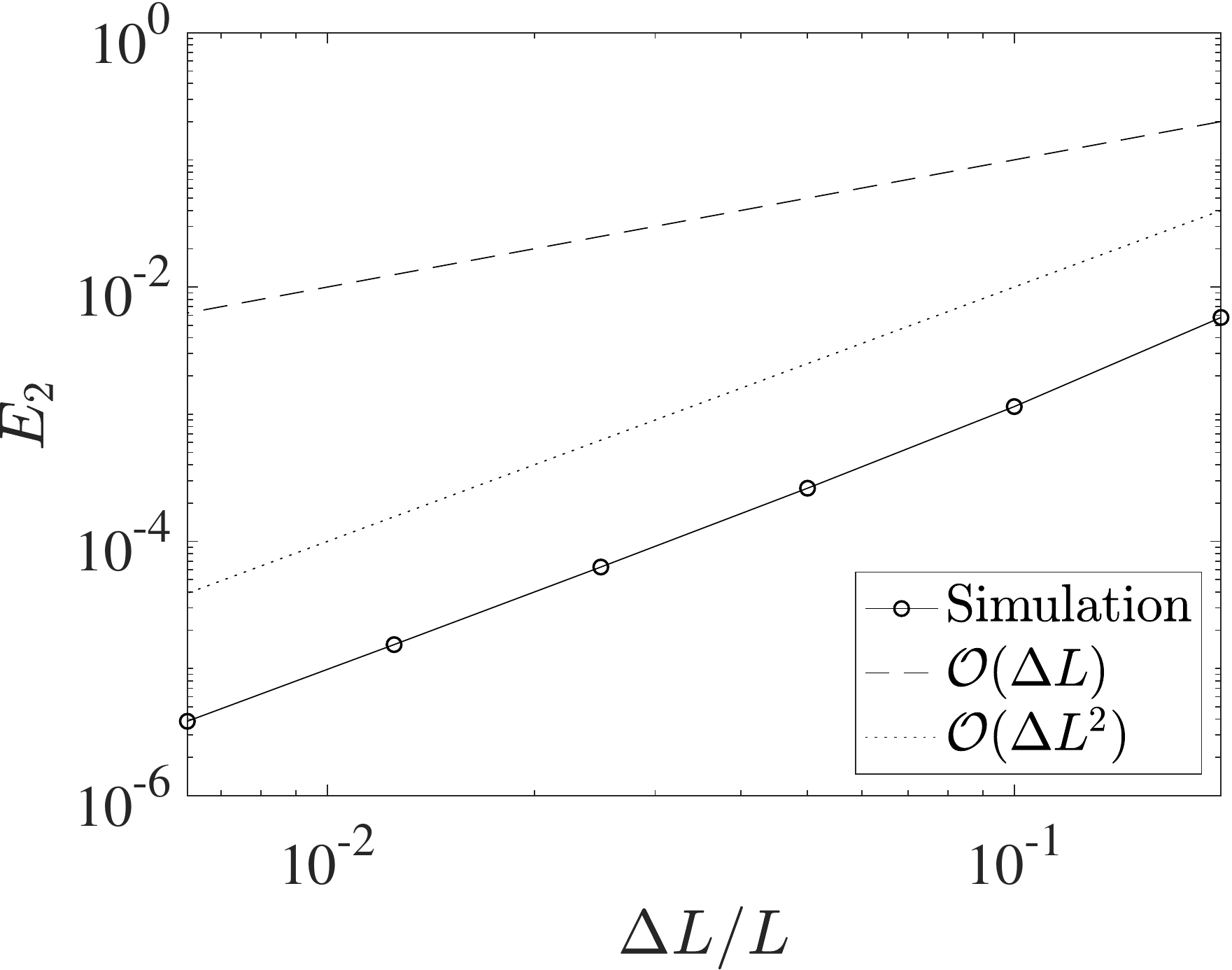}
    \caption{Error between the simulated steady-state filament shape and the analytical solution given by \citet{landau_theory_1986} for different values of segment separation $\Delta L$.}
    \label{fig:spatial_resolution_study}
\end{figure}
To demonstrate the second-order convergence of the finite difference scheme used to discretise the force and moment balances, \cref{eqn:DifferencingForce,eqn:DifferencingMoments,equation:inex_constraints}, we consider an initially horizontal filament that is clamped at one end $(s = 0)$ and subject to a vertical force of magnitude $F = 1.93 K_B/L^2$ at the free end $(s = L)$.  The filament bends in the direction of the force, and after some time, reaches an equilibrium shape described by $\theta(s)$, the angle the filament makes with the vertical.  After integrating the continuous force and moment balances, one finds that the equilibrium angle satisfies \citep[\S 19]{landau_theory_1986}
\begin{equation}
s = \sqrt{\frac{K_B}{2F}}\int_{\theta(s)}^{\pi/2} \frac{\d \phi}{\sqrt{\cos(\theta(L)) - \cos(\phi)}}.
\end{equation}
\cref{fig:spatial_resolution_study} shows the
$L^2$ error,
\begin{equation}
    E_2 = \sqrt{\Delta L \sum_{n=1}^{N}\left(\theta^n - \theta(s_n)\right)^2},
\end{equation}
where $\theta^n$ is the angle for segment $n$ given by the finite difference scheme, as a function of $\Delta L$.  We observe second-order convergence to the analytical solution as we decrease $\Delta L$.

\subsection{Convergence in time}
\begin{figure}
    \centering
    \begin{subfigure}[t]{0.39\textwidth}
        \centering
        \includegraphics[height=6.1cm]{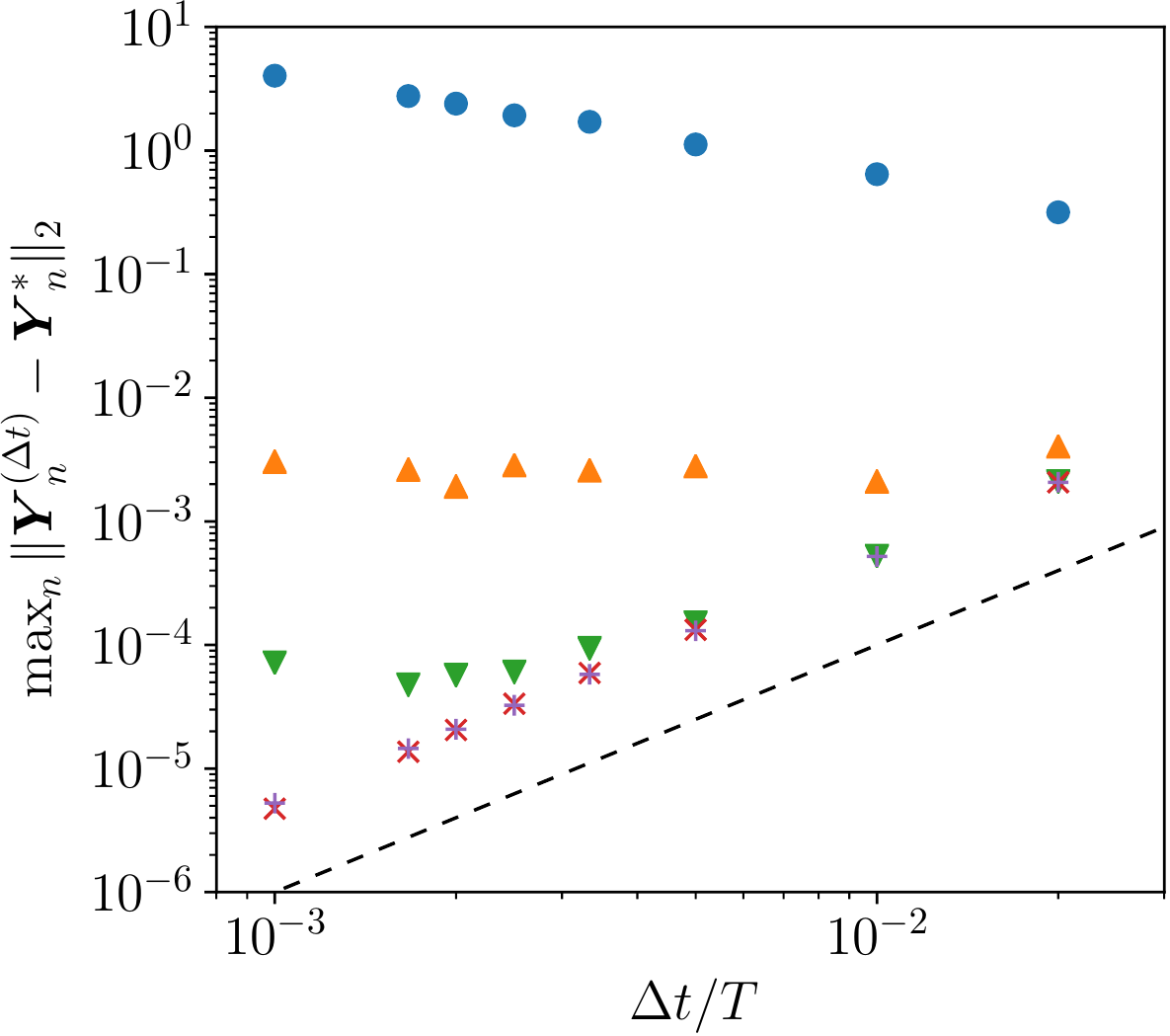}
        \caption{2D convergence: $M=2$ settling filaments}
    \end{subfigure}
    \begin{subfigure}[t]{0.59\textwidth}
        \centering
        \includegraphics[height=6.1cm]{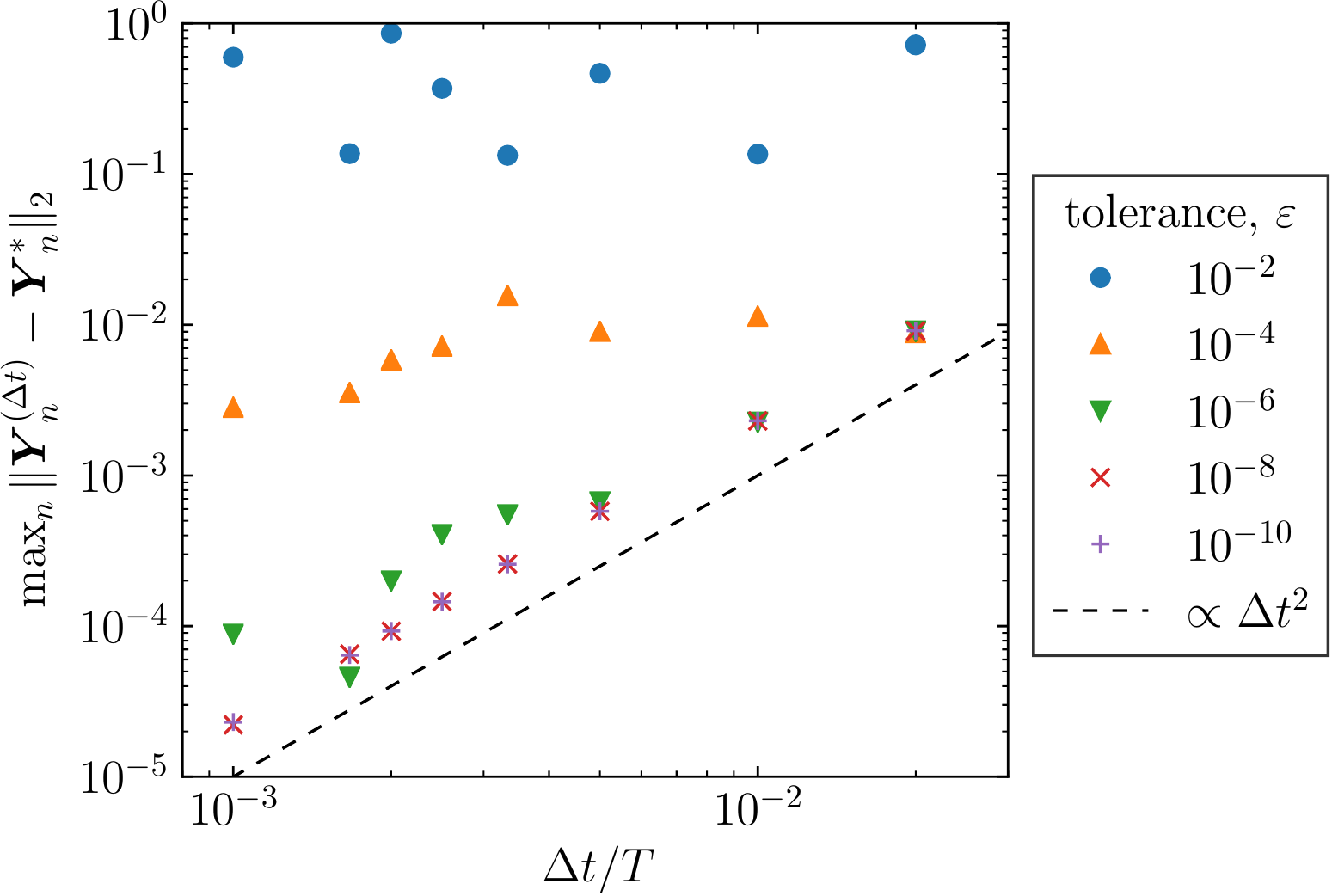}
        \caption{3D convergence: $M=4$ settling filaments}
    \end{subfigure}
    \caption{Maximum filament position error at $t_f = 20T$ for two (in 2D) or four (in 3D) settling filaments with $N=30$ and $B=10^3$. Units are such that the RPY radius of each segment is $a = 1$. The error is estimated by comparing to the very accurate solution $\v{Y}^*(s,t_f)$ computed with tolerance $\varepsilon=10^{-12}$ and $\Delta t/T = 10^{-4}$. Simulations were performed using BDF2 and in both cases, we obtained the expected convergence rate for appropriately chosen tolerances.}
    \label{fig:convergence-2d-and-3d}
\end{figure}

Here, we confirm the second-order convergence of the BDF2 schemes for two- and three-dimensional filament motion.  To test the 2D scheme, we consider a pair of filaments that are initially straight, oriented vertically, and separated horizontally.  The filaments are then allowed to settle as a result of a constant vertical force per unit length applied to each.   For the 3D scheme, we consider the same settling problem, but for four initially straight filaments that are aligned vertically and arranged such that when viewed from above, they are at the corners of a square (cf.\ \cref{fig:GTpanel}). In all 2D and 3D simulations, we use pairwise hydrodynamics via RPY and the dimensionless elasto-gravitational number is $B = L^3 W/K_B = 1000$, where $W$ is the force per unit length acting on each filament.  Each filament consists of $N=30$ segments and the simulations are run for $t_f = 20T$.  The settling time is denoted as $T = \eta L/W$.  With these parameters, the filaments bend as they settle and move apart from one another until they each reach a horseshoe-like shape (cf. \cref{fig:sed-at-different-times}).

\cref{fig:convergence-2d-and-3d} shows the error based on the maximum Euclidean distance of the positions $\v{Y}^{(\Delta t)}(s,t_f)$ to the very accurate solution $\v{Y}^*(s,t_f)$ that was computed using $10^4$ timesteps per settling time and a tolerance of $10^{-12}$.  We remind the reader that the positions along the length of the filament are computed using the values of the angles (in 2D) or quaternions (in 3D), and thus their accuracy depends on that of the BDF scheme.  For both the 2D and 3D schemes, we do observe second-order accuracy provided that the tolerance for the Broyden's iterations is sufficiently low.  Thus, a key parameter in ensuring accuracy of the solution is not only the timestep, but the tolerance to which the resulting system of equations is solved.

\subsection{Accuracy of the hydrodynamic model}\label{sec:accuracy-of-hydrodynamic-model}

\begin{figure}
    \centering
    \begin{subfigure}[t]{0.5\textwidth}
        \centering
        \includegraphics[width=\textwidth]{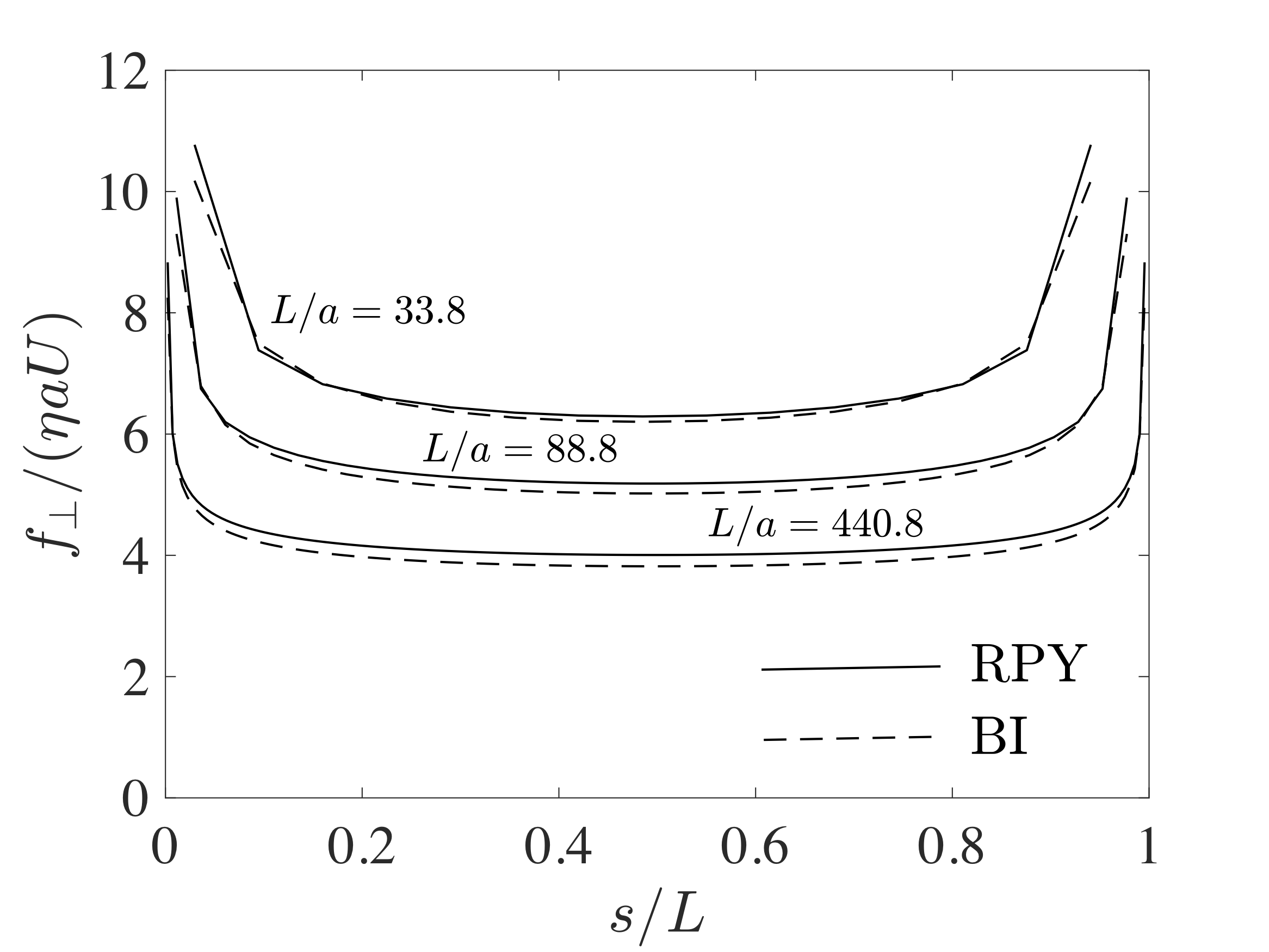}
        \caption{}
    \end{subfigure}
    \begin{subfigure}[t]{0.5\textwidth}
        \centering
        \includegraphics[width=\textwidth]{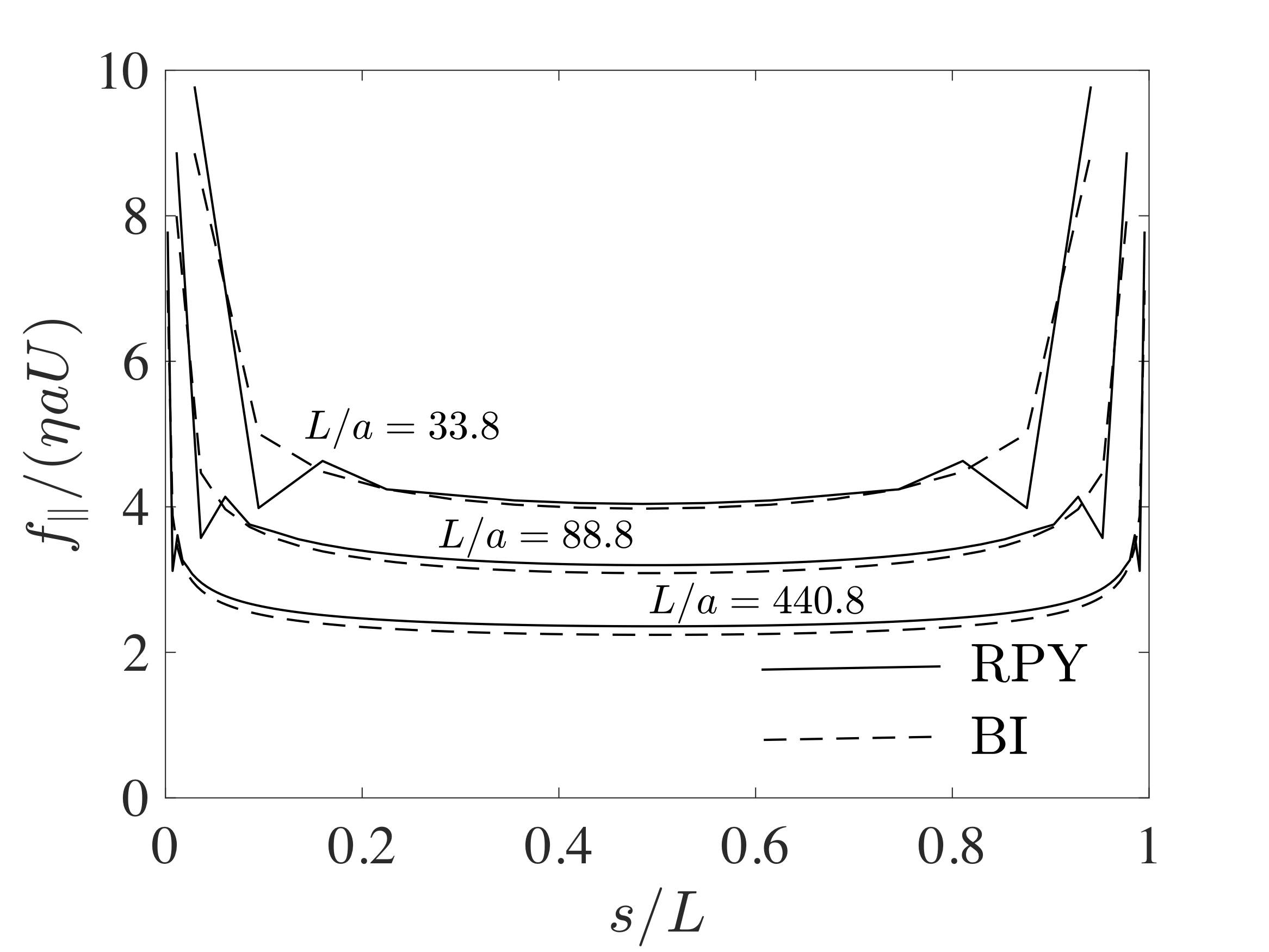}
        \caption{}
    \end{subfigure}
    \begin{subfigure}[t]{0.5\textwidth}
        \centering
        \includegraphics[width=\textwidth]{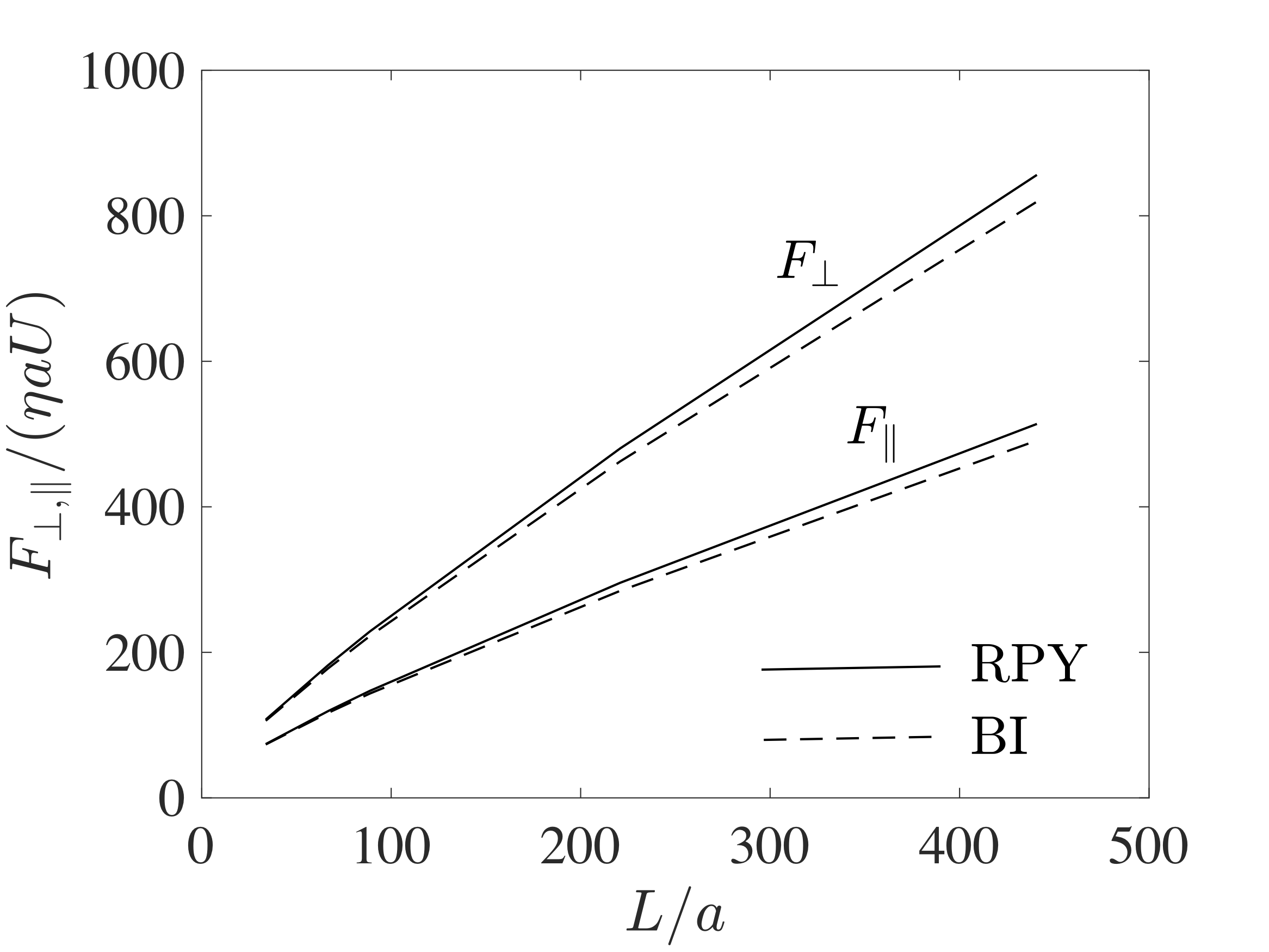}
        \caption{}
    \end{subfigure}
    \caption{A comparison between the RPY model and boundary integral (BI) computations for (a) the segment forces as a function of $s$ for translation perpendicular to the filament axis, (b) the segment forces as a function of $s$ for translation parallel to the filament axis, and (c) the total force on the filament as a function of $L/a$ for both perpendicular and parallel translation.}
    \label{fig:RPYvBI}
\end{figure}
In this section, we examine the accuracy of the hydrodynamic models that we employ in this work to solve the mobility problem and obtain the translational and angular velocity of the segments.  In this test, we consider the simplest case of a translating rigid, straight filament with a circular cross-section of radius $a$.  We compute the force on each segment using the RPY model described above and compare these values with those given by a second-order discretisation of the boundary integral (BI) equations for rigid body motion \citep{keaveny2011applying}.  For the boundary integral solver, we take the surface of the filament to be that of a cylinder with radius $a$ that has spherical caps.  The spacing between the segments with the RPY model is $\Delta L = 2.2a$.  We only compare the RPY model with the BI values since both solvers are configured for an unbounded fluid, while our implementation of FCM requires the imposition of periodic boundary conditions.  We do expect, though, that for an unbounded fluid, FCM will yield force values similar to those of RPY as the expressions \citep{maxey_localized_2001,lomholt_force-coupling_2003} for the FCM fluid velocity and particle mobility are asymptotic to their RPY counterparts.

\cref{fig:RPYvBI}(a) shows the force on the filament segments as a function of $s$ given by the RPY model and BI computations for the case where the filament translates with speed $U$ in a direction perpendicular to its axis.  The segment force values are shown for filaments with different aspect ratios, $L/a$.  For the BI solutions, the segment forces are computed by integrating numerically the surface tractions over sections of the cylinder of length $\Delta L$, corresponding directly to those in the RPY model.  The numerical integration is performed using the trapezoidal rule.  Similarly, \cref{fig:RPYvBI}(b) shows the force on the segments given by the two methods when the filament translates with speed $U$ in the direction of its axis, while \cref{fig:RPYvBI}(c) shows the total force as a function of aspect ratio.  We see that while the RPY model slightly overestimates the segment and total force values with a maximum relative error of approximately 4\%, it does reproduce the general dependence of the segment force on $s$ given by the BI computations.  We also observe that the error is largely independent of the filament aspect ratio.

%----------------------------------------------------
\section{Simulations}\label{sec:simulations}
In this section, we present results from several simulations that range from a single tethered filament whose position is fixed at one end, to many interacting filaments moving freely in the surrounding fluid.
\subsection{Implementations}
In the simulations that follow, we used one of three implementations of the method:
\begin{enumerate}
\item \textbf{MATLAB with RPY}: For the single filament simulations, we have used a MATLAB implementation of the method in which the hydrodynamic interactions are accounted for through the RPY mobility matrices.  We have made this implementation available on GitHub \citep{schoeller_github_2019}.

For our simulations involving tethered filament arrays, we have accelerated the RPY evaluation by using MATLAB's MEX functionality and performing the pairwise computation in C.

\item \textbf{MATLAB with FCM}: For the two-dimensional simulations in periodic domains involving many sedimenting filaments or active swimmers, we use a MATLAB--MEX implementation of the method with FCM resolving the hydrodynamic interactions between the segments.  Specifically, since with FCM we must solve the Stokes equations, we use C functions to loop through the grid points when assembling the FCM fluid forcing, and to loop through the wave numbers when inverting the Stokes operator.  The remaining aspects of the computation, including the FFTs, are performed in MATLAB.

\item \textbf{C++ with RPY:} The final simulations of sedimenting filament clouds were performed with a more powerful C++ implementation of the method that takes advantage of the Armadillo linear algebra library \citep{sanderson_armadillo_2016,sanderson_user-friendly_2018}.  In addition, the RPY computation, though still performed pairwise, was parallelised using OpenMP.
\end{enumerate}
\subsection{Tethered filaments}
\subsubsection{Rotational dynamics of a single tethered filament} \label{sec:single_tethered_filament}

\begin{figure}
    \centering
    \begin{subfigure}[t]{0.35\textwidth}
    \includegraphics[width=0.95\textwidth]{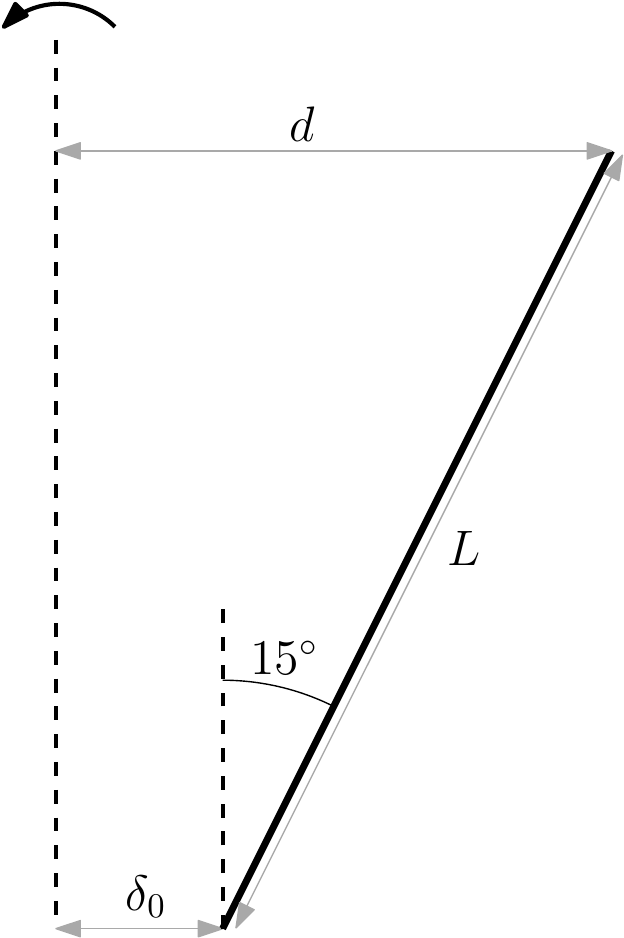}
    \caption{A diagram of the rotating filament simulation. The distance of the base from the axis of rotation, $\delta_0$, is a fixed value across all of our simulations.  The distance of the filament end from the axis is $d$.}
    \label{fig:tethered_filament_diagram}
    \end{subfigure}
    \begin{subfigure}[t]{0.6\textwidth}
    \includegraphics[width=0.95\textwidth]{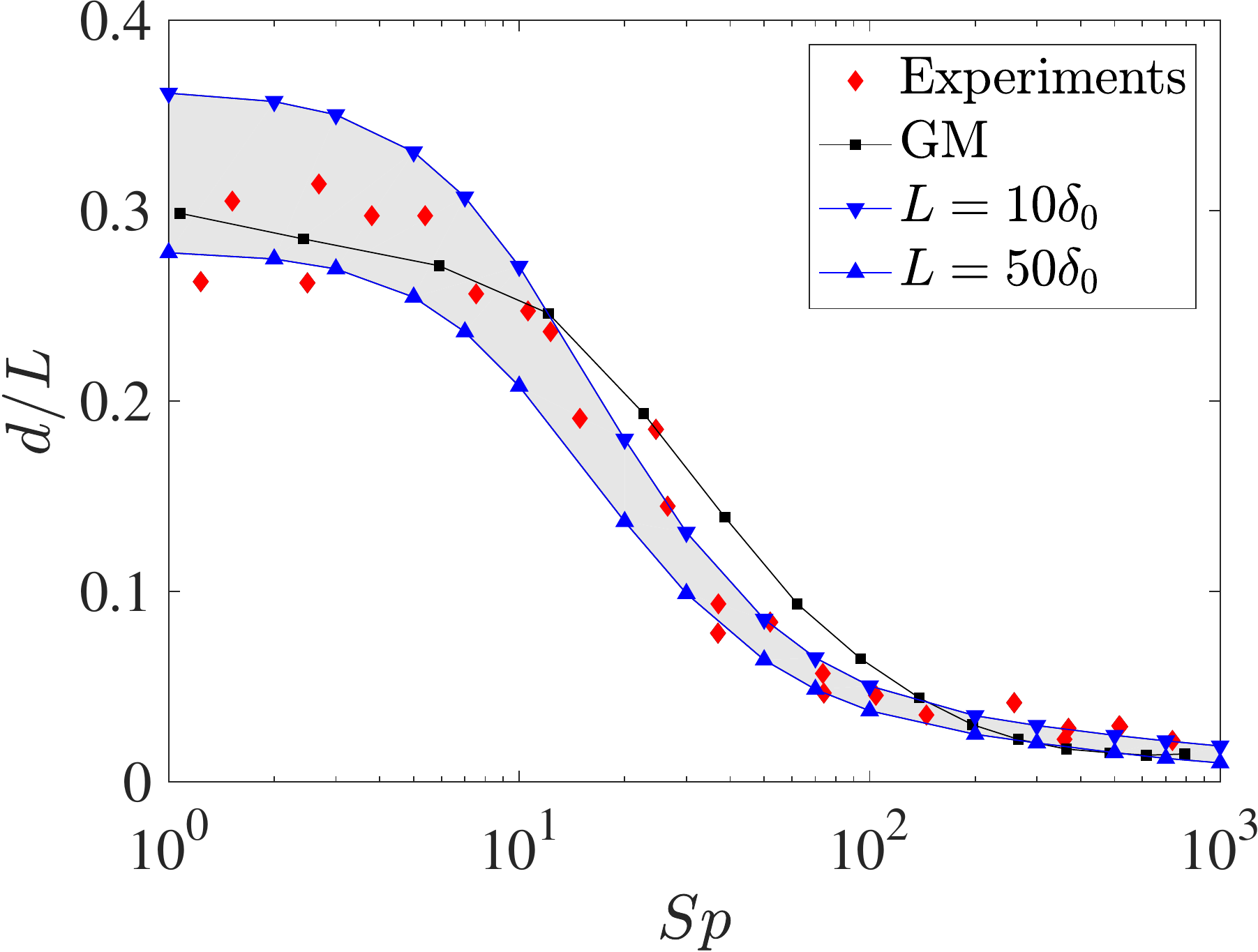}
    \caption{The normalised end distance, $d/L$, as a function of $Sp$. Results from our simulations are compared to those of the Gears Model (GM) \citep{delmotte_general_2015} and the experimental results of \citep{coq_rotational_2008}. The two blue curves correspond to values obtained using the longest and shortest filaments in both our simulations and the experiments.  The enclosed shaded area indicates the values at intermediate filament lengths.}
    \label{fig:tethered_filament_plot}
    \end{subfigure}
    \caption{Deflection of a tethered rotating filament.}
\end{figure}

In \citet{coq_rotational_2008,coq_helical_2009}, an elastic filament is immersed in a tank of pure glycerine. The filament is connected to a motor at its base such that at rest, the filament forms an angle of \ang{15} with the motor axis.  The base of the filament is displaced from the motor axis by $\delta_0 = \SI{2}{mm}$ (see \cref{fig:tethered_filament_diagram}). The filament has radius $a = \SI{435}{\micro m}$, and its length is varied between \SI{2}{cm} and \SI{10}{cm}.  A dimensionless Sperm number, $\Sp$, that characterises the ratio of viscous to elastic forces is introduced as
\begin{equation}
    \Sp = \frac{\zeta_\bot \omega L^4}{K_B},
\end{equation}
where $\zeta_\bot = 4\pi\eta/(\log(L/a) + 1/2)$ is the transverse drag coefficient for a straight rod, $\omega$ is the rotation rate of the motor, and $K_B$ is the bending modulus of the elastic filament. By varying $\omega$, a range of $\Sp$ can be explored and the steady-state distance, $d$, of the free end from the rotation axis is recorded. At lower $\Sp$ the filament barely deforms from its straight configuration and undergoes a nearly rigid body rotation such that $d \approx \delta_0 + L\sin(\ang{15})$.  At larger $\Sp$, the filament slowly collapses onto the rotation axis and $d \approx 0$.

In order to simulate this experiment, the system of equations \cref{equation:3d-nonlinear-system} is supplemented with the additional constraints such that for timestep $j$,
\begin{equation} \label{eq:tethering_constraint}
    \v{Y}_1 = \left(\begin{matrix}\delta_0 \cos(j\omega\Delta t) \\ \delta_0 \sin(j\omega\Delta t) \\ 0\end{matrix}\right),
\end{equation}
to tether the filament base to the motor, and
\begin{equation} \label{eq:tethered_rotation_constraint}
    \v{u}_1^j = \left(\begin{matrix}0 \\ 0 \\ \omega\Delta t\end{matrix}\right),
\end{equation}
to rotate the filament base with the motor. These constraints are associated with two new vector Lagrange multipliers $\v{\lambda}_1$ and $\v{\lambda}_2$, and consequently the first segment experiences additional constraint forces and torques given by
\begin{align} \label{eq:tethering_constraint_force_and_torque}
    \v{F}^C &= \v{\lambda}_1, \\ \v{T}^C &= \t{D}^\top \v{\lambda}_2,
\end{align}
where $\t{D}$ is the matrix such that $\t{D}\v{v} = \dexpinv_{\v{u}_1^j}(\v{v})$. The total differential algebraic system is modified by substituting these new constraints, \cref{eq:tethering_constraint,eq:tethered_rotation_constraint}, into the update equations for the first segment.

To compare our simulations to the experimental results, we match the ratio $\delta_0 / a = 4.5977$ and the length range $10\delta_0 \leq L \leq 50\delta_0$ by varying the number of segments, $N$. For each length considered, the simulation values of $\omega$ or $K_B$ are varied to explore the range of $Sp$.  In addition, the simulations are run using the RPY mobility matrices (see \cref{sec:RPY}) to capture segment hydrodynamic interactions.

In \cref{fig:tethered_filament_plot}, we present the filament end distance, $d/L$, as a function of $Sp$ from our simulations for the range of filament lengths.  Along with this data, we show the values measured in the experiments in \citet{coq_rotational_2008}.  Though the exact filament lengths associated with the data points were not indicated in \citet{coq_rotational_2008}, the values of $d/L$ measured in the experiments lie broadly within the range given by our simulations.  For comparison, we have also included similar results given by the gears model (GM) from \citet{delmotte_general_2015}.  While these results largely coincide with those from our simulations and the experiments, there are some differences that may be attributed to the interpretation of the offset distance, $\delta_0$, as discussed in \citet{delmotte_general_2015}.

\subsubsection{Coordination in cilia arrays}

In this section, we extend our study of a single tethered elastic filament to the simulation of a model ciliary array. Cilia are biologically occurring elastic filaments that undergo time-periodic motion and tend to exhibit metachronal waves and coordination \citep{brennen_fluid_1977,guo_bistability_2018}. A simplified model of cilia dynamics involves the ``geometric switch'' \citep{guo_bistability_2018,kim_pumping_2006}, wherein a driving torque is applied to the base of an elastic filament until some fixed angle $\theta_\text{max}$ is reached with respect to the vertical.  At this point, the direction of the torque is reversed until the filament reaches an angle of $-\theta_\text{max}$, when it switches again and so on. In \citet{kim_pumping_2006}, both the direction and the magnitude of the driving torque change once the critical angle is reached (i.e.\ $\tau_{\text{slow}} \mapsto \tau_{\text{fast}} = -\alpha\tau_{\text{slow}} \mapsto \tau_{\text{slow}} \mapsto \cdots$ with $\alpha \geq 1$), allowing for distinct ``effective'' and ``recovery'' strokes. The resulting filament motion, which is non-reciprocal due to its elasticity, allows for the net pumping of the fluid.

\begin{figure}
    \centering
    \begin{subfigure}[t]{0.45\textwidth}
    \includegraphics[width=0.95\textwidth]{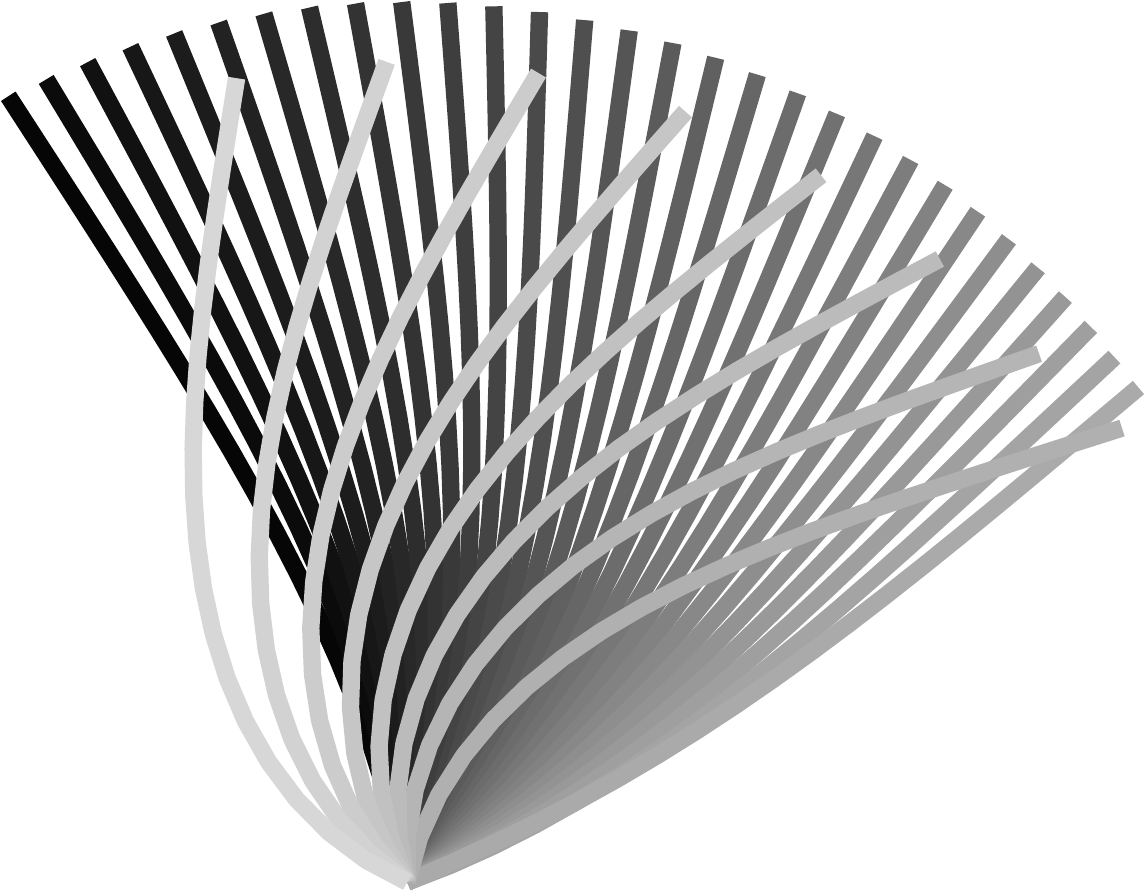}
    \caption{Dynamics of a model cilium over the course of one period. The colour fades from black to white with time. Snapshots are taken at fixed time intervals and the fewer lines for the recovery stroke indicate its faster speed.}
    \label{fig:cilia_steady_state_shape}
    \end{subfigure}
    \begin{subfigure}[t]{0.5\textwidth}
    \includegraphics[width=0.95\textwidth]{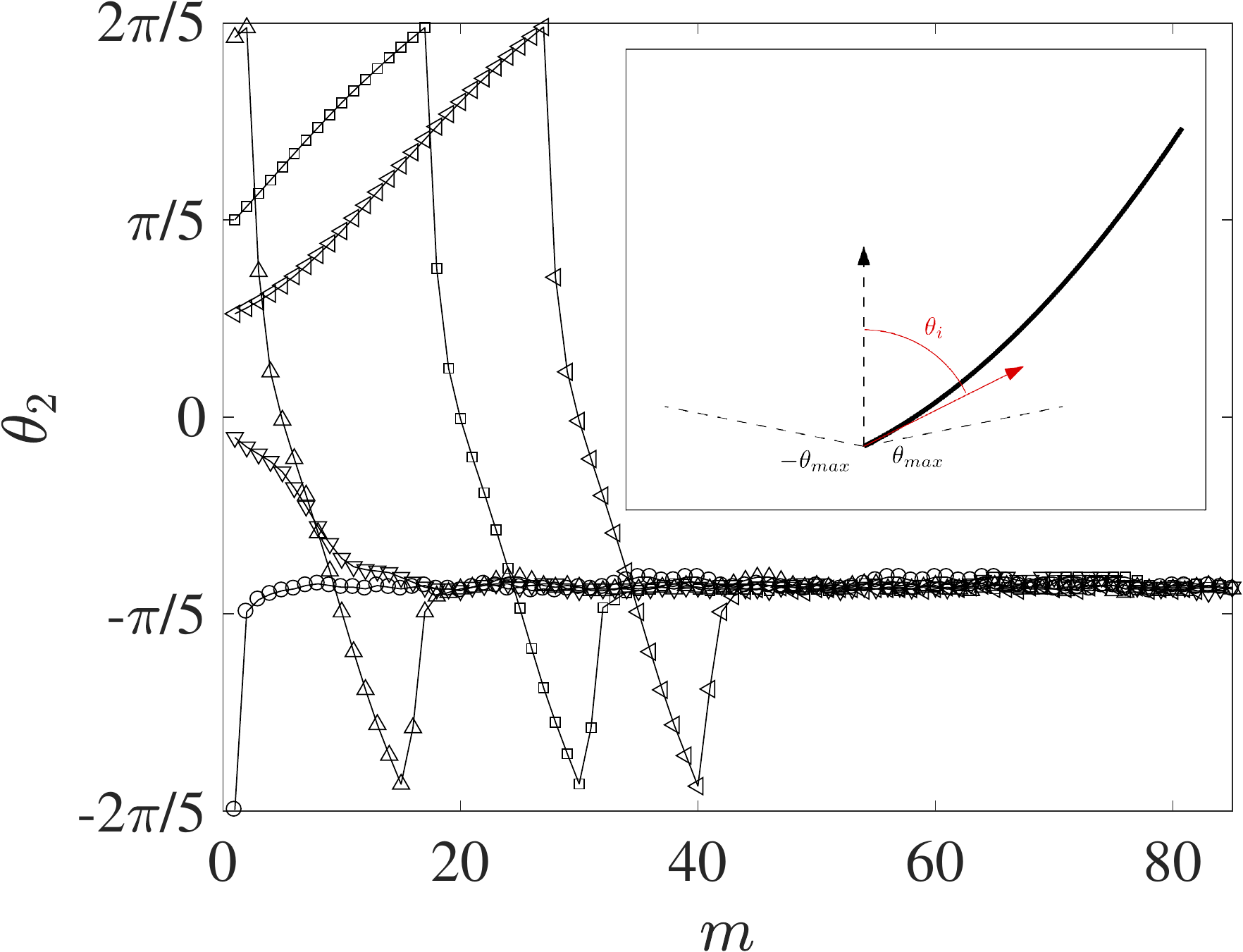}
    \caption{The long term phase-locking of a pair of model cilia for different initial phase differences. The base angle, $\theta_2$, of filament $2$ at the times, indexed by $m$, the base angle of filament $1$ reaches $\theta_1 = -\theta_\text{max} = -2\pi/5$ and switches from recovery to effective stroke.  These simulations used a torque magnitude ratio of $\alpha = 3$ and the cilia have a base-to-base distance of $L$.}
    \label{fig:cilia_pair_angle_sync_plot}
    \end{subfigure}
    \caption{Dynamics and phase-locking of model cilia.}
\end{figure}

Using our methodology, we first simulate a single model cilium comprised of $N = 20$ segments.  The position of the base segment is constrained to remain fixed to where it is attached to a no-slip wall.  The hydrodynamic effects of the wall are included through the RPY tensor for a half-space given in \citet{swan_simulation_2007}.  In addition, for numerical stability of our implicit scheme, we allow for a smooth transition between the torque values once $\theta_\text{max}$ or $-\theta_\text{max}$ are reached.  This is accomplished by introducing a transition period of several timesteps during which the driving torque is taken from a sigmoid curve connecting $\tau_{\text{slow}}$ to $\tau_{\text{fast}}$.  \cref{fig:cilia_steady_state_shape} shows the shape of the cilium over a period of oscillation given by our simulations.  The differences in filament shape and speed due to the different values of $\tau_{\text{fast}}$ and $\tau_{\text{slow}}$ can be clearly observed.

Since the switching between $\tau_{\text{fast}}$ and $\tau_{\text{slow}}$ is determined by the tangent vector at the model cilium's base, \citet{kim_pumping_2006} observed that hydrodynamically interacting pairs of model cilia can phase-lock regardless of their initial phase difference.  Using our model, we performed similar simulations of two interacting model cilia that are separated by a distance $L$ at the base.  Each model cilium is again comprised of $N=20$ segments and has torque ratio $\alpha = 3$ and $\theta_\text{max} = 2\pi/5$.  The torque applied at the base of the model cilia is perpendicular to both the base-to-base separation vector and the surface normal.  \Cref{fig:cilia_pair_angle_sync_plot} shows $\theta_2$, the angle between the surface normal and base tangent for the second filament, when the first filament switches from the fast to the slow stroke.  Similar to the kind of coordination observed in \citet{kim_pumping_2006}, we find that regardless of the initial phase difference, $\theta_2$ reaches the same value of $\theta_2 \approx -\pi/5$ for each simulation.

\begin{figure}
    \centering
    \includegraphics[width=0.7\textwidth]{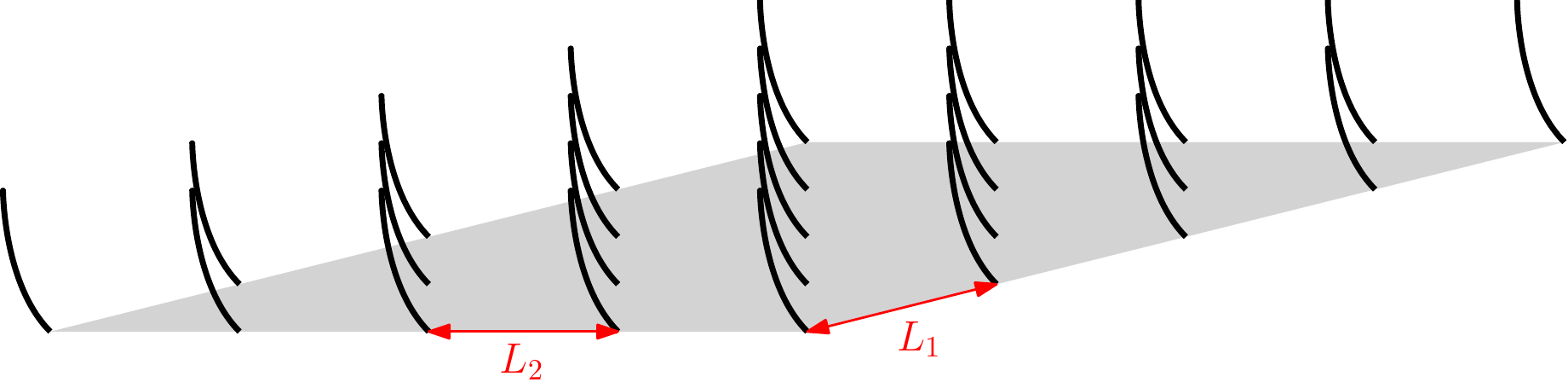}
    \caption{A diagram of a model cilia array, including the definitions of the lengths $L_1$ and $L_2$.}
    \label{fig:cilia_array_diagram}
\end{figure}

Expanding from the pair simulations, we now consider arrays of these cilia and investigate the effect of inter-filament spacing on coordination.  We arranged $N_c^2$ filaments on an $N_c \times N_c$ grid at a base-to-base separation of $L_1$ in the plane of motion and $L_2$ in the orthogonal plane (see \cref{fig:cilia_array_diagram}).  Fixing $L_1 = L$, we vary the ratio $L_2/L_1$ and explore its effect on cilia self-organisation.  Building from the coordination parameter $Q$ of \citet{guo_bistability_2018}, we introduce the time-dependent parameter
\begin{equation}
    Q(t) = \frac{1}{t_0}\int_{t - t_0}^t\left(\frac{\sum_{i=1}^{N_c^2}\sum_{j=1}^{N_c^2}\alpha_i(t')\alpha_j(t') - N_c^2}{N_c^2 (N_c^2 - 1)}\right)\d t'
\end{equation}
to measure the coordination of $N_c^2$ cilia, where $\alpha_i(t') \in \{-1,1\}$ is the sign of the driving torque on cilium $i$ at time $t'$.  We set the moving window size, $t_0$, to be $t_0 = 5T_c$, where $T_c$ is the period of an isolated cilium.  When $Q = 1$, there is perfect synchrony.  Values of $Q(t)$ close to zero indicate a lack of coordination amongst the cilia.

\begin{figure}
    \centering
    \includegraphics[width=0.7\textwidth]{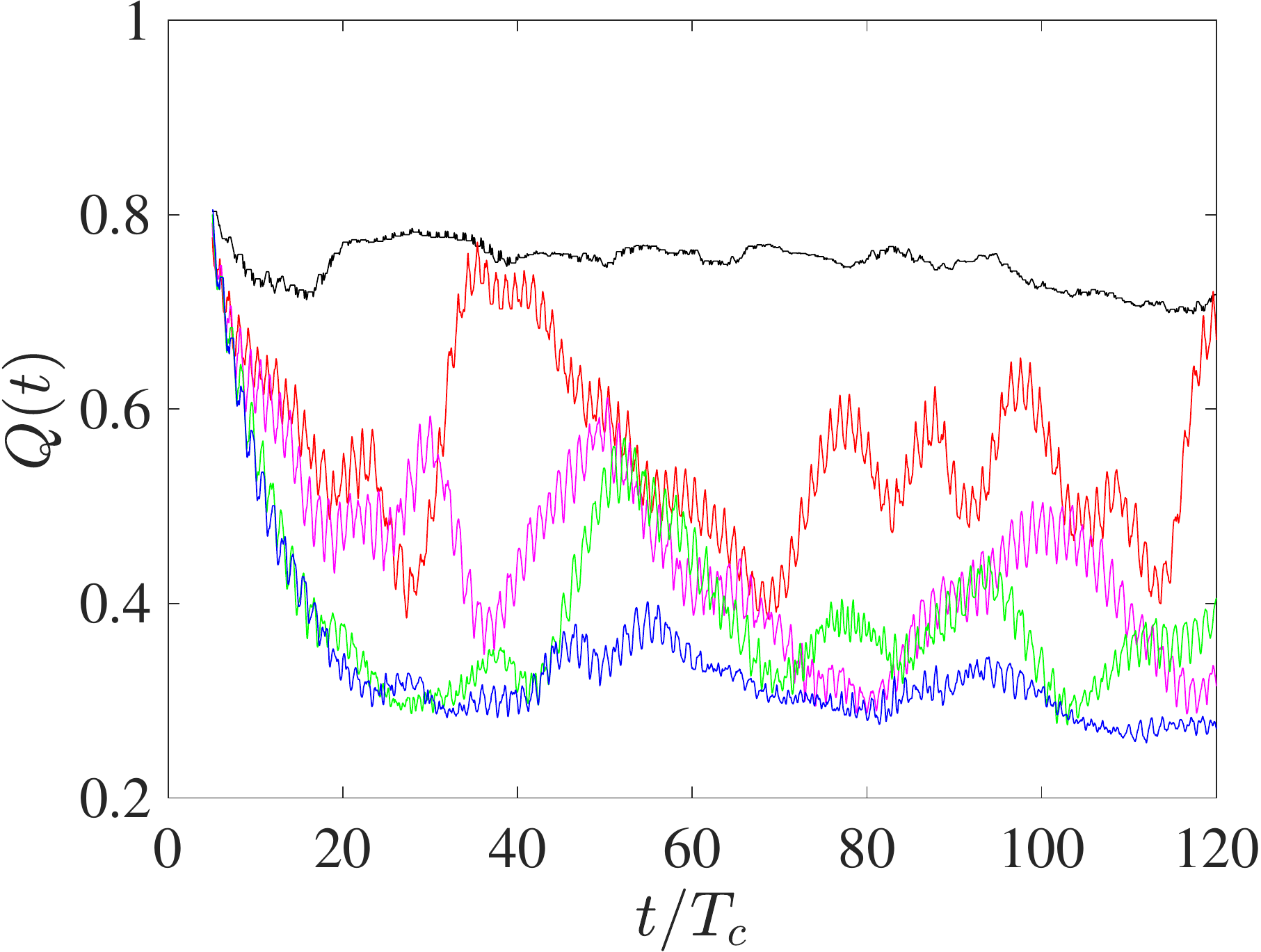}
    \caption{The coordination parameter, $Q$, for differently spaced cilia arrays over time. We fix the distance $L_1 = L$ and vary $L_2$ to examine the behaviour of arrays with $L_2/L_1 = $
    0.5  (\legendline{black}),
    0.75 (\legendline{red}),
    1    (\legendline{magenta}),
    1.25 (\legendline{green}) and
    1.5  (\legendline{blue}).
    }
    \label{fig:cilia_array_moving_window}
\end{figure}

\begin{figure}
    \centering
    \begin{subfigure}[t]{0.55\textwidth}
    \includegraphics[width=0.95\textwidth]{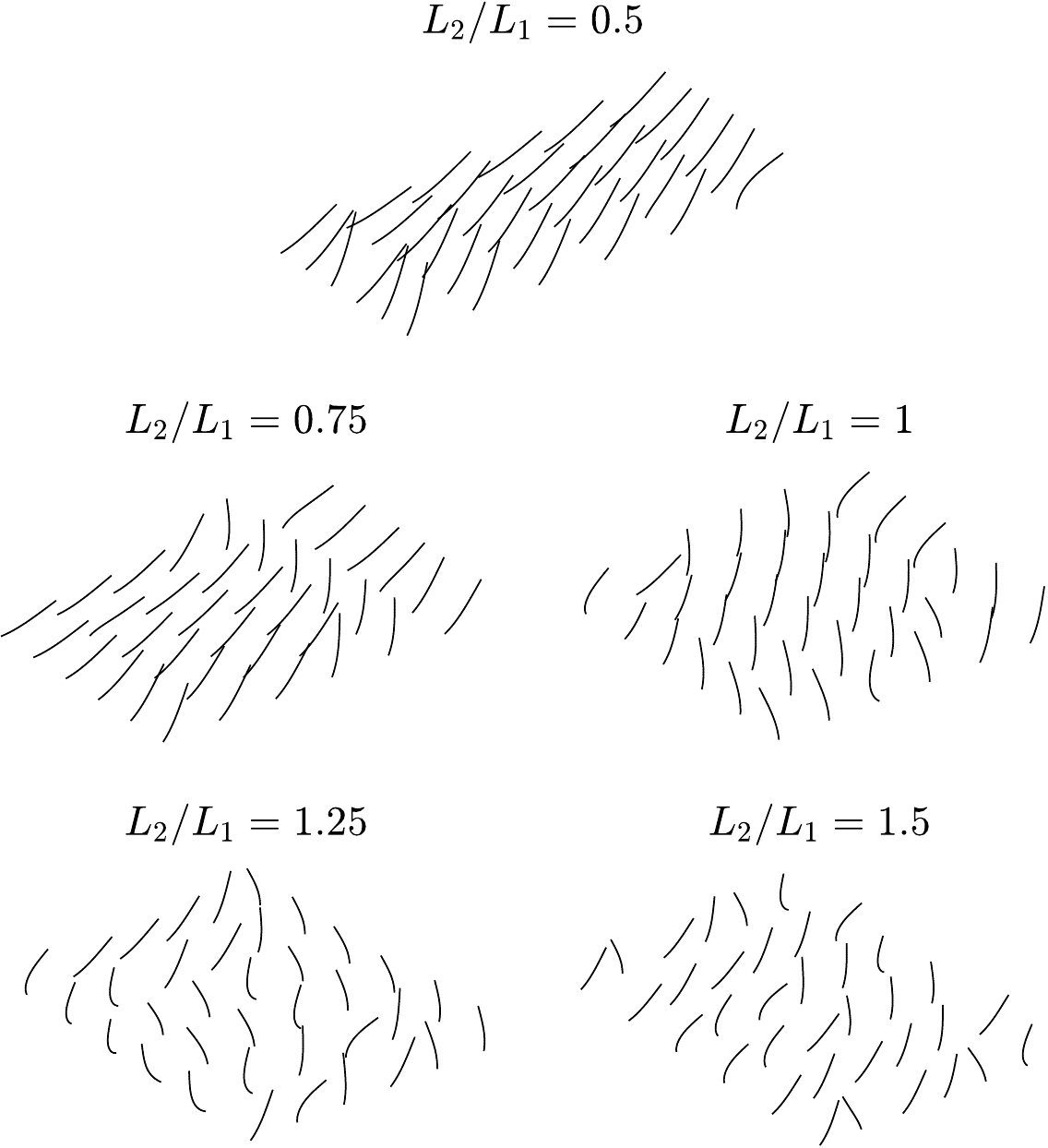}
    \caption{Cilia array snapshots illustrating different levels of coordination.}
    \label{fig:array_snapshot}
    \end{subfigure}
    \begin{subfigure}[t]{0.4\textwidth}
    \includegraphics[width=0.95\textwidth]{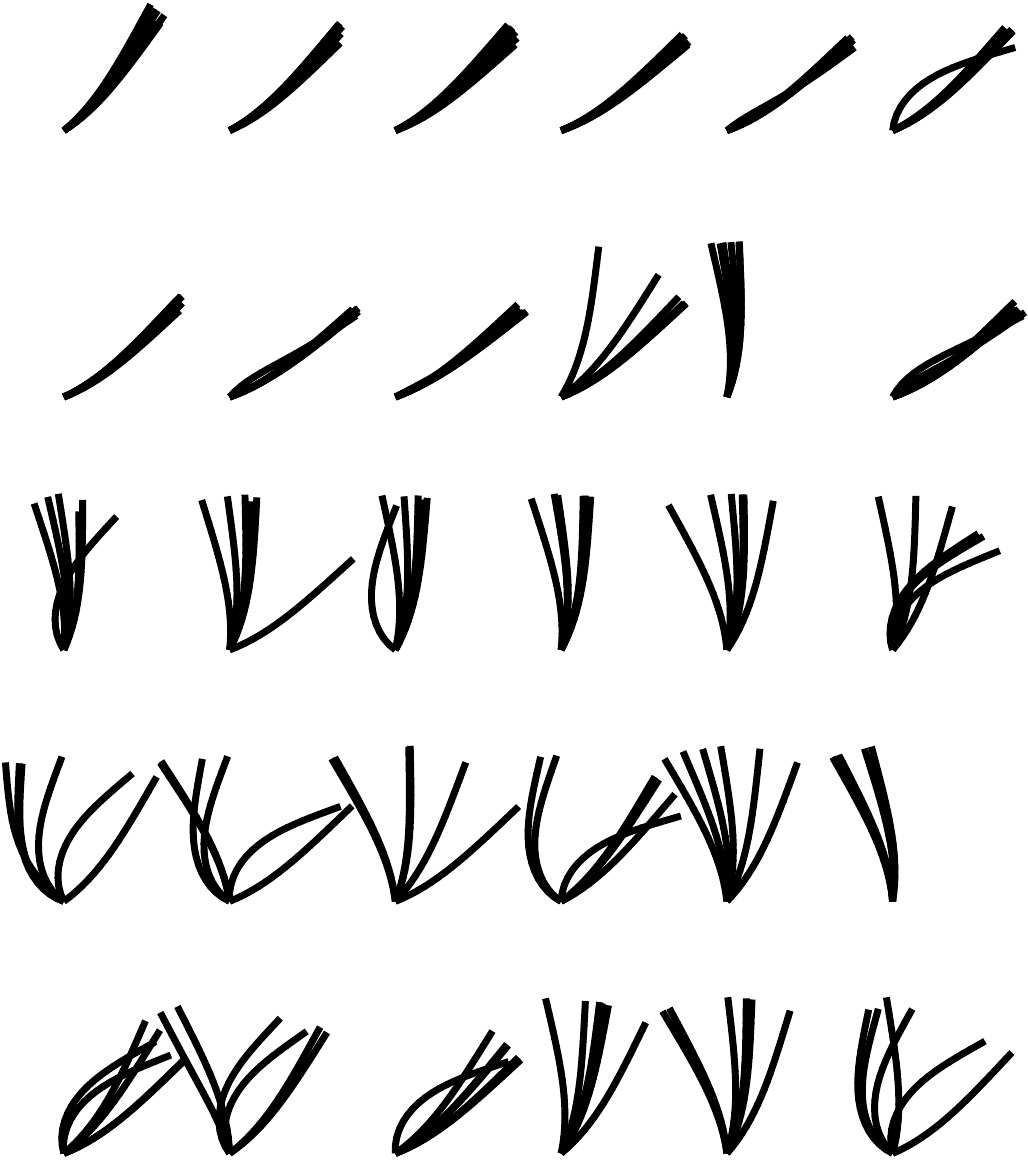}
    \caption{A side view of the arrays in \cref{fig:array_snapshot}. $L_2/L_1$ increases from top to bottom.}
    \label{fig:array_snapshot_side_on}
    \end{subfigure}
    \caption{Cilia array simulations at $t/\Delta t = 6.2 \times 10^4$.}
    \label{fig:array_images}
\end{figure}

In \cref{fig:cilia_array_moving_window}, we show $Q$ over time for arrays of $N_c^2 = 36$ cilia with different values of $L_2/L_1$ ranging from $L_2/L_1 = 0.5$ to $1.5$.  Each array was initialised with all cilia in-phase and the simulations are run to a final time of approximately $120 T_c$. Overall, we find that denser arrays provide a greater level of coordination, as indicated by the higher values of $Q$.  The decrease in coordination with $L_2/L_1$ is also evident in \cref{fig:array_images}, where we show snapshots of the array near the final time for the different cases of $L_2/L_1$.  At the extremes of the $L_2/L_1$ range, $Q(t)$ quickly reaches a value about which it exhibits fairly small fluctuations.  For intermediate $L_2/L_1$, however, we see more complex behaviour where $Q(t)$ fluctuates over a wide range with intermittent periods of coordinated and uncoordinated motion.

\subsection{Bidisperse suspension of undulatory swimmers}

In this section, we employ the 2D version of our computational model to simulate a suspension of slender undulatory swimmers moving in the $xy$-plane.  The suspension consists of two populations (indexed by $p=1$ and $p=2$) that have different swimming gaits.  Similar simulations \citep{agrawal_self-organization_2018} have been performed using multiparticle collision dynamics for two swimmer populations with different frequencies, but the same waveform.

All swimmers have length $L$ and are formed of $N = 30$ segments. The motion of the swimmers is driven by a time-dependent preferred curvature,
\begin{equation}
    \kappa_\nz^{(p)}(s,t) = K_0 \sin\left(\frac{2\pi k^{(p)}}{L} s -\omega t + \phi\right),
\end{equation}
in \cref{equation:3d_moment_frame}, where $K_0$ is the amplitude, $k^{(p)}$ is the wavenumber for population $p$, $\omega=2\pi/T$ is the undulation frequency with $T$ as the undulation period, and $\phi$ is the phase.  The two different swimming gaits are prescribed by specifying different $k^{(1)}$ and $k^{(2)}$.

\begin{figure}
\centering
    \includegraphics[width=0.65\textwidth]{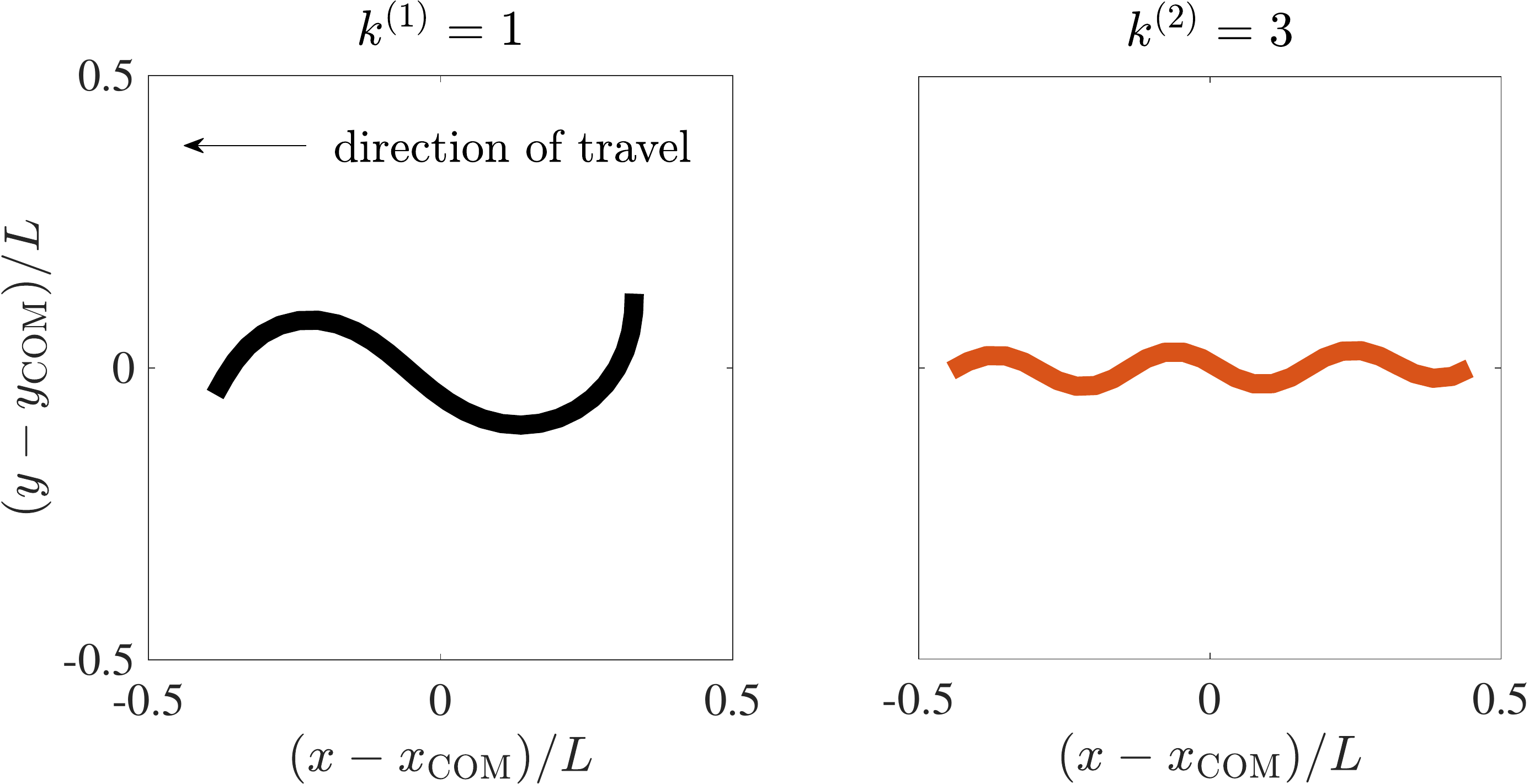}
    \caption{Swimmer shapes for the two different wavenumbers. Figures are centred at the filament centre of mass $(x_{\text{COM}},y_{\text{COM}})$ and the axes are normalised by the swimmer length, $L$.}
    \label{fig:isolated-swimmers}
\end{figure}

In our simulations, the curvature amplitude is $K_0=10.61/L$, while the undulation frequency is set such that the ratio of viscous to elastic forces is $(4\pi\omega\eta/K_B)^{1/4}L = 10$. The phase, $\phi$, for each swimmer is drawn randomly from a uniform distribution.  To achieve sufficiently different waveforms, we set $k^{(1)}=1$ and $k^{(2)}=3$. \Cref{fig:isolated-swimmers} shows waveforms for isolated $p=1$ and $p=2$ swimmers with these parameters.  With the swimming speeds $V_1/(L\omega) = 0.01$ and $V_2/(L\omega) = 0.0024$, the swimmers from population $p=2$ travel at less than a quarter of the speed of those from $p=1$.

In our simulations, we consider a suspension of 100 swimmers with 50 in population $p=1$ and 50 in $p=2$.  The swimmers are all initially straight and distributed uniformly and isotropically in the centre-plane of a periodic domain of size $4.7L\times4.7L\times0.29L$. FCM is used to resolve the hydrodynamic interactions with the fluid flow being solved on a $1024\times1024\times64$ grid. Steric interactions are implemented as described in \cref{sec:steric-interactions}.

\begin{figure}
\centering
    \includegraphics[width=0.8\textwidth]{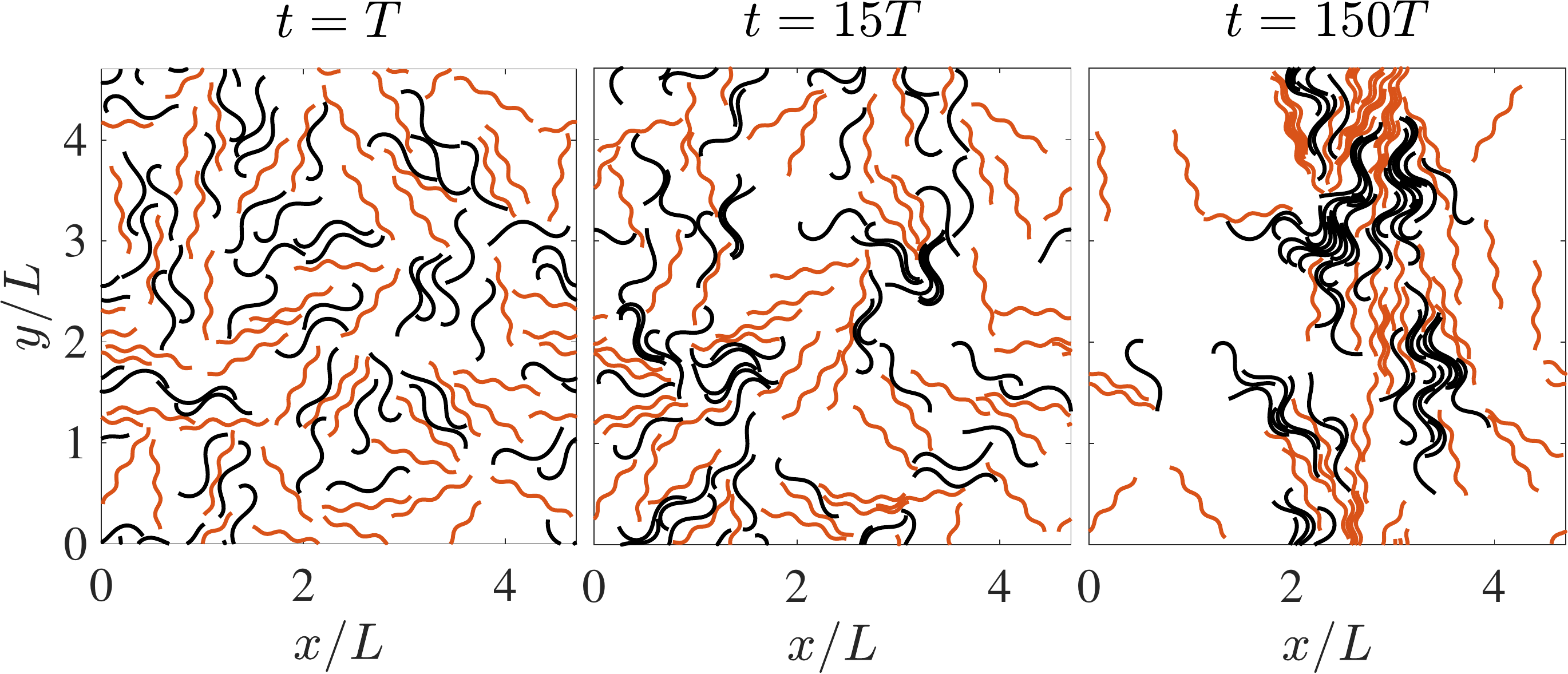}
    \caption{Snapshots at three different times of 100 active filaments in a periodic domain of size $4.7L\times4.7L\times0.29L$. Filaments with $k^{(1)}=1$ are coloured black, while those with $k^{(2)}=3$ are coloured red, as in \cref{fig:isolated-swimmers}. Filaments are initially straight and randomly oriented. }
    \label{fig:swim-clusters}
\end{figure}

The configuration of the suspension at three times is shown in \cref{fig:swim-clusters}. After 150 undulation periods, we see that all swimmers have clustered, with swimmers tending to cluster with those from their own population, i.e.\ with a similar waveform.  Clustering is also observed in planar monodisperse suspensions of model sperm cells \citep{schoeller_flagellar_2018,yang_cooperation_2008} and slender undulatory swimmers \citep{yang_swarm_2010}, as well as in planar bidisperse suspensions \citep{agrawal_self-organization_2018} of undulatory swimmers.  We first observe swimmers from $p=1$ forming clusters with each other, while the slower swimmers from $p=2$ remain dispersed in the surrounding fluid.  Once $p=1$ clusters are established, we begin to see $p=2$ swimmers joining these clusters, but tending to remain close to other $p=2$ swimmers.  Interestingly, filament softness has contributed to the development of these slightly mixed clusters. We sometimes observed high-wavenumber $p=2$ swimmers deforming to match the shape of neighbouring low-wavenumber $p=1$ swimmers.

\subsection{Sedimentation}
In this section, we use our methodology to study filament sedimentation.  We both revisit the case of a single filament falling under gravity, as well as explore cases of multiple filaments, ranging from small ensembles to clouds and suspensions.  We address these situations in cases where filament motion is restricted to a plane, as well as those where their motion is completely three-dimensional.

\subsubsection{Settling filament}\label{sec:sim-single-settling-filament}
We first revisit the case of a single filament settling under gravity and measure its deformation and effective drag coefficient.  This problem, first investigated numerically by \citet{cosentino_lagomarsino_hydrodynamic_2005} and more recently in experiments by \citet{marchetti_deformation_2018}, has also been used previously as a test problem for filament models \citep{keaveny_dynamics_2008,delmotte_general_2015}.

\begin{figure}
    \includegraphics[width=0.95\textwidth]{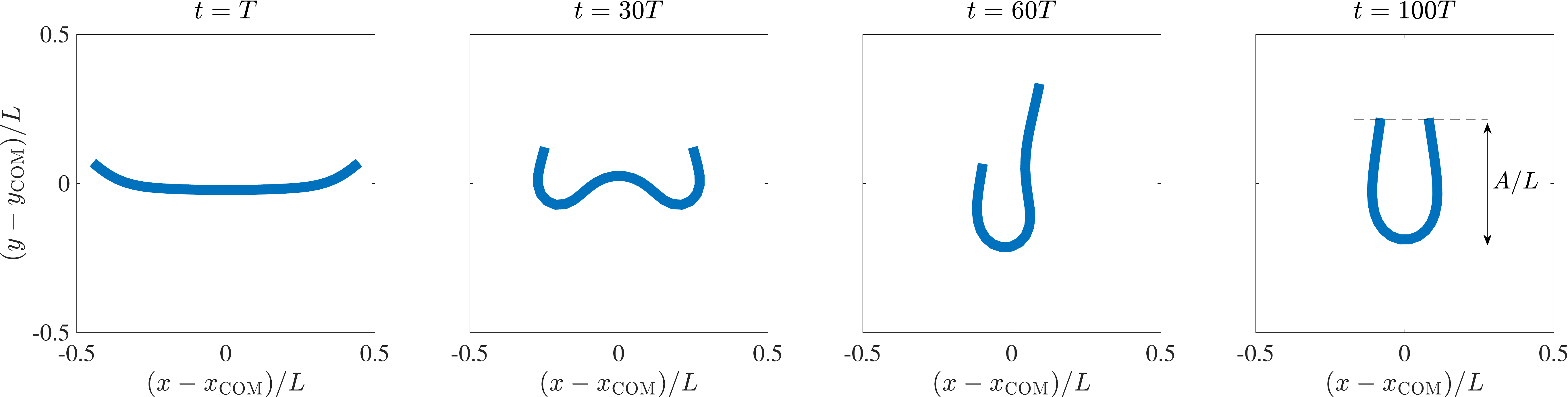}
    \caption{A filament with $N = 31$ and $B=10^4$, is placed horizontally and allowed to sediment vertically under a constant force per unit length. Upon reaching its final steady shape, the filament's normal deflection, $A$, is measured. Figures are centred at the filament centre of mass $(x_{\text{COM}},y_{\text{COM}})$ and are normalised by the length of the filament, $L$. }
    \label{fig:sed-at-different-times}
\end{figure}

In our simulations, we consider an initially straight and horizontal filament formed of $N=31$ segments in an unbounded fluid with the hydrodynamic interactions resolved using the RPY tensor with hydrodynamic radius, $a$.  As in \cref{sec:numerical-tests-and-method-choice}, a constant force per unit length of magnitude $W$ is applied vertically to the filament.  Using the parameters arising in the problem, we can define the settling time as $T=\eta L/W$, as well as the elasto-gravitational number $B=L^3 W/K_B$.  Running the simulation with timestep size $\Delta t = T/30$, the filament is allowed to sediment until it reaches a steady-state shape and steady-state settling speed, $V_s$.  The filament at various points in time is shown in \cref{fig:sed-at-different-times} for the case where $B = 10^4$.  Once the steady state is reached, we measure the normal deflection $A$, defined as the distance between the highest and lowest points of the filament (see \cref{fig:sed-at-different-times}), and the effective drag coefficient $\gamma/\gamma^0$, where $\gamma = WL/V_s$ and $\gamma^0$ is the value of $\gamma$ in the limit $B \rightarrow 0$ corresponding to a straight filament.

\begin{figure}
    \centering
    \begin{subfigure}[t]{0.45\textwidth}
    \includegraphics[width=0.95\textwidth]{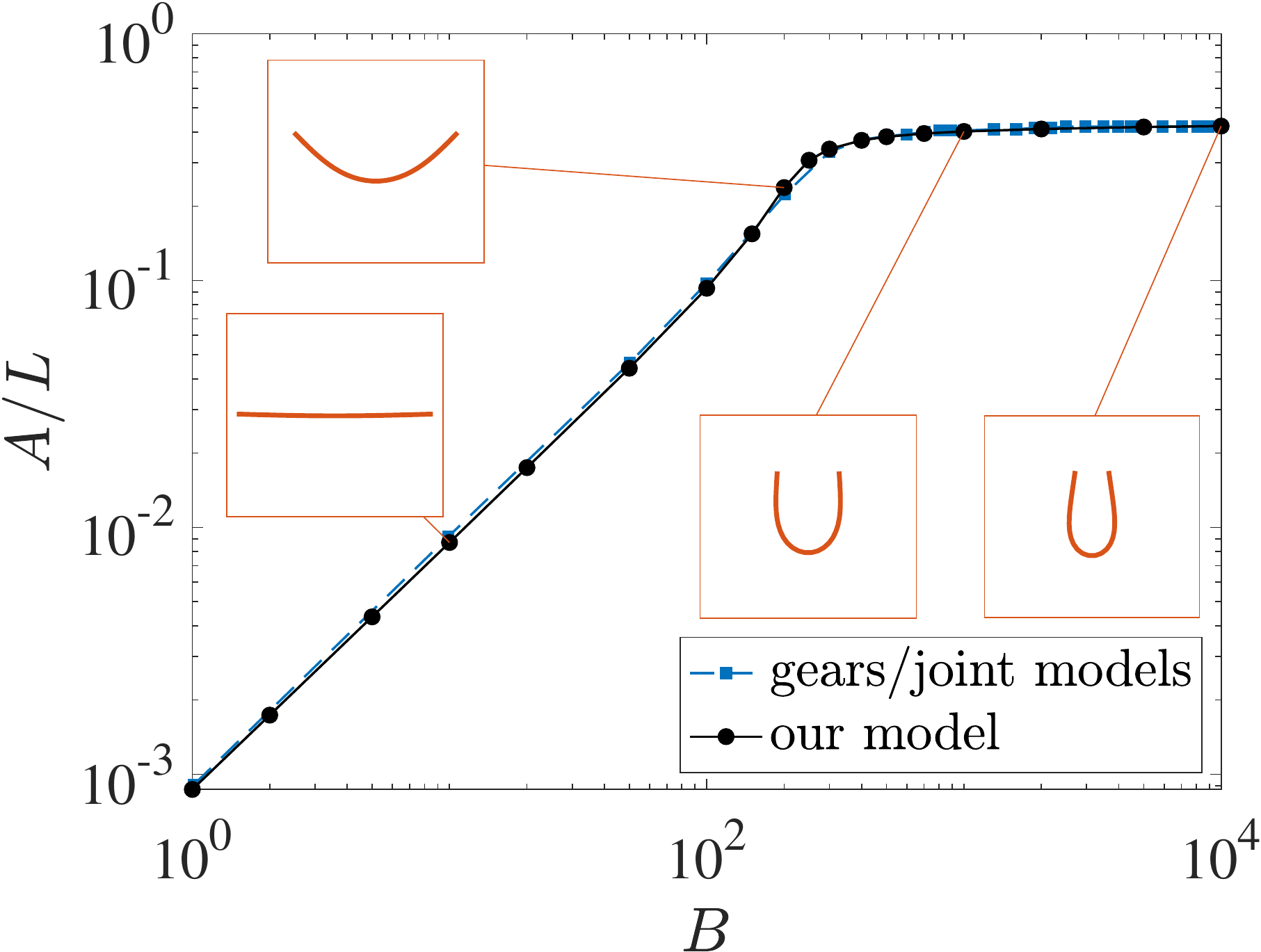}
    \caption{Normalised deflection, $A/L$, as a function of the elasto-gravitational number, $B$. The final filament shapes at certain values of $B$ are shown in the inset plots.}
    \label{fig:a-vs-b}
    \end{subfigure}
    \begin{subfigure}[t]{0.45\textwidth}
    \includegraphics[width=0.95\textwidth]{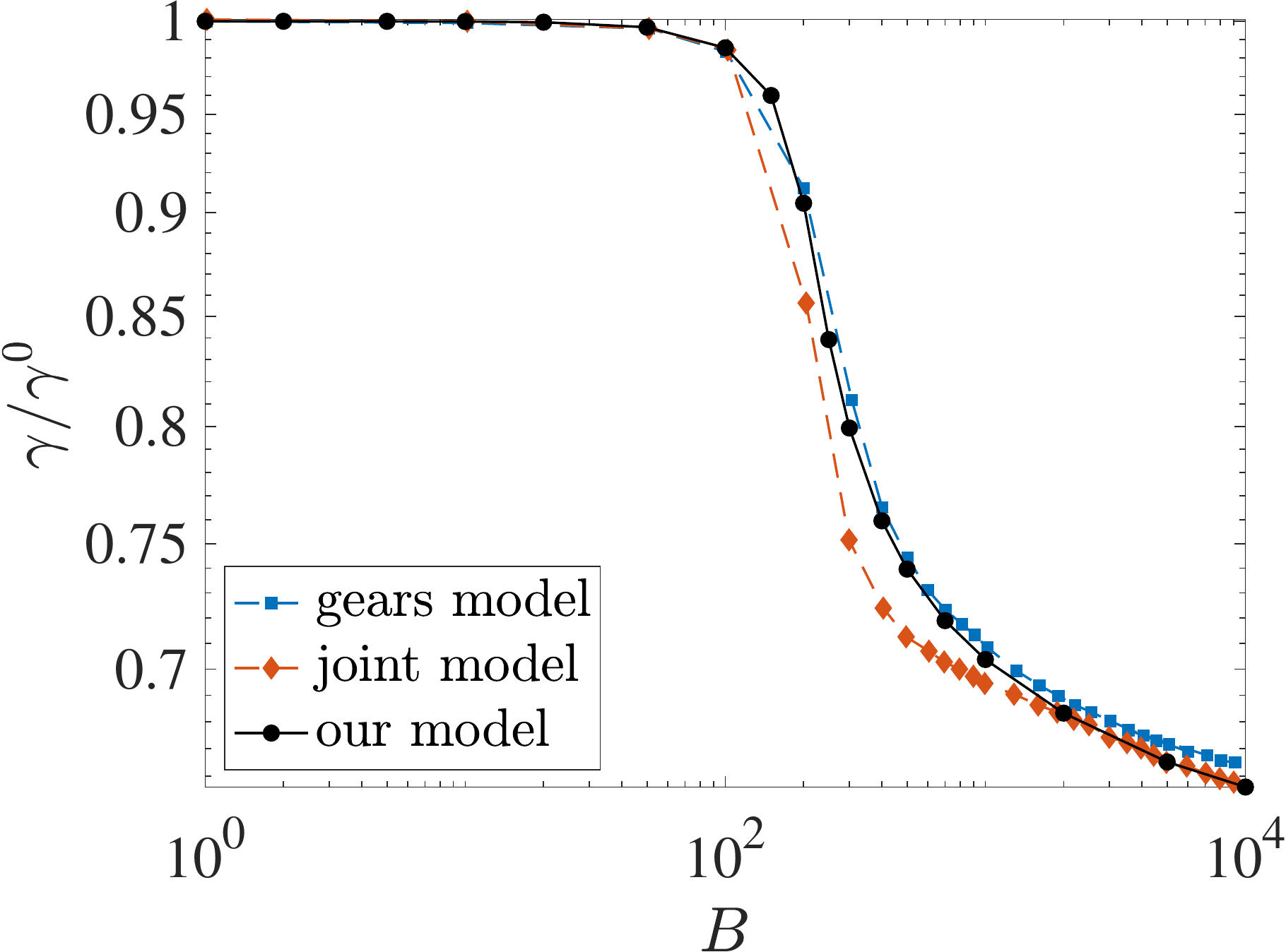}
    \caption{Normalised effective drag coefficient $\gamma/\gamma^0$, as a function of the dimensionless elasto-gravitational number, $B$.  Here, $\gamma^0$ is the drag coefficient for a rigid rod (i.e.\ at $B\rightarrow 0$).}
    \label{fig:drag-vs-b}
    \end{subfigure}
    \caption{Normal deflection and effective drag coefficients for a single sedimenting filament as a function of $B$. Results are compared to the gears and joint model results from \citet{delmotte_general_2015}. Simulations were performed with a filament formed of $N = 31$ segments in an infinite domain.}
    \label{fig:a-and-drag-vs-b}
\end{figure}

The values of $A/L$ and $\gamma/\gamma^0$ as a function of $B$ are presented in \cref{fig:a-and-drag-vs-b}.  We compare our results with those presented in \citet{delmotte_general_2015} for both their gears model, where filaments are formed of 34 particles with no separation between the particles, and a joint model, formed of 31 particles with gaps of $0.2a$ between the particles.  As we have $\Delta L = 2.2a$, our spacing is equivalent to that used in the joint model.  In all cases, the RPY mobility matrices are used to capture hydrodynamics interactions.

All models show excellent agreement in their respective measurements of normal deflection, $A/L$.  In all cases, the models predict the same initial linear increase of $A/L$ with $B$, followed by a plateau to the same value at high $B$.  For the drag coefficient, $\gamma/\gamma^0$, the differences, though still small, are more pronounced.  In particular, our measurements find good agreement with the gears model for $B\lesssim 10^3$ as the sudden decay in $\gamma/\gamma^0$ given by both models coincide.  For higher $B$, however, we observe better agreement with the joint model, which yield a similar decay with increasing $B$. We suspect that the agreement with the gears model for lower $B$ is due to the better resolution of the bending moments given by these models.  At higher $B$, we attribute the agreement between our results and those given by the joint model to the similar viscous force on the filament that results from setting $\Delta L = 2.2a$.

\subsubsection{Sedimentation of large 2D suspensions}\label{sec:sedimentation-2d-large}
Here, we investigate the sedimentation of a large, planar suspension of filaments in a periodic domain.  This set of simulations was inspired by \citet{gustavsson_gravity_2009, tornberg_numerical_2006}, where a similar simulation was also performed in a periodic domain, but for a suspension of rigid filaments that can move in all three dimensions.  We note as they do, that sedimentation in periodic domains, while a canonical problem, is quite different from that in unbounded or bounded systems and can exhibit dependencies on domain sizes.  Nevertheless, these simulations do provide a context with which to explore the effects of filament flexibility on sedimentation.

The simulations presented here are larger versions of the test problem described in \cref{sec:numerical-tests-and-method-choice}.  We also run these simulations for longer times.  Specifically, we consider monolayers of 100, 500, and 1000 filaments, each formed of $N=15$ segments.  The dimensionless elasto-gravitational number is set at two values, $B = 1$ and $1000$, to examine the effect of deformation on suspension dynamics as the filaments settle for as long as $500T$.  The timestep is chosen such that $\Delta t = T/300$.  The filaments are initially straight and are distributed uniformly and isotropically in the centre-plane of a periodic domain of size $18.8L \times 18.8L \times 1.2 L$. FCM is used to resolve the hydrodynamic interactions between the filaments, with the Stokes equations being solved on a grid of $2048\times2048\times128$ points. Steric interactions are implemented as in \cref{sec:steric-interactions}.

\begin{figure}
    \centering
    \includegraphics[width=0.6\textwidth]{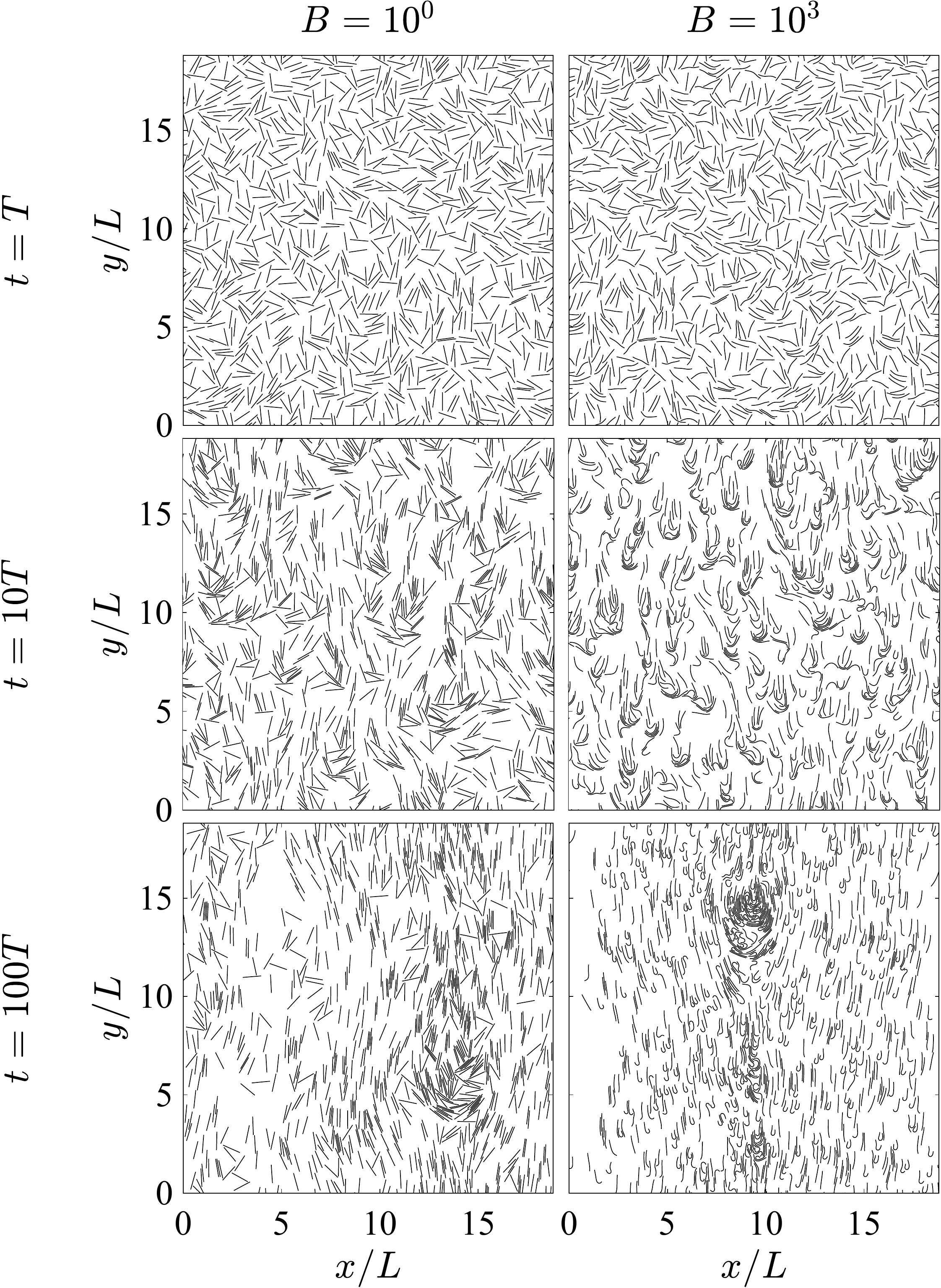}
    \caption{Snapshots of 1000 filaments sedimenting in a periodic domain of size $18.8L \times 18.8L \times 1.2 L$, at three different times for two different values of $B$. Filaments are initially straight and randomly oriented.}
    \label{fig:m1000-snapshots}
\end{figure}

Snapshots of the sedimenting suspensions for $M=1000$ with $B=1$ and $B=1000$ are shown in \cref{fig:m1000-snapshots}. In general, we find that both stiff ($B=1$) and flexible ($B=1000$) filaments over time form a single large cluster surrounded by isolated filaments that are found to tumble, but are most often oriented in the direction of gravity.  We do see, however, that for the flexible filaments, the cluster formed is of much higher density, and the filaments surrounding it can buckle and flex, as well as tumble, due to the shear induced by the falling cluster.

\begin{figure}
    \centering
    \includegraphics[width=0.8\textwidth]{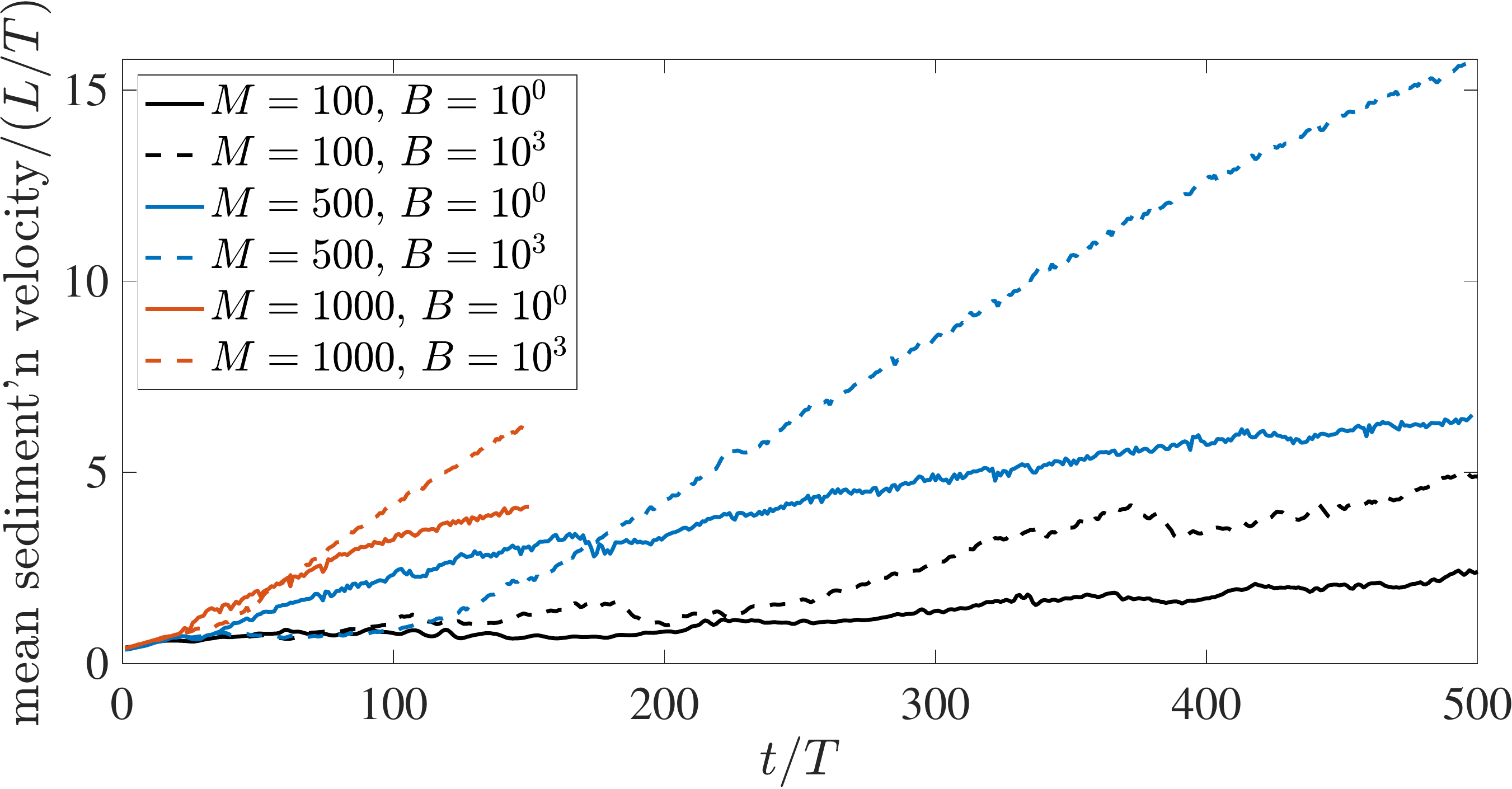}
    \caption{Mean sedimentation velocity, $\bar{w}$, of a suspension of $M$ ($N=15$) filaments in a periodic domain of size $18.8L \times 18.8L \times 1.2 L$. Simulations performed for two values of $B$.}
    \label{fig:sed-velocity-over-time}
\end{figure}

Following \citet{gustavsson_gravity_2009}, for each of the simulations that we have performed, we compute the mean sedimentation velocity,
\begin{align}
\bar{w}(t) = \sum_{m=1}^M w_m(t),
\end{align}
where $w_m(t)$ is the vertical centre-of-mass velocity of filament $m$.  The resulting values over time are shown in \cref{fig:sed-velocity-over-time}. We observe that for each concentration, the settling velocity and its growth rate increase with filament flexibility (higher $B$).  We suspect that this is a result of two factors.  First, as we observed in \cref{sec:sim-single-settling-filament}, flexibility leads to a reduction in the filaments' effective drag coefficient.  Second, we also see in \cref{fig:m1000-snapshots} that deformable filaments form denser clusters.  This increase in local filament density in turn enhances interactions, yielding higher sedimentation velocities.

In previous work \citep{mackaplow_numerical_1998,gustavsson_gravity_2009} on rigid fibre sedimentation in periodic domains, the onset of a plateau in the mean settling velocity coincided with the formation of a single cluster in the centre of the domain that is surrounded by an upwelling of clear fluid.  Additionally, observed fluctuations about the plateau value corresponded to cluster break-up and reformation events that occur periodically.  In our case, we find that even after $500T$, the mean settling velocity continues grow with time.  For each case of $M$, the setting velocity growth rate is higher for the more flexible filaments, but perhaps more surprising is that the growth rate appears to be independent of $M$.  Indeed, increasing the filament density by setting $M=1000$, we see that over the $150T$ that for both $B$, the mean sedimentation velocity parallels that for $M=500$.  The continued growth may be attributed to the observation that there many filaments have yet to join the cluster.  Additionally, we suspect that the continued growth in sedimentation velocity as compared to the fully three-dimensional cases \citep{mackaplow_numerical_1998,gustavsson_gravity_2009} is due to both the larger lateral distances in our simulation that the filaments must traverse, as well as their restriction to move only in two dimensions.  We suspect also that the reduced out-of-plane thickness enhances the hydrodynamic interactions between the filaments, allowing for the higher growth rates in our simulations.

\subsubsection{Tumbling, settling and buckling of a filament square}

\citet{gustavsson_gravity_2009} have shown that rigid fibres evenly distributed along a circle can periodically tumble as they settle.  Similar periodic orbits have been seen both numerically \citep{claeys_suspensions_1993} and experimentally \citep{,jung_periodic_2006} for ellipsoids and other rigid structures released from an initially symmetric configuration.  Here, we explore how the introduction of filament flexibility can affect these dynamics.

\begin{figure}
  \centering
  \includegraphics[width=.8\textwidth]{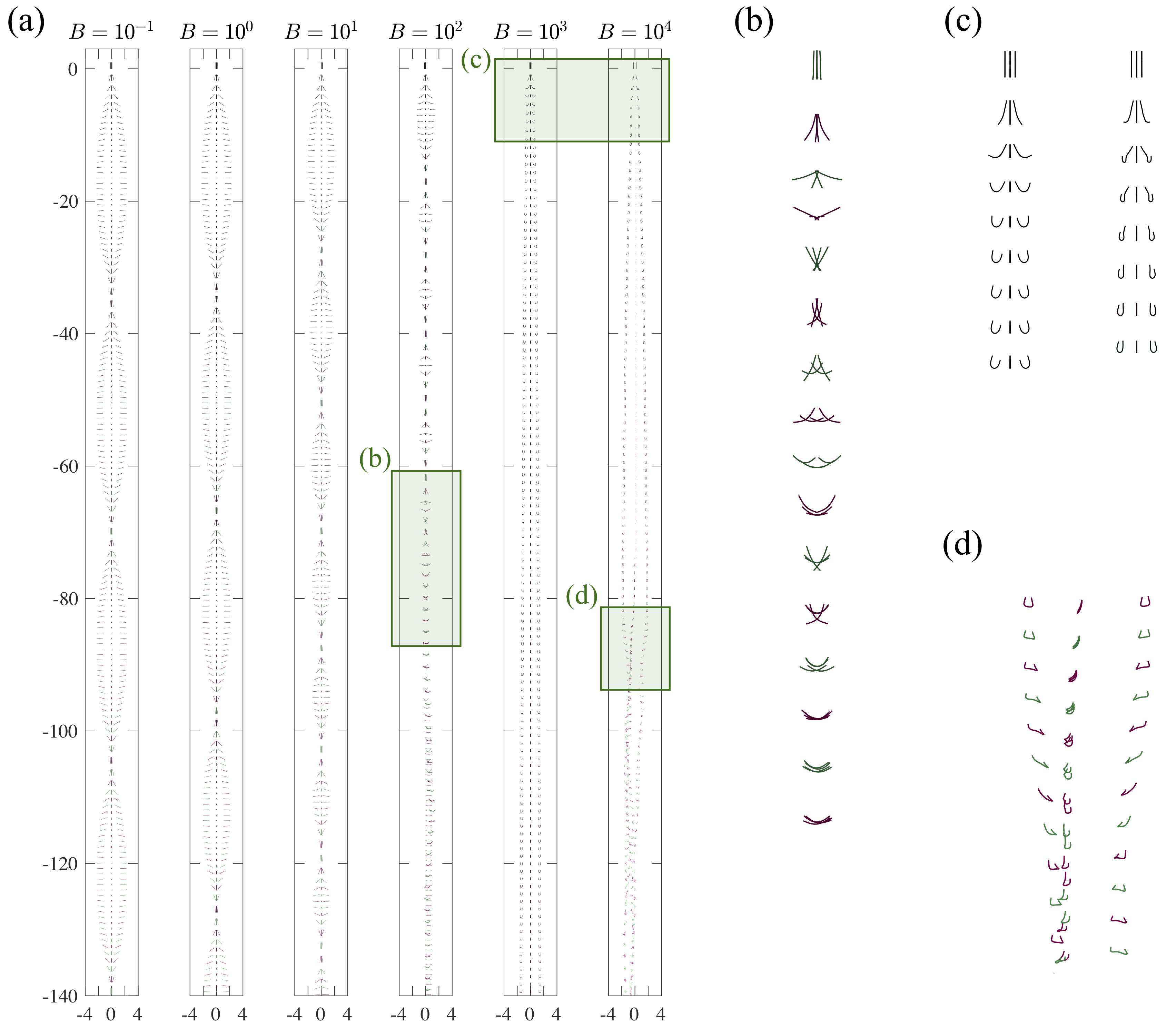}
  \caption{Settling dynamics of four, initially vertical filaments for different values of $B$.  The filaments are shown at every $2T$, with the colours alternating between green and magenta. (a) Stiffer filaments ($B\ll 10^2$) tumble periodically.  Though bending is not appreciable for these cases, the tumbling period decreases as $B$ increases.  Softer filaments ($B = 10^3$) deform considerably and do not tumble, but may buckle if $B\gg10^3$. (b) For $B=10^2$, filaments interact through steric forces, leading to symmetry breaking. (c) For $B=10^3$ (left) filaments immediately bend at their centres, while for $B=10^4$ (right), the filaments first buckle at their lower end and this bend propagates to their centres as they fall. (d) For $B=10^4$, the configuration achieved after the filaments deform is unstable.  As perturbations grow, symmetry is broken and more complex trajectories emerge.}
  \label{fig:GTpanel}
\end{figure}

In our simulations, we consider four filaments whose centres coincide with the corners of a horizontal square of side length $D/L = 0.268$.  The filaments are initially aligned vertically in the direction of gravity.  We discretise the filaments into $N=30$ segments and, to allow for an unbounded fluid, we resolve their hydrodynamic interactions using the RPY mobility matrix.  Steric interactions are implemented as in \cref{sec:steric-interactions}.  The timestep is chosen such that $\Delta t = T/300$.  The elasto-gravitational number, $B$, is varied from $B = 10^{-1}$ (very stiff filaments) to  $B = 10^{4}$ (very soft filaments).  The long-term behaviour reveals significant differences between these cases, including symmetry breaking (\cref{fig:GTpanel}).  We note that similar observations have been made previously \citet{llopis_sedimentation_2007, saggiorato_conformations_2015}.

At the lowest value of $B=10^{-1}$, we observe periodic tumbling (\cref{fig:GTpanel}a) where the filaments first separate and then reapproach when they return to their initial configuration. This matches the periodic orbits seen in \citet{gustavsson_gravity_2009, jung_periodic_2006}.  For $B = 1$ and $B=10$, while we again observe the filaments returning very close to their initial configuration, we see that the time it takes to return to this configuration decreases with $B$, as does the distance between the filaments.  For the case $B=10^2$, when returning after tumbling, the filaments, whose deformation is now apparent, approach each other so closely that they collide and experience steric interactions (\cref{fig:GTpanel}b).  At this point, the filaments cease tumbling and instead begin to settle as a group, rearranging their relative positions as they move downwards.  At $B=10^3$, we see that filaments do not exhibit any tumbling and instead adopt a horseshoe-like shape and a steady configuration as they settle, see \cref{fig:GTpanel}c.  Finally, for $B = 10^4$, while we see that at early times the dynamics are similar to those for $B=10^3$, we observe also that this state is unstable.  The accumulation of round-off errors are sufficient to trigger the instability (\cref{fig:GTpanel}d) and the onset of more complex filament behaviour.

\subsubsection{Sedimentation of flexible filament clouds}
As a final demonstration of our methodology, we consider the dynamics of a cloud of flexible filaments settling under gravity.  For clouds of rigid, spherical particles \citep{nitsche_break-up_1997,metzger_falling_2007}, as the cloud settles, the particles within the cloud move along trajectories reminiscent of the streamlines within a settling spherical drop \citep{batchelor_introduction_1967}.  As the cloud settles, it sheds particles at the cloud axis, until it becomes a torus, leading to the cloud dividing.  While similar studies have been performed for clouds of rigid fibres \citep{park_cloud_2010,nazockdast_fast_2017}, showing how particle anisotropy accelerates cloud break-up, the role of flexibility, to the best of our knowledge, has not been explored.

\begin{figure}
  \centering
  \includegraphics[width=\textwidth]{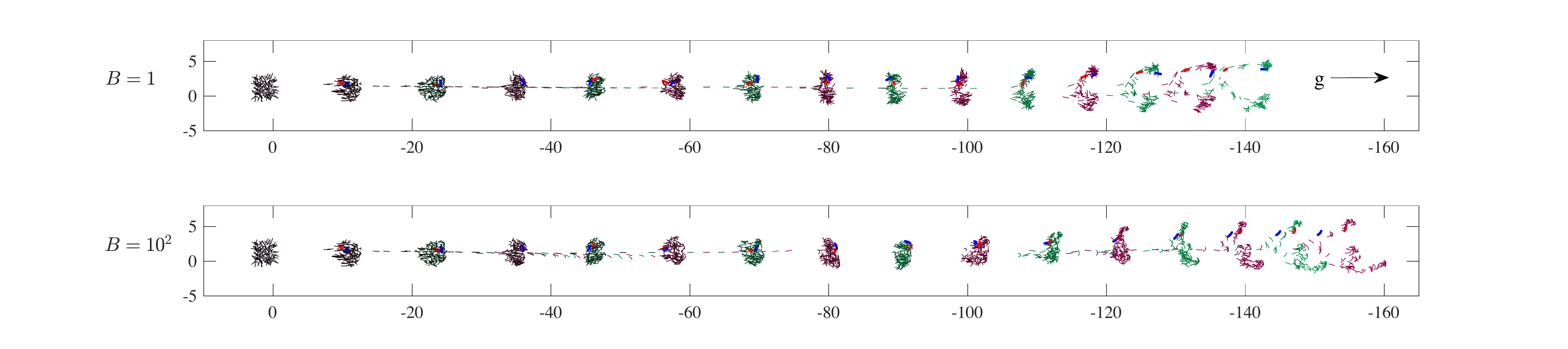}
  \caption{Sedimentation and break-up of a dense, three-dimensional cloud of $M=120$ filaments with $B = 1$ and $B = 10^2$. Filament positions are initially in a cube of linear size $2.5 L$. The cloud is shown every $3T$ with the colour alternating between magenta and green and fading with time. Two filaments are highlighted (red and blue). The arrow (g) indicates the direction of gravity.}
  \label{pic:CloudB100}
\end{figure}

\begin{figure}
  \centering
  \includegraphics[width=\textwidth]{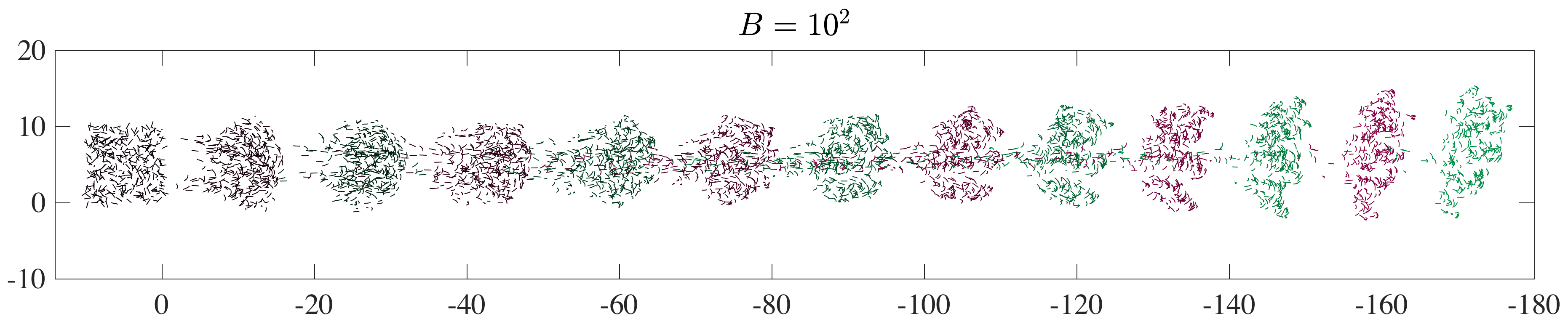}
  \caption{Sedimentation of a three-dimensional cloud of $M = 360$ filaments with $B=10^2$. Initially, all filaments have at least one end in a cube with side $10L$. The cloud is shown every $4T$ with the colour alternating between magenta and green and fading with time.}
  \label{pic:CloudBigB100}
\end{figure}

In our fully three-dimensional cloud simulations, the filaments are distributed uniformly and isotropically in cubes  (\cref{pic:CloudB100,pic:CloudBigB100}) of side length $2.5 L$ and $10L$ using the initialisation method describe in \cref{sec:isoinit}.  The filaments are discretised into $N=20$ segments and the timestep size is $\Delta t = T/200$.  The hydrodynamic interactions in the unbounded domain are resolved using the RPY motility matrix and steric interactions are implemented as in \cref{sec:steric-interactions}.  To highlight the effect of filament flexibility, simulations were performed with elasto-gravitational numbers $B = 1$ and $B = 10^2$.

\cref{pic:CloudB100} shows snapshots from the cloud simulations with $M=120$ filaments for both $B=1$ and $B=10^2$.  The initial conditions for these simulations are identical.  At short times, we observe that both clouds exhibit very similar behaviour. Later in the simulations, however, marked differences arise. The cloud of stiff filaments ($B=1$) splits into two clusters as a result of filament depletion at its centre.  This behaviour matches results from \citet{park_cloud_2010}. The cloud consisting of more flexible filaments ($B=10^2$) breaks up later, and in a much more complicated way.  We see that it disintegrates into several clusters, rather than splitting in two.  We suspect that this could be connected to individual filaments' orbits within clouds, as well as their sedimentation velocities.  We have also performed a larger, though less dense, cloud simulation with $M = 360$ filaments and $B=10^2$.  As shown in \cref{pic:CloudBigB100}, we find that the features of the cloud's initial shape persist as the cloud overturns and only starts to disappear once the cloud begins to break up.

\section{Discussion and conclusions}

In this paper, we have presented a scalable, efficient and flexible method for performing large-scale simulations of passive and active filaments and their suspensions.  In particular, we have shown how to describe the fully 3D filament deformations due to bending and twisting through the use of quaternions, and in doing so, demonstrated how quasi-Newton methods coupled with an implicit, geometric multi-step scheme can be used to overcome the numerical stiffness associated with deformation while ensuring constraints on filament motion are satisfied.  Our framework can be easily integrated with many existing models and methods for resolving hydrodynamic interactions.  In addition, the second-order spatial discretisation could be substituted for a higher-order finite difference method on the equispaced grid, though new expressions for the approximate Jacobian would need to be established given the change in stencil.  Our usage of iterative schemes for the nonlinear systems that arise ensures that efficient, matrix-free methods for the hydrodynamics are also applicable, which we have also demonstrated.  We have shown that our method can be used to study many relevant problems in biofluid dynamics, as well as suspension mechanics, potentially at scales previously inaccessible.  We encourage the reader to experiment with the MATLAB/Octave implementation \citep{schoeller_github_2019} of the method to study similar problems or expand upon it to explore new directions.

While we have developed and explored the method exclusively in the limit where fluid inertia is negligible, it would be of interest to understand how our framework could be adapted for use with fluid solvers where inertia is included \citep{wiens_simulating_2015}, to study, for example, filament suspensions in turbulent flows \citep{rosti_flexible_2018}.  Additionally, it could be of interest to explore filament motion in non-Newtonian fluids, such as viscoelastic fluids, especially with respect to modelling flagellar motion and cell swimming \citep{teran_viscoelastic_2010}.  Our method could be used in the context where the filaments are components in more complex physical systems, such as networks, or, by incorporating Brownian motion \citep{keaveny_fluctuating_2014,delong_brownian_2015}, suspended polymers in solution, or entangled polymers in melts.

\section*{Acknowledgements}
EEK and AKT gratefully acknowledge support from EPSRC Grant EP/P013651/1. SFS gratefully acknowledges funding by an Imperial College President's PhD scholarship.  TAW is thankful for funding through an EPSRC Studentship (Ref: 1832024).  All authors also thank Smitha Maretvadakethope for a detailed reading of the manuscript.

\appendix
\section{Quaternions}

\subsection{Quaternion initialisation} \label{appendix:initial_quaternion}

In this section we detail how to initialise quaternion encoding the rotation from the standard basis $\{\uv{e}_x, \uv{e}_y, \uv{e}_z\}$ to the material frame $\{\uv{\nx}, \uv{\ny}, \uv{\nz}\}$. Since $\uv{\nz} = \uv{\nx}\times\uv{\ny}$, we need only provide initial conditions for the tangent and one of normal vectors. The main idea is to determine the quaternion which rotates a given unit vector to another and then apply this twice in succession by first mapping $\uv{e}_x$ to $\uv{\nx}$ and then mapping the image of $\uv{e}_y$ under this first rotation to $\uv{\ny}$.

Let $\uv{a}$ and $\uv{b}$ be unit vectors. Recall that $\uv{a}\cdot\uv{b} = \cos \theta$, $\uv{a}\times\uv{b} = \sin\theta\,\uv{u}$, where $\uv{u}$ is normal to the plane spanned by $\uv{a}$ and $\uv{b}$, and $\theta$ is the angle between the two vectors as measured anticlockwise about $\uv{u}$. Recalling \cref{equation:quaternion_polar_form}, we see that the quaternion $(\uv{a}\cdot\uv{b},\; \uv{a}\times\uv{b})$ represents the rotation of $2\theta$ about the vector $\uv{u}$.  Since we need only rotate $\uv{a}$ by $\theta$ to align with $\uv{b}$, we seek the quaternion $\q{q}
$ for which $\q{q}^2 = \q{q}\qprod\q{q} = (\uv{a}\cdot\uv{b},\; \uv{a}\times\uv{b})$. By expanding and rearranging $\q{q}^2 = \q{p}$ we find that a square root of a quaternion may be defined as
\begin{equation}
\left(\cdot\right)^{1/2} : \left(p_0,\v{p}\right) \mapsto \twopartdef{\left(\sqrt{\frac{p_0 + 1}{2}},\frac{\v{p}}{\sqrt{2\left(p_0 + 1\right)}}\right)}{p_0 \neq -1,}{\left(0,0,0,1\right)}{p_0 = -1.}
\end{equation}
It is readily checked that for all $\q{p} \in \mathbb{U}$: (a) $\q{p}^{1/2} \in \mathbb{U}$, and (b) $\q{p}^{1/2}\qprod\q{p}^{1/2} = \q{p}$. With this defined, we produce the quaternion initial conditions as follows:
\begin{enumerate}
\item Let $\uv{\nx} = (\hat{\nx}_1,\hat{\nx}_2,\hat{\nx}_3)$ and $\uv{\ny} = (\hat{\ny}_1,\hat{\ny}_2,\hat{\ny}_3)$ be the initial tangent and normal vectors to the filament centreline.
\item Evaluate the quaternion which maps $\uv{e}_x \mapsto \uv{\nx}$ as $\q{p}_1 = (\uv{e}_x\cdot\uv{\nx},\;\uv{e}_x\times\uv{\nx})^{1/2} = (\hat{\nx}_1,0,-\hat{\nx}_3,\hat{\nx}_2)^{1/2}$.
\item Let $\uv{n}' = \t{R}(\q{p}_1)\uv{e}_y$ be the image of the $y$-axis direction under this first rotation.
\item Evaluate the rotation mapping $\uv{n}' \mapsto \uv{\ny}$ as $\q{p}_2 = (\uv{n}'\cdot\uv{\ny},\;\uv{n}'\times\uv{\ny})^{1/2}$
\item Assign the quaternion an initial value of $\q{p}_2\qprod\q{p}_1$.
\end{enumerate}

\subsection{Internal moments in terms of quaternions} \label{appendix:quaternion_elasticity}

In this section, we derive an expression for the internal moments, \cref{equation:3d_moment_frame}, expressed solely in terms of the unit orientation quaternion. Recall \cref{equation:3d_moment_frame}:
\begin{equation}
  	\v{M} = K_B \left(\uv{\ny}\left(\uv{\nx}\cdot\pd{\uv{\nz}}{s} - \kappa_\ny \right) + \uv{\nz}\biggl(\uv{\ny}\cdot\pd{\uv{\nx}}{s}- \kappa_\nz\biggr)\right)+ K_T \uv{\nx} \left(\uv{\nz}\cdot \pd{\uv{\ny}}{s} -\gamma_0 \right).
\end{equation}
Since $\uv{\nx} = \t{R}(\q{q})\uv{e}_x,\;\uv{\ny} = \t{R}(\q{q})\uv{e}_y,\;\uv{\nz} = \t{R}(\q{q})\uv{e}_z$, we have that
\begin{equation} \label{equation:R_outside_moment_expression}
\v{M} = \t{R}\left(\q{q}\right)\left(\begin{matrix}
K_T (\uv{\nz}\cdot \pd{\uv{\ny}}{s} -\gamma_0 ) \\
K_B (\uv{\nx}\cdot \pd{\uv{\nz}}{s} - \kappa_\ny )\\
K_B (\uv{\ny}\cdot \pd{\uv{\nx}}{s} - \kappa_\nz )
\end{matrix}\right).
\end{equation}
Next, considering $\uv{\nx}(s,t)$, a function of arclength $s$ and time $t$, we observe that
\begin{subequations}\label{equation:quaternion_derivative_simplification}
\begin{align}
\left(0,\pd{\uv{\nx}}{s}\right) &= \pd{}{s}\left(0,\uv{\nx}\right) \\ &= \pd{}{s}\left[\q{q}\qprod\left(0,\uv{e}_x\right)\qprod\q{q}^*\right] \\ &= \pd{\q{q}}{s}\qprod\q{q}^*\qprod\left(0,\uv{\nx}\right) - \left(0,\uv{\nx}\right)\qprod\pd{\q{q}}{s}\qprod\q{q}^*,
\end{align}
\end{subequations}
where we have used $\q{q}\qprod\pd{\q{q}^*}{s} = -\pd{\q{q}}{s}\qprod\q{q}^*$, a consequence of $\q{q}\qprod\q{q}^* = \q{q}^*\qprod\q{q} = \q{I}_q$. Now, since $\|\q{q}\| = 1$, we have that $\pd{\q{q}}{s}\qprod\q{q}^*$ is a pure quaternion (i.e.\ one with zero scalar part) and thus may be unambiguously identified with its vector part $[\pd{\q{q}}{s}\qprod\q{q}^*]_{\Reals^3}$. Observing that for pure quaternions we have $(0,\v{v})\qprod(0,\v{w}) - (0,\v{w})\qprod(0,\v{v}) = (0,2\v{v}\times\v{w})$, the above yields
\begin{equation}
	\pd{\uv{\nx}}{s} = 2\left[\pd{\q{q}}{s}\qprod\q{q}^*\right]_{\Reals^3}\times\uv{\nx}.
\end{equation}
Analogous expressions hold for $\uv{\ny}$ and $\uv{\nz}$. Using the invariance under circular shift of the scalar triple product, we find that
\begin{align}
    \uv{\ny}\cdot\pd{\uv{\nx}}{s} &= 2\left[\pd{\q{q}}{s}\qprod\q{q}^*\right]_{\Reals^3}\cdot\uv{\nz},\\
    \uv{\nz}\cdot\pd{\uv{\ny}}{s} &= 2\left[\pd{\q{q}}{s}\qprod\q{q}^*\right]_{\Reals^3}\cdot\uv{\nx},\\
    \uv{\nx}\cdot\pd{\uv{\nz}}{s} &= 2\left[\pd{\q{q}}{s}\qprod\q{q}^*\right]_{\Reals^3}\cdot\uv{\ny}.
\end{align}
Thus, we simplify,
\begin{subequations}
\begin{align}
2\pd{\q{q}}{s}\qprod\q{q}^* &= \left(0,\left(\uv{\nz}\cdot\pd{\uv{\ny}}{s}\right)\uv{\nx} + \left(\uv{\nx}\cdot\pd{\uv{\nz}}{s}\right)\uv{\ny} + \left(\uv{\ny}\cdot\pd{\uv{\nx}}{s}\right)\uv{\nz}\right) \\ &= \q{q}\qprod\left(0,\uv{\nz}\cdot\pd{\uv{\ny}}{s},\uv{\nx}\cdot\pd{\uv{\nz}}{s},\uv{\ny}\cdot\pd{\uv{\nx}}{s}\right)\qprod\q{q}^*;
\end{align}
\end{subequations}
which yields
\begin{equation}
2\q{q}^*\qprod\pd{\q{q}}{s} = \left(0,\uv{\nz}\cdot\pd{\uv{\ny}}{s},\uv{\nx}\cdot\pd{\uv{\nz}}{s},\uv{\ny}\cdot\pd{\uv{\nx}}{s}\right).
\end{equation}
Defining $\t{D} = \diag(K_T,K_B,K_B)$ we find that \cref{equation:R_outside_moment_expression} reduces to
\begin{equation}
\v{M} = \t{R}\left(\q{q}\right)\t{D}\left(2\left[\q{q}^*\qprod\pd{\q{q}}{s}\right]_{\Reals^3} - \left(\begin{matrix}\gamma_0 \\ \kappa_\ny \\ \kappa_\nz\end{matrix}\right)\right).
\end{equation}

\subsection{The quaternion ODE} \label{appendix:quaternion_ode}

In this section, we show that if a frame vector $\v{r}$ satisfies $\fd{\v{r}}{t} = \v{\Omega}\times\v{r}$, where $\v{r}$ is defined through a unit quaternion $\q{q}$ and an initial vector $\v{r}_0$ (i.e.\ $(0,\v{r}) = \q{q}\qprod(0,\v{r}_0)\qprod\q{q}^*$), then the quaternion $\q{q}$ must satisfy $\fd{\q{q}}{t} = \frac{1}{2}(0,\v{\Omega})\qprod\q{q}$. Employing the same simplifications as in \cref{equation:quaternion_derivative_simplification}, we observe that
\begin{equation}
\fd{}{t}\left(0,\v{r}\right) = \fd{\q{q}}{t}\qprod\q{q}^*\qprod\left(0,\v{r}\right) - \left(0,\v{r}\right)\qprod\fd{\q{q}}{t}\qprod\q{q}^*.
\end{equation}
This implies
\begin{equation}
\fd{\v{r}}{t} = 2\left[\fd{\q{q}}{t}\qprod\q{q}^*\right]_{\Reals^3}\times\v{r}.
\end{equation}
Given that this holds true for all the vectors of the frame, we obtain $2[\fd{\q{q}}{t} \qprod \q{q}^*]_{\Reals^3} = \v{\Omega}$, and since $\fd{\q{q}}{t}\qprod\q{q}^*$ is a pure quaternion we have
\begin{equation}
\left(0,\v{\Omega}\right) = 2\fd{\q{q}}{t}\qprod\q{q}^*.
\end{equation}
Dividing by 2 and right-multiplying by $\q{q}$ yields the desired result.

\subsection{BDF2 in the Lie algebra} \label{appendix:lie_algebra_bdf2}

In this section, we derive the form of the BDF2 update equation which applies to the Lie algebra elements.  This derivation follows the more general framework laid out in \citet{faltinsen_multistep_2001}. Let $\v{u}_{(k)}^m$ denote the Lie algebra element corresponding to the quaternion at timestep $m$ expressed in the coordinate system from timestep $k$, that is,
\begin{equation}
    \q{q}^m = \exp\left(\v{u}_{(k)}^m\right)\qprod\q{q}^k.
    \label{equation:qm-to-qk}
\end{equation} To integrate \cref{equation:lie_algebra_ode} using BDF2 we need to solve
\begin{equation} \label{equation:first_bdf2_lie_algebra}
\v{u}^{n+1}_{(n)} = \frac{4}{3}\v{u}^{n}_{(n)} - \frac{1}{3}\v{u}^{n-1}_{(n)} + \frac{2}{3} \Delta t \dexpinv_{\v{u}^{n+1}_{(n)}}\left(\v{\Omega}^{n+1}\right),
\end{equation}
given $\v{u}^{n}_{(n-1)}$ and $\v{u}^{n-1}_{(n-1)}$. The coordinate transform between tangent spaces needs to be such that a given Lie algebra element represents the same quaternion independent of the current coordinate system, hence
\begin{equation} \label{equation:coordinate_transform_condition_lie_algebra}
\exp\left(\v{u}^{m}_{(k+1)}\right)\qprod\q{q}^{k+1} = \q{q}^{m} = \exp\left(\v{u}^{m}_{(k)}\right)\qprod\q{q}^{k}
\end{equation}
needs to hold.

To this end, we use the Baker--Campbell--Hausdorff formula $\BCH$ which is defined to satisfy
\begin{equation}
    \exp\left(\BCH\left(\v{X},\v{Y}\right)\right) = \exp\left(\v{X}\right)\qprod\exp\left(\v{Y}\right).
    \label{equation:BCH-satisfies-this}
\end{equation}
While in our case ($\mathfrak{so}(3)$ with BDF2), there is a closed-form expression for the Baker--Campbell--Hausdorff formula \citep{engo_bch-formula_2001}, it will only be necessary to use the identity $\BCH(\v{X},-\v{X}) = \vu{0}$.

For \cref{equation:coordinate_transform_condition_lie_algebra} to hold, the Lie algebra elements need to be transferred from one tangent space to the next.
It follows from the condition $\exp(\v{u}^{m}_{(k+1)})\qprod\q{q}^{k+1} = \exp(\v{u}^{m}_{(k)})\qprod\q{q}^{k}$ that this  can be achieved via
\begin{equation}
\v{u}^{m}_{(k+1)} = \BCH\left(\v{u}^{m}_{(k)},-\v{u}^{k+1}_{(k)}\right).
\end{equation}
Substituting this into \cref{equation:first_bdf2_lie_algebra} we obtain
\begin{equation}
\v{u}^{n+1}_{(n)} = \frac{4}{3}\BCH\left(\v{u}^{n}_{(n-1)},-\v{u}^{n}_{(n-1)}\right) - \frac{1}{3}\BCH\left(\v{u}^{n-1}_{(n-1)},-\v{u}^{n}_{(n-1)}\right) + \frac{2}{3} \Delta t \dexpinv_{\v{u}^{n+1}_{(n)}}\left(\v{\Omega}^{n+1}\right).
\end{equation}
Observing in the right-hand side of this expression that
\begin{equation}
\BCH\left(\v{u}^{n}_{(n-1)},-\v{u}^{n}_{(n-1)}\right) = \vu{0},
\end{equation}
and, since $\v{u}^{n-1}_{(n-1)} = \vu{0}$ from \cref{equation:qm-to-qk}, that
\begin{equation}
 \BCH\left(\v{u}^{n-1}_{(n-1)},-\v{u}^{n}_{(n-1)}\right) = -\v{u}^{n}_{(n-1)},
\end{equation}
we obtain
\begin{equation}
\v{u}^{n+1}_{(n)} = \frac{1}{3}\v{u}^{n}_{(n-1)} + \frac{2}{3} \Delta t \dexpinv_{\v{u}^{n+1}_{(n)}}\left(\v{\Omega}^{n+1}\right).
\end{equation}
Noticing that the Lie algebra elements only appear here expressed according to the coordinate system in which they were first produced, we may unambiguously write the BDF2 update equation as
\begin{equation}
\v{u}^{n+1} = \frac{1}{3}\v{u}^{n} + \frac{2}{3} \Delta t \dexpinv_{\v{u}^{n+1}}\left(\v{\Omega}^{n+1}\right).
\end{equation}

\subsection{Filament initialisation in cloud simulations}\label{sec:isoinit}
The isotropic initialisation of straight filaments works as follows. We generate uniform random points in $[0,S]^3$, where $S$ is the linear box size.  These points set the positions of the filament segments at the $s=0$ ends.  We then generate four-dimensional random vectors drawn from the standard normal distribution. After normalising, these vectors provide the initial quaternions for all segments comprising each filament.  We then initialise the remaining segment positions for each filament in straight lines along directions based on the initial quaternions.  If there is any overlap between filaments, one is discarded and we reinitialise its position and quaternion.  This process generates a uniform distribution of filament orientations \citep{karney_quaternions_2007}.

\section{Form of the approximate Jacobian}\label{sec:form-of-jacobian}
In this section, we detail how we generate an analytic approximation to the Jacobian, $\pd{\v{f}}{\v{X}}$, by including only local physics for each particle -- a local Stokes drag model, without hydrodynamic or steric interactions. The resulting approximate Jacobian retains the essence of the full Jacobian, but can be rapidly generated and posesses a block diagonal structure.

The supplementary code \citep{schoeller_github_2019} implements the generation of the Jacobian. Here, we demonstrate how the expressions are derived.

\subsection{In 2D}
The system $\v{f}(\v{X})=\vu{0}$ as described in \cref{eqn:RA-2d-system} is formed of the position update for particle $n=1$ in each filament, the orientation updates for all $n$, $1\leq n \leq N$, and the modified constraints for $1< n\leq N$:

\begin{subequations}
\begin{align}
\vu{0} =
(\v{f}^1)_{n=1}^{j+1}\coloneqq\;&\v{Y}_n^{j+1} -\frac{4}{3} \v{Y}_n^{j} + \frac{1}{3} \v{Y}_n^{j-1} -\frac{2}{3} \Delta t \,\v{V}_n^{j+1},%  &= \vu{0},
\label{eqn:2d-system-repeated-1}\\
0 =
(f^2)_{1 \leq n \leq N}^{j+1}\coloneqq\;&\theta_n^{j+1} -\frac{4}{3} \theta_n^{j} + \frac{1}{3} \theta_n^{j-1} -\frac{2}{3} \Delta t \,\Omega_n^{j+1},% &= 0,
\label{eqn:2d-system-repeated-2} \\
\vu{0} =
(\v{f}^3)_{1 < n \leq N}^{j+1}\coloneqq\;&\v{Y}_1^{j+1} + \frac{\Delta L}{2}\sum_{m = 2}^n \left(\uv{t}_{m-1}^{j+1} + \uv{t}_m^{j+1}\right) - \frac{4}{3}\v{Y}_n^{j} + \frac{1}{3} \v{Y}_n^{j-1}   -\frac{2}{3}\Delta t \, \v{V}^{j+1}_n.% &= \vu{0}.
\label{eqn:2d-system-repeated-3}
\end{align}
\end{subequations}
This system is repeated for each filament.

In general, the velocities and angular velocities of the filament segments are found by solving the mobility problem,
\begin{equation}
	\begin{pmatrix}
	\v{V}\\\v{\Omega}
	\end{pmatrix}
    = \mathcal{M}\cdot
    \begin{pmatrix}
	\v{F}\\\v{T}
	\end{pmatrix},
\end{equation}
where $\v{V}^\top = (\v{V}_1^\top,\dots,\v{V}_{NM}^\top)$ is the vector of all segments' velocity components, similarly $\v{\Omega}$ of all angular velocity components, $\v{F}$ of all force components and $\v{T}$ of all torque components acting on the filaments' segments.

In the approximate Jacobian, we implement a local Stokes drag model, where the mobility matrix is diagonal, giving $\v{V}_n = \v{F}_n/6\pi\eta a$ and $\v{\Omega}_n = \v{T}_n/8\pi\eta a^3$. The only force we apply is the constraint force, \cref{eqn:constraint-force}, and the torques applied are the elastic torque, \cref{eqn:elastic-torque}, and the constraint torque, \cref{eqn:constraint-torque}.

Recall that for planar motion, without loss of generality, we take the third frame vector $\uv{\nz}$ to be aligned with the fixed vector $\uv{e}_z$. This direction is always normal to the plane of motion.

Together, we have
\begin{subequations}
\begin{align}
\vu{0} = (\v{f}_0^1)_{n=1}^{j+1}\coloneqq\;&\v{Y}_1^{j+1} -\frac{4}{3} \v{Y}_1^{j} + \frac{1}{3} \v{Y}_1^{j-1} -\frac{2}{3} \frac{\Delta t}{6\pi\eta a}\left(\v{\Lambda}^{j+1}_{3/2}-\v{\Lambda}^{j+1}_{1/2}\right), \label{eqn:single-filament-full-equations-2d-1}\\% &= \vu{0},\\
0 = (f_0^2)_{1 \leq n \leq N}^{j+1}\coloneqq\;&\theta_n^{j+1} -\frac{4}{3} \theta_n^{j} + \frac{1}{3} \theta_n^{j-1} -\frac{2}{3} \frac{\Delta t}{8\pi\eta a^3}\times\\
&\uv{e}_z\cdot \left[ K_B\left(\frac{\uv{t}^{j+1}_n\times\left(\uv{t}^{j+1}_{n-1}+\uv{t}^{j+1}_{n+1}\right)}{\Delta L} - \uv{e}_z\kappa_\nz\right)    - \frac{\Delta L}{2}\uv{t}^{j+1}_n\times\left(\v{\Lambda}^{j+1}_{n+1/2}+\v{\Lambda}^{j+1}_{n-1/2}\right) \right], \nonumber\\%  &= 0,\\
\vu{0} = (\v{f}_0^3)_{1 < n \leq N}^{j+1}\coloneqq\;& \v{Y}_1^{j+1} + \frac{\Delta L}{2}\sum_{m = 2}^n \left(\uv{t}_{m-1}^{j+1} + \uv{t}_m^{j+1}\right) - \frac{4}{3}\v{Y}_n^{j} + \frac{1}{3} \v{Y}_n^{j-1}   -\frac{2}{3}\frac{\Delta t}{6\pi\eta a}\left(\v{\Lambda}^{j+1}_{n+1/2}-\v{\Lambda}^{j+1}_{n-1/2}\right),% &= \vu{0},
\end{align}%
\label{eqn:single-filament-full-equations-2d}%
\end{subequations}
repeated for each filament. Since the approximate Jacobian assumes no interaction between filaments, we can build the approximate Jacobian for a multifilament system from $M$ independent blocks, placed along the diagonal, with each block holding a single-filament Jacobian. In the following we discuss the derivation of a single filament's approximate Jacobian without loss of generality.

Having written out the system $\v{f}_0 = (\v{f}^1_0, \v{f}^2_0, \v{f}^3_0)$ in \cref{eqn:single-filament-full-equations-2d}, it can be differentiated with respect to the state variable, $\v{X}^{j+1} = (\v{Y}^{j+1}_1,\theta^{j+1}_{1\ldots N},\v{\Lambda}^{j+1}_{3/2\ldots N-1/2})$, to form the Jacobian
\begin{equation}
    \t{J}_0 = \pd{\v{f}_0}{\v{X}^{j+1}} = \begin{pmatrix}
    \t{I} & \tu{0} & \t{J}^{13} \\
    \tu{0} & \t{J}^{22} & \t{J}^{23} \\
    \t{I} & \t{J}^{32} & \t{J}^{33}
    \end{pmatrix}.
    \label{eqn:jacobian-labelled}
\end{equation}
The remaining expressions for the submatrices relating segments $m$ and $n$ are given by, after some algebraic manipulation and dropping the time (superscript) indices for clarity,
\begingroup
\allowdisplaybreaks
\begin{subequations}
\begin{align}
\t{J}^{13}_{1n} =& \pd{(\v{f}^1_0)_1}{\v{\Lambda}_{n-1/2}} &&= \frac{2}{3}\frac{\Delta t}{6\pi\eta a}(\delta_{1n} - \delta_{2n})\t{I},\label{eq:j13-2d}\\
{J}^{22}_{mn}  =& \pd{(f^2_0)_m}{\theta_n} &&= \delta_{mn}-\frac{2}{3}\frac{\Delta t}{8\pi\eta a^3}\left[\frac{K_B}{\Delta L}%
\Big(-\delta_{mn}[\cos(\theta_m - \theta_{m-1}) + \cos(\theta_{m+1}-\theta_m)] \right.\nonumber \\
&&& \qquad\qquad + \delta_{(m-1)n} \cos(\theta_m - \theta_{m-1}) + \delta_{(m+1)n}\cos(\theta_{m+1}-\theta_m) \Big) \nonumber\\
&&& \left. - \frac{\Delta L}{2} \delta_{mn}\left(
\sin\theta_m [\Lambda^y_{m+1/2}+\Lambda^y_{m-1/2}] +\cos \theta_m [\Lambda^x_{m+1/2}+\Lambda^x_{m-1/2}]
\right) \right],  \\
\v{J}^{23}_{mn} =& \pd{(f^2_0)_m}{\v{\Lambda}_{n-1/2}} &&= \frac{2}{3}\frac{\Delta t}{8\pi\eta a^3}\frac{\Delta L}{2}(-\sin\theta_m \uv{e}_x + \cos\theta_m \uv{e}_y)(\delta_{mn}+\delta_{(m+1)n}), \\
\v{J}^{32}_{mn} =& \pd{(\v{f}^3_0)_m}{\theta_n} &&= \frac{\Delta L}{2}\sum_{r=2}^m\left[ (-\sin\theta_{r-1}\delta_{(r-1)n} - \sin\theta_r \delta_{rn})\uv{e}_x + (\cos\theta_{r-1}\delta_{(r-1)n} + \cos\theta_r \delta_{rn})\uv{e}_y  \right],\\
\t{J}^{33}_{mn} =& \pd{(\v{f}^3_0)_m}{\v{\Lambda}_{n-1/2}} &&= \frac{2}{3}\frac{\Delta t}{6\pi\eta a}(\delta_{mn} - \delta_{(m+1)n})\t{I},\label{eq:j33-2d}
\end{align}
\end{subequations}
where $\v{\Lambda} = (\Lambda^x,\Lambda^y,0)$ and $\delta_{ij}$ is the Kronecker delta.
\endgroup

\subsection{In 3D using Lie algebra elements}

Here we discuss the differences in the construction of the approximate Jacobian in 3D compared to 2D. Again we consider the system $\v{f}(\v{X})=\vu{0}$, this time represented by \cref{eqn:RA-3d-system}. The standard BDF2 update for the first segment in the filament is handled analogously to the 2D case, \cref{eqn:single-filament-full-equations-2d-1}, so we restrict our attention to the Lie algebra update equations and the modified constraints. Substituting in the Stokes drag approximation, each filament-level sub-block of the Jacobian is based on the equations
\begin{subequations}
\begin{align} \label{eq:Lie_algebra_update_with_drag}
    \vu{0}=(\v{f}^2_0)_{1\leq n \leq N}^{j+1} \coloneqq\;& \v{u}_n^{j+1} - \frac{1}{3}\v{u}_n^{j} - \frac{\Delta t}{12\pi\eta a^3} \dexpinv_{\v{u}_n^{j+1}}\left(\v{T}_n^{j+1}\right),\\
    \vu{0}=(\v{f}^3_0)_{1<n\leq N}^{j+1} \coloneqq\;& \v{Y}_1^{j+1} + \frac{\Delta L}{2}\sum_{m = 2}^n \left(\uv{t}_{m-1}^{j+1} + \uv{t}_m^{j+1}\right) - \frac{4}{3}\v{Y}_n^{j} + \frac{1}{3} \v{Y}_n^{j-1} -\frac{\Delta t}{9\pi\eta a}\left(\v{\Lambda}^{j+1}_{n+1/2}-\v{\Lambda}^{j+1}_{n-1/2}\right),
\end{align}
\end{subequations}
where the Lie algebra update equations are for all segments $1\leq n \leq N$, and the modified constraints are for segments $1< n\leq N$.

For the sake of legibility, we discard any terms with a time index other than $j+1$, as they will have zero derivative in all cases, and drop the time superscript. Hence, to produce the Jacobian it suffices to consider the equations
\begin{subequations}\label{equation:Lie_algebra_update_with_drag_simplified}
\begin{align}
    \vu{0}=(\v{f}^2_0)_{1 \leq n \leq N} \coloneqq\;& \v{u}_n - \frac{\Delta t}{12\pi\eta a^3} \dexpinv_{\v{u}_n}\left(\v{T}_n\right), \\
    \vu{0}=(\v{f}^3_0)_{1 < n \leq N} \coloneqq\;& \v{Y}_1 + \frac{\Delta L}{2}\sum_{m = 2}^n \left(\uv{t}_{m-1} + \uv{t}_m\right) - \frac{\Delta t}{9\pi\eta a}\left(\v{\Lambda}_{n+1/2}-\v{\Lambda}_{n-1/2}\right).
\end{align}
\end{subequations}

The Jacobian then corresponds to the system $\v{f}_0 = (\v{f}^1_0, \v{f}^2_0, \v{f}^3_0)$ in \cref{eqn:single-filament-full-equations-2d-1,equation:Lie_algebra_update_with_drag_simplified}, differentiated with respect to the state variable, $\v{X} = (\v{Y}_1,\v{u}_{1\ldots N},\v{\Lambda}_{3/2\ldots N-1/2})$, forming
\begin{equation}
    \t{J}_0 = \pd{\v{f}_0}{\v{X}} = \begin{pmatrix}
    \t{I} & \tu{0} & \t{J}^{13} \\
    \tu{0} & \t{J}^{22} & \t{J}^{23} \\
    \t{I} & \t{J}^{32} & \t{J}^{33}
    \end{pmatrix}.
    \label{eqn:jacobian-labelled-3d}
\end{equation}
The expressions for $\t{J}^{13}$ and $\t{J}^{33}$ are the same as in the 2D case, \cref{eq:j13-2d,eq:j33-2d}. We therefore consider the remaining three blocks.

We first observe that
\begin{equation}
    \t{J}^{32}_{mn} = \pd{(\v{f}_0^3)_m}{\v{u}_n} = \frac{\Delta L}{2}\sum_{k=2}^m\left(\delta_{(k-1)n}\pd{\uv{t}_{k-1}}{\v{u}_n} + \delta_{kn}\pd{\uv{t}_{k}}{\v{u}_n}\right),
\end{equation}
where $\delta_{ij}$ is the Kronecker delta. By taking \cref{equation:lie_algebra_ode} and replacing $\dexpinv_{\v{u}}(\v{\Omega})$ by the first term of its Taylor series expansion,
\begin{equation}
    \dexpinv_{\v{u}}\left(\v{v}\right) \approx \v{v} - \frac{1}{2}\v{u}\times\v{v} + \frac{1}{12}\v{u}\times\left(\v{u}\times\v{v}\right),
    \label{eq:lie-algebra-taylor-expansion}
\end{equation}
we can approximate $\fd{\v{u}_n}{t}\approx\v{\Omega}$. Combining this with $\fd{\uv{t}_n}{t}=\v{\Omega}\times\uv{t}_n$ from \cref{equation:angular-velocity}, we have
\begin{equation}
    \pd{\uv{t}_{n}}{\v{u}_n} \approx \left[\times\uv{t}_n\right],
\end{equation}
where we have introduced the notation $[\v{v}\times]$ for the skew-symmetric matrix satisfying $[\v{v}\times]\v{x} = \v{v}\times\v{x}$ (and $[\times\v{v}] = -[\v{v}\times] = [\v{v}\times]^\top$ satisfying $[\times\v{v}]\v{x} = \v{x}\times\v{v}$). Hence,
\begin{equation}
    \t{J}^{32}_{mn} = \pd{(\v{f}_0^3)_m}{\v{u}_n} \approx \frac{\Delta L}{2}\sum_{k=2}^m\left(\delta_{(k-1)n} + \delta_{n,k}\right)\left[\times\uv{t}_n\right].
\end{equation}
For the final blocks, we employ all three terms written in the Taylor series expansion in \cref{eq:lie-algebra-taylor-expansion} to
reduce \cref{eq:Lie_algebra_update_with_drag} to
\begin{equation}
     (\v{f}_0^2)_m \approx \v{u}_m - \frac{\Delta t}{12\pi\eta a^3}\left(\v{T}_m - \frac{1}{2}\v{u}_m\times\v{T}_m + \frac{1}{12}\v{u}_m\times\left(\v{u}_m\times\v{T}_m\right)\right).
\end{equation}
From this we can see that the matrices of interest take the forms
\begin{align}
    \t{J}^{23}_{mn} = \pd{(\v{f}_0^2)_m}{\v{\Lambda}_{n-1/2}} \approx& -\frac{\Delta t}{12\pi\eta a^3}\left(\t{I} - \frac{1}{2}[\v{u}_m\times] + \frac{1}{12}[\v{u}_m\times]^2\right)\pd{\v{T}_m}{\v{\Lambda}_{n-1/2}},\\
    \t{J}^{22}_{mn} = \pd{(\v{f}_0^2)_m}{\v{u}_n} \approx& -\frac{\Delta t}{12\pi\eta a^3}\left(\t{I} - \frac{1}{2}[\v{u}_m\times] + \frac{1}{12}[\v{u}_m\times]^2\right)\pd{\v{T}_m}{\v{u}_n},\\
\intertext{for $m \neq n$, and otherwise,}
        \t{J}^{22}_{mm} = \pd{(\v{f}_0^2)_m}{\v{u}_m} \approx& \, \t{I} - \frac{\Delta t}{12\pi\eta a^3}\left\{\frac{\p \v{T}_m}{\p \v{u}_m} - \frac{1}{2}\left(\left[\times\v{T}_m\right] + \left[\v{u}_m\times\right]\frac{\p \v{T}_m}{\p \v{u}_m}\right)\right. \nonumber \\
        & + \left.\frac{1}{12}\left(\left[\times\left(\v{u}_m\times\v{T}_m\right)\right] + \left[\v{u}_m\times\right]\left[\times\v{T}_m\right] + \left[\v{u}_m\times\right]^2\frac{\p \v{T}_m}{\p \v{u}_m}\right)\right\}.
\end{align}
The problem of constructing these matrices thus reduces to that of evaluating the derivatives of the torque on segment $m$. Segment $m$ experiences elastic and constraint torques
\begin{equation}
    \v{T}_m = \frac{\Delta L}{2}\uv{t}_m\times\left(\v{\Lambda}_{m+1/2} + \v{\Lambda}_{m-1/2}\right) + \v{M}_{m+1/2} - \v{M}_{m-1/2},
\end{equation}
where $\v{M}_{m+1/2}$ is the elastic moment between segments $m$ and $m+1$. Since the elastic moments do not depend on the Lagrange multipliers, we have
\begin{equation}
    \pd{\v{T}_m}{\v{\Lambda}_{n-1/2}} \approx \frac{\Delta L}{2}\left(\delta_{(m+1)n} + \delta_{mn}\right)\left[\uv{t}_m\times\right].
\end{equation}
Now considering the derivatives with respect to the Lie algebra elements, we have
\begin{equation}
    \pd{\v{T}_m}{\v{u}_n} \approx \frac{\Delta L}{2}\delta_{mn}\left[\times\left(\v{\Lambda}_{m+1/2} + \v{\Lambda}_{m-1/2}\right)\right]\left[\times\uv{t}_m\right] + \pd{\v{M}_{m+1/2}}{\v{u}_n} - \pd{\v{M}_{m-1/2}}{\v{u}_n}.
\end{equation}
If we use a frame vector approximation for the elastic moment,
    \begin{equation}
        \v{M}_{m+1/2} \approx K_B\left(\frac{\uv{t}_m \times \uv{t}_{m+1}}{\Delta L} - \frac{\kappa_\ny}{2}\left(\uv{\ny}_m + \uv{\ny}_{m+1}\right) - \frac{\kappa_\nz}{2}\left(\uv{\nz}_m + \uv{\nz}_{m+1}\right)\right)%\\
        + \frac{K_T\beta}{2}\left(\uv{t}_m + \uv{t}_{m+1}\right),
    \end{equation}
    where $\beta = \frac{1}{2\Delta L}(\uv{\ny}_{m+1}\cdot\uv{\nz}_m - \uv{\ny}_{m}\cdot\uv{\nz}_{m+1}) - \gamma_0$, then we can write
    \begin{equation} \begin{split}
        \frac{\p \v{M}_{m+1/2}}{\p \v{u}_m} \approx\,& K_B\left(\frac{1}{\Delta L}\left[\times\uv{t}_{m+1}\right]\left[\times\uv{t}_m\right] - \frac{\kappa_\ny}{2}\left[\times\uv{\ny}_m\right] - \frac{\kappa_\nz}{2}\left[\times\uv{\nz}_m\right]\right)\\ &+ \frac{K_T}{2}\left[\beta\left[\times\uv{t}_m\right] + \frac{1}{2\Delta L}\left(\uv{t}_m + \uv{t}_{m+1}\right)\left(\uv{\nz}_m\times\uv{\ny}_{m+1} - \uv{\ny}_m\times\uv{\nz}_{m+1}\right)\right],
    \end{split} \end{equation}
    where the last term is a dyadic product. Similarly,
    \begin{equation} \begin{split}
        \frac{\p \v{M}_{m+1/2}}{\p \v{u}_{m+1}} \approx\,& K_B\left(\frac{1}{\Delta L}\left[\uv{t}_m\times\right]\left[\times\uv{t}_{m+1}\right] - \frac{\kappa_\ny}{2}\left[\times\uv{\ny}_{m+1}\right] - \frac{\kappa_\nz}{2}\left[\times\uv{\nz}_{m+1}\right]\right)\\ &+ \frac{K_T}{2}\left[\beta\left[\times\uv{t}_{m+1}\right] - \frac{1}{2\Delta L}\left(\uv{t}_m + \uv{t}_{m+1}\right)\left(\uv{\nz}_m\times\uv{\ny}_{m+1} - \uv{\ny}_m\times\uv{\nz}_{m+1}\right)\right].
    \end{split} \end{equation}
The derivative with respect to any other Lie algebra element is zero, and thus we have provided all of the elements necessary to form an approximate Jacobian. We find that level of approximation employed here works well in our implementation.

\small
\bibliography{references.bib}

\begin{thebibliography}{104}
\expandafter\ifx\csname natexlab\endcsname\relax\def\natexlab#1{#1}\fi

\bibitem[Agrawal \& Babu(2018)]{agrawal_self-organization_2018}
{\sc Agrawal, A. \& Babu, S.~B.} 2018 Self-organization in a bimotility mixture
  of model microswimmers. {\em Physical Review E\/} {\bf 97}~(2), 020401.

\bibitem[Allen \& Tildesley(2017)]{allen_computer_2017}
{\sc Allen, M.~P. \& Tildesley, D.~J.} 2017 {\em Computer {{Simulation}} of
  {{Liquids}}\/}, 2nd edn. {Oxford University Press}, Oxford.

\bibitem[Ascher \& Petzold(1998)]{ascher_computer_1998}
{\sc Ascher, U.~M. \& Petzold, L.~R.} 1998 {\em Computer Methods for Ordinary
  Differential Equations and Differential-Algebraic Equations\/}, vol.~61.
  {SIAM}.

\bibitem[Baker {\em et~al.\/}(2009)Baker, Bonnecaze \&
  Zaman]{baker_extracellular_2009}
{\sc Baker, E.~L., Bonnecaze, R.~T. \& Zaman, M.~H.} 2009 Extracellular matrix
  stiffness and architecture govern intracellular rheology in cancer. {\em
  Biophysical Journal\/} {\bf 97}~(4), 1013--1021.

\bibitem[Batchelor(1967)]{batchelor_introduction_1967}
{\sc Batchelor, G.~K.} 1967 {\em An {{Introduction}} to {{Fluid Dynamics}}\/}.
  {Cambridge University Press}.

\bibitem[Brady \& Bossis(1988)]{brady_stokesian_1988}
{\sc Brady, J.~F. \& Bossis, G.} 1988 Stokesian {{Dynamics}}. {\em Annual
  Review of Fluid Mechanics\/} {\bf 20}~(1), 111--157.

\bibitem[Brennen \& Winet(1977)]{brennen_fluid_1977}
{\sc Brennen, C.~E. \& Winet, H.} 1977 Fluid mechanics of propulsion by cilia
  and flagella. {\em Annual Review of Fluid Mechanics\/} {\bf 9}, 339--398.

\bibitem[Broyden(1965)]{broyden_class_1965}
{\sc Broyden, C.~G.} 1965 A class of methods for solving nonlinear simultaneous
  equations. {\em Mathematics of Computation\/} {\bf 19}~(92), 577--593.

\bibitem[Chelakkot {\em et~al.\/}(2010)Chelakkot, Winkler \&
  Gompper]{chelakkot_migration_2010}
{\sc Chelakkot, R., Winkler, R.~G. \& Gompper, G.} 2010 Migration of
  semiflexible polymers in microcapillary flow. {\em Europhysics Letters\/}
  {\bf 91}~(1), 14001.

\bibitem[Claeys \& Brady(1993)]{claeys_suspensions_1993}
{\sc Claeys, I.~L. \& Brady, J.~F.} 1993 Suspensions of prolate spheroids in
  {{Stokes}} flow. {{Part}} 1. {{Dynamics}} of a finite number of particles in
  an unbounded fluid. {\em Journal of Fluid Mechanics\/} {\bf 251}, 411--442.

\bibitem[Coq {\em et~al.\/}(2009)Coq, {du Roure}, Fermigier \&
  Bartolo]{coq_helical_2009}
{\sc Coq, N., {du Roure}, O., Fermigier, M. \& Bartolo, D.} 2009 Helical
  beating of an actuated elastic filament. {\em Journal of Physics: Condensed
  Matter\/} {\bf 21}~(20), 204109.

\bibitem[Coq {\em et~al.\/}(2008)Coq, {du Roure}, Marthelot, Bartolo \&
  Fermigier]{coq_rotational_2008}
{\sc Coq, N., {du Roure}, O., Marthelot, J., Bartolo, D. \& Fermigier, M.} 2008
  Rotational dynamics of a soft filament: {{Wrapping}} transition and
  propulsive forces. {\em Physics of Fluids\/} {\bf 20}~(5), 051703.

\bibitem[Cortez(2001)]{cortez_method_2001}
{\sc Cortez, R.} 2001 The method of regularized {{Stokeslets}}. {\em SIAM
  Journal on Scientific Computing\/} {\bf 23}~(4), 1204--1225.

\bibitem[Cortez(2018)]{cortez2018regularized}
{\sc Cortez, R.} 2018 Regularized stokeslet segments. {\em Journal of
  Computational Physics\/} {\bf 375}, 783--796.

\bibitem[Cortez {\em et~al.\/}(2005)Cortez, Fauci \&
  Medovikov]{cortez_method_2005}
{\sc Cortez, R., Fauci, L.~J. \& Medovikov, A.~A.} 2005 The method of
  regularized {{Stokeslets}} in three dimensions: {{Analysis}}, validation, and
  application to helical swimming. {\em Physics of Fluids\/} {\bf 17}~(031504).

\bibitem[Cosentino~Lagomarsino {\em et~al.\/}(2005)Cosentino~Lagomarsino,
  Pagonabarraga \& Lowe]{cosentino_lagomarsino_hydrodynamic_2005}
{\sc Cosentino~Lagomarsino, M., Pagonabarraga, I. \& Lowe, C.~P.} 2005
  Hydrodynamic induced deformation and orientation of a microscopic elastic
  filament. {\em Physical Review Letters\/} {\bf 94}~(14), 148104.

\bibitem[Dance {\em et~al.\/}(2004)Dance, Climent \&
  Maxey]{dance_collision_2004}
{\sc Dance, S.~L., Climent, E. \& Maxey, M.~R.} 2004 Collision barrier effects
  on the bulk flow in a random suspension. {\em Physics of Fluids\/} {\bf
  16}~(3), 828--831.

\bibitem[Delmotte {\em et~al.\/}(2015)Delmotte, Climent \&
  Plourabou\'e]{delmotte_general_2015}
{\sc Delmotte, B., Climent, E. \& Plourabou\'e, F.} 2015 A general formulation
  of bead models applied to flexible fibers and active filaments at low
  {{Reynolds}} number. {\em Journal of Computational Physics\/} {\bf 286},
  14--37.

\bibitem[Delong {\em et~al.\/}(2015)Delong, Balboa~Usabiaga \&
  Donev]{delong_brownian_2015}
{\sc Delong, S., Balboa~Usabiaga, F. \& Donev, A.} 2015 Brownian dynamics of
  confined rigid bodies. {\em The Journal of Chemical Physics\/} {\bf
  143}~(14), 144107.

\bibitem[Derakhshandeh {\em et~al.\/}(2011)Derakhshandeh, Kerekes,
  Hatzikiriakos \& Bennington]{derakhshandeh_rheology_2011}
{\sc Derakhshandeh, B., Kerekes, R.~J., Hatzikiriakos, S.~G. \& Bennington, C.
  P.~J.} 2011 Rheology of pulp fibre suspensions: {{A}} critical review. {\em
  Chemical Engineering Science\/} {\bf 66}~(15), 3460--3470.

\bibitem[{du Roure} {\em et~al.\/}(2017){du Roure}, Lindner, Nazockdast \&
  Shelley]{du_roure_dynamics_2017}
{\sc {du Roure}, O., Lindner, A., Nazockdast, E.~N. \& Shelley, M.~J.} 2017
  Dynamics of flexible fibers in viscous flows and fluids. {\em Annual Review
  of Fluid Mechanics\/} {\bf 51}~(1), 539--572.

\bibitem[Dunn \& Parberry(2011)]{dunn_3d_2011}
{\sc Dunn, F. \& Parberry, I.} 2011 {\em {{3D Math Primer}} for {{Graphics}}
  and {{Game Development}}\/}, 2nd edn. {A K Peters/CRC Press}, Boca Raton,
  Florida.

\bibitem[Elgeti \& Gompper(2013)]{elgeti_emergence_2013}
{\sc Elgeti, J. \& Gompper, G.} 2013 Emergence of metachronal waves in cilia
  arrays. {\em Proceedings of the National Academy of Sciences of the United
  States of America\/} {\bf 110}~(12), 4470--4475.

\bibitem[Elgeti {\em et~al.\/}(2015)Elgeti, Winkler \&
  Gompper]{elgeti_physics_2015}
{\sc Elgeti, J., Winkler, R.~G. \& Gompper, G.} 2015 Physics of microswimmers
  -- single particle motion and collective behavior: A review. {\em Reports on
  Progress in Physics\/} {\bf 78}~(5), 056601.

\bibitem[Eng\o(2001)]{engo_bch-formula_2001}
{\sc Eng\o, K.} 2001 On the {{BCH}}-formula in $\mathfrak{so}(3)$. {\em BIT
  Numerical Mathematics\/} {\bf 41}~(3), 629--632.

\bibitem[Faltinsen {\em et~al.\/}(2001)Faltinsen, Marthinsen \&
  {Munthe-Kaas}]{faltinsen_multistep_2001}
{\sc Faltinsen, S., Marthinsen, A. \& {Munthe-Kaas}, H.~Z.} 2001 Multistep
  methods integrating ordinary differential equations on manifolds. {\em
  Applied Numerical Mathematics\/} {\bf 39}~(3), 349--365.

\bibitem[Faubel {\em et~al.\/}(2016)Faubel, Westendorf, Bodenschatz \&
  Eichele]{faubel_cilia-based_2016}
{\sc Faubel, R., Westendorf, C., Bodenschatz, E. \& Eichele, G.} 2016
  Cilia-based flow network in the brain ventricles. {\em Science\/} {\bf
  353}~(6295), 176--178.

\bibitem[Fauci \& Peskin(1988)]{fauci_computational_1988}
{\sc Fauci, L.~J. \& Peskin, C.~S.} 1988 A computational model of aquatic
  animal locomotion. {\em Journal of Computational Physics\/} {\bf 77}~(1),
  85--108.

\bibitem[Guo {\em et~al.\/}(2018)Guo, Fauci, Shelley \&
  Kanso]{guo_bistability_2018}
{\sc Guo, H., Fauci, L.~J., Shelley, M.~J. \& Kanso, E.} 2018 Bistability in
  the synchronization of actuated microfilaments. {\em Journal of Fluid
  Mechanics\/} {\bf 836}, 304--323.

\bibitem[Gustavsson \& Tornberg(2009)]{gustavsson_gravity_2009}
{\sc Gustavsson, K. \& Tornberg, A.-K.} 2009 Gravity induced sedimentation of
  slender fibers. {\em Physics of Fluids\/} {\bf 21}~(12), 123301.

\bibitem[{Hall-McNair} {\em et~al.\/}(2019){Hall-McNair}, Gallagher,
  {Montenegro-Johnson}, Gad\^elha \& Smith]{hall-mcnair_efficient_2019}
{\sc {Hall-McNair}, A.~L., Gallagher, M.~T., {Montenegro-Johnson}, T.~D.,
  Gad\^elha, H. \& Smith, D.~J.} 2019 Efficient {{Implementation}} of
  {{Elastohydrodynamics}} via {{Integral Operators}}. {\em arXiv:1903.03427
  [physics]\/} .

\bibitem[Heck {\em et~al.\/}(2017)Heck, Smeets, Vanmaercke, Bhattacharya,
  Odenthal, Ramon, Van~Oosterwyck \& Van~Liedekerke]{heck_modeling_2017}
{\sc Heck, T., Smeets, B., Vanmaercke, S., Bhattacharya, P., Odenthal, T.,
  Ramon, H., Van~Oosterwyck, H. \& Van~Liedekerke, P.} 2017 Modeling
  extracellular matrix viscoelasticity using smoothed particle hydrodynamics
  with improved boundary treatment. {\em Computer Methods in Applied Mechanics
  and Engineering\/} {\bf 322}, 515--540.

\bibitem[Hwang {\em et~al.\/}(1969)Hwang, Litt \&
  Forsman]{hwang_rheological_1969}
{\sc Hwang, S.~H., Litt, M. \& Forsman, W.~C.} 1969 Rheological properties of
  mucus. {\em Rheologica Acta\/} {\bf 8}~(4), 438--448.

\bibitem[Iserles {\em et~al.\/}(2000)Iserles, {Munthe-Kaas}, N\o{}rsett \&
  Zanna]{iserles_lie-group_2000}
{\sc Iserles, A., {Munthe-Kaas}, H.~Z., N\o{}rsett, S.~P. \& Zanna, A.} 2000
  Lie-group methods. {\em Acta Numerica\/} {\bf 9}, 215--365.

\bibitem[Jung {\em et~al.\/}(2006)Jung, Spagnolie, Parikh, Shelley \&
  Tornberg]{jung_periodic_2006}
{\sc Jung, S., Spagnolie, S.~E., Parikh, K., Shelley, M.~J. \& Tornberg, A.-K.}
  2006 Periodic sedimentation in a {{Stokesian}} fluid. {\em Physical Review
  E\/} {\bf 74}~(3), 035302.

\bibitem[Kamal \& Keaveny(2018)]{kamal_enhanced_2018}
{\sc Kamal, A. \& Keaveny, E.~E.} 2018 Enhanced locomotion, effective diffusion
  and trapping of undulatory micro-swimmers in heterogeneous environments. {\em
  Journal of The Royal Society Interface\/} {\bf 15}~(148), 20180592.

\bibitem[Karney(2007)]{karney_quaternions_2007}
{\sc Karney, C. F.~F.} 2007 Quaternions in molecular modeling. {\em Journal of
  Molecular Graphics and Modelling\/} {\bf 25}~(5), 595--604.

\bibitem[Keaveny(2008)]{keaveny_dynamics_2008}
{\sc Keaveny, E.~E.} 2008 {\em Dynamics of structures in active suspensions of
  paramagnetic particles and applications to artificial micro-swimmers\/}.
  {{PhD}} Thesis, Brown University, United States.

\bibitem[Keaveny(2014)]{keaveny_fluctuating_2014}
{\sc Keaveny, E.~E.} 2014 Fluctuating force-coupling method for simulations of
  colloidal suspensions. {\em Journal of Computational Physics\/} {\bf 269},
  61--79.

\bibitem[Keaveny \& Shelley(2011)]{keaveny2011applying}
{\sc Keaveny, E.~E. \& Shelley, M.~J.} 2011 Applying a second-kind boundary
  integral equation for surface tractions in stokes flow. {\em Journal of
  Computational Physics\/} {\bf 230}~(5), 2141--2159.

\bibitem[Kim \& Netz(2006)]{kim_pumping_2006}
{\sc Kim, Y.~W. \& Netz, R.~R.} 2006 Pumping fluids with periodically beating
  grafted elastic filaments. {\em Physical Review Letters\/} {\bf 96}~(15),
  158101.

\bibitem[Kirchhoff(1859)]{kirchhoff_uber_1859}
{\sc Kirchhoff, G.} 1859 {\"Uber das Gleichgewicht und die Bewegung eines
  unendlich d\"unnen elastischen Stabes}. {\em Journal f\"ur die Reine und
  Angewandte Mathematik\/} {\bf 56}, 285--313.

\bibitem[Knoll \& Keyes(2004)]{knoll_jacobian-free_2004}
{\sc Knoll, D.~A. \& Keyes, D.~E.} 2004 Jacobian-free
  {{Newton}}\textendash{{Krylov}} methods: A survey of approaches and
  applications. {\em Journal of Computational Physics\/} {\bf 193}~(2),
  357--397.

\bibitem[Kvaalen(1991)]{kvaalen_faster_1991}
{\sc Kvaalen, E.} 1991 A faster {{Broyden}} method. {\em BIT Numerical
  Mathematics\/} {\bf 31}~(2), 369--372.

\bibitem[Lai {\em et~al.\/}(2009)Lai, Wang, Wirtz \& Hanes]{lai_micro-_2009}
{\sc Lai, S.~K., Wang, Y.-Y., Wirtz, D. \& Hanes, J.} 2009 Micro- and
  macrorheology of mucus. {\em Advanced Drug Delivery Reviews\/} {\bf 61}~(2),
  86--100.

\bibitem[Landau \& Lifshitz(1986)]{landau_theory_1986}
{\sc Landau, L.~D. \& Lifshitz, E.~M.} 1986 {\em Theory of {{Elasticity}}\/},
  3rd edn., vol.~7. {Elsevier}.

\bibitem[Larson(1999)]{larson_structure_1999}
{\sc Larson, R.~G.} 1999 {\em The {{Structure}} and {{Rheology}} of {{Complex
  Fluids}}\/}, 1st edn. {Oxford University Press}, New York.

\bibitem[Lauga \& Powers(2009)]{lauga_hydrodynamics_2009}
{\sc Lauga, E. \& Powers, T.~R.} 2009 The hydrodynamics of swimming
  microorganisms. {\em Reports on Progress in Physics\/} {\bf 72}~(9), 096601.

\bibitem[Li {\em et~al.\/}(2013)Li, Manikantan, Saintillan \&
  Spagnolie]{li_sedimentation_2013}
{\sc Li, L., Manikantan, H., Saintillan, D. \& Spagnolie, S.~E.} 2013 The
  sedimentation of flexible filaments. {\em Journal of Fluid Mechanics\/} {\bf
  735}, 705--736.

\bibitem[Liang {\em et~al.\/}(2013)Liang, Gimbutas, Greengard, Huang \&
  Jiang]{liang_fast_2013}
{\sc Liang, Z., Gimbutas, Z., Greengard, L., Huang, J. \& Jiang, S.} 2013 A
  fast multipole method for the
  {{Rotne}}\textendash{{Prager}}\textendash{{Yamakawa}} tensor and its
  applications. {\em Journal of Computational Physics\/} {\bf 234}, 133--139.

\bibitem[Lim(2010)]{lim_dynamics_2010}
{\sc Lim, S.} 2010 Dynamics of an open elastic rod with intrinsic curvature and
  twist in a viscous fluid. {\em Physics of Fluids\/} {\bf 22}~(2), 024104.

\bibitem[Lim {\em et~al.\/}(2008)Lim, Ferent, Wang \&
  Peskin]{lim_dynamics_2008}
{\sc Lim, S., Ferent, A., Wang, X.~S. \& Peskin, C.~S.} 2008 Dynamics of a
  closed rod with twist and bend in fluid. {\em SIAM Journal on Scientific
  Computing\/} {\bf 31}~(1), 273--302.

\bibitem[Liu {\em et~al.\/}(2009)Liu, Keaveny, Maxey \&
  Karniadakis]{liu_force-coupling_2009}
{\sc Liu, D., Keaveny, E.~E., Maxey, M.~R. \& Karniadakis, G.} 2009
  Force-coupling method for flows with ellipsoidal particles. {\em Journal of
  Computational Physics\/} {\bf 228}~(10), 3559--3581.

\bibitem[Llopis {\em et~al.\/}(2007)Llopis, Pagonabarraga,
  Cosentino~Lagomarsino \& Lowe]{llopis_sedimentation_2007}
{\sc Llopis, I., Pagonabarraga, I., Cosentino~Lagomarsino, M. \& Lowe, C.~P.}
  2007 Sedimentation of pairs of hydrodynamically interacting semiflexible
  filaments. {\em Physical Review E\/} {\bf 76}~(6), 061901.

\bibitem[Lomholt \& Maxey(2003)]{lomholt_force-coupling_2003}
{\sc Lomholt, S. \& Maxey, M.~R.} 2003 Force-coupling method for particulate
  two-phase flow: {{Stokes}} flow. {\em Journal of Computational Physics\/}
  {\bf 184}, 381--405.

\bibitem[Mackaplow \& Shaqfeh(1998)]{mackaplow_numerical_1998}
{\sc Mackaplow, M.~B. \& Shaqfeh, E. S.~G.} 1998 A numerical study of the
  sedimentation of fibre suspensions. {\em Journal of Fluid Mechanics\/} {\bf
  376}, 149--182.

\bibitem[Majmudar {\em et~al.\/}(2012)Majmudar, Keaveny, Zhang \&
  Shelley]{majmudar_experiments_2012}
{\sc Majmudar, T., Keaveny, E.~E., Zhang, J. \& Shelley, M.~J.} 2012
  Experiments and theory of undulatory locomotion in a simple structured
  medium. {\em Journal of The Royal Society Interface\/} {\bf 9}~(73),
  1809--1823.

\bibitem[Marchetti {\em et~al.\/}(2018)Marchetti, Raspa, Lindner, {du Roure},
  Bergougnoux, Guazzelli \& Duprat]{marchetti_deformation_2018}
{\sc Marchetti, B., Raspa, V., Lindner, A., {du Roure}, O., Bergougnoux, L.,
  Guazzelli, E. \& Duprat, C.} 2018 Deformation of a flexible fiber settling in
  a quiescent viscous fluid. {\em Physical Review Fluids\/} {\bf 3}~(10),
  104102.

\bibitem[Maxey \& Patel(2001)]{maxey_localized_2001}
{\sc Maxey, M.~R. \& Patel, B.~K.} 2001 Localized force representations for
  particles sedimenting in {{Stokes}} fow. {\em International Journal of
  Multiphase Flow\/} {\bf 27}, 1603--1626.

\bibitem[Metzger {\em et~al.\/}(2007)Metzger, Nicolas \&
  Guazzelli]{metzger_falling_2007}
{\sc Metzger, B., Nicolas, M. \& Guazzelli, E.} 2007 Falling clouds of
  particles in viscous fluids. {\em Journal of Fluid Mechanics\/} {\bf 580},
  283--301.

\bibitem[Moreau {\em et~al.\/}(2018)Moreau, Giraldi \&
  Gad\^elha]{moreau_asymptotic_2018}
{\sc Moreau, C., Giraldi, L. \& Gad\^elha, H.} 2018 The asymptotic
  coarse-graining formulation of slender-rods, bio-filaments and flagella. {\em
  Journal of The Royal Society Interface\/} {\bf 15}~(144), 20180235.

\bibitem[Nazockdast {\em et~al.\/}(2017)Nazockdast, Rahimian, Zorin \&
  Shelley]{nazockdast_fast_2017}
{\sc Nazockdast, E.~N., Rahimian, A., Zorin, D. \& Shelley, M.~J.} 2017 A fast
  platform for simulating semi-flexible fiber suspensions applied to cell
  mechanics. {\em Journal of Computational Physics\/} {\bf 329}, 173--209.

\bibitem[Nitsche \& Batchelor(1997)]{nitsche_break-up_1997}
{\sc Nitsche, J.~M. \& Batchelor, G.~K.} 1997 Break-up of a falling drop
  containing dispersed particles. {\em Journal of Fluid Mechanics\/} {\bf 340},
  161--175.

\bibitem[Olson {\em et~al.\/}(2013)Olson, Lim \& Cortez]{olson_modeling_2013}
{\sc Olson, S.~D., Lim, S. \& Cortez, R.} 2013 Modeling the dynamics of an
  elastic rod with intrinsic curvature and twist using a regularized {{Stokes}}
  formulation. {\em Journal of Computational Physics\/} {\bf 238}, 169--187.

\bibitem[Park \& Chung(2005)]{park_geometric_2005}
{\sc Park, J. \& Chung, W.-K.} 2005 Geometric integration on {{Euclidean}}
  group with application to articulated multibody systems. {\em IEEE
  Transactions on Robotics\/} {\bf 21}~(5), 850--863.

\bibitem[Park {\em et~al.\/}(2010)Park, Metzger, Guazzelli \&
  Butler]{park_cloud_2010}
{\sc Park, J., Metzger, B., Guazzelli, E. \& Butler, J.~E.} 2010 A cloud of
  rigid fibres sedimenting in a viscous fluid. {\em Journal of Fluid
  Mechanics\/} {\bf 648}, 351--362.

\bibitem[Peskin(2002)]{peskin_immersed_2002}
{\sc Peskin, C.~S.} 2002 The immersed boundary method. {\em Acta Numerica\/}
  {\bf 11}, 479--517.

\bibitem[Pettersson {\em et~al.\/}(2017)Pettersson, Hellwig, Gustafsson \&
  Stenstr\"om]{pettersson_measurement_2017}
{\sc Pettersson, T., Hellwig, J., Gustafsson, P.-J. \& Stenstr\"om, S.} 2017
  Measurement of the flexibility of wet cellulose fibres using atomic force
  microscopy. {\em Cellulose\/} {\bf 24}~(10), 4139--4149.

\bibitem[Powers(2010)]{powers_dynamics_2010}
{\sc Powers, T.~R.} 2010 Dynamics of filaments and membranes in a viscous
  fluid. {\em Reviews of Modern Physics\/} {\bf 82}~(2), 1607--1631.

\bibitem[Pozrikidis(1992)]{pozrikidis_boundary_1992}
{\sc Pozrikidis, C.} 1992 {\em Boundary {{Integral}} and {{Singularity
  Methods}} for {{Linearized Viscous Flow}}\/}. {Cambridge University Press}.

\bibitem[Purcell(1977)]{purcell_life_1977}
{\sc Purcell, E.~M.} 1977 Life at low {{Reynolds}} number. {\em American
  Journal of Physics\/} {\bf 45}~(1).

\bibitem[Quraishi {\em et~al.\/}(1998)Quraishi, Jones \&
  Mason]{quraishi_rheology_1998}
{\sc Quraishi, M.~S., Jones, N.~S. \& Mason, J.} 1998 The rheology of nasal
  mucus: A review. {\em Clinical Otolaryngology \& Allied Sciences\/} {\bf
  23}~(5), 403--413.

\bibitem[Ross \& Klingenberg(1997)]{ross_dynamic_1997}
{\sc Ross, R.~F. \& Klingenberg, D.~J.} 1997 Dynamic simulation of flexible
  fibers composed of linked rigid bodies. {\em The Journal of Chemical
  Physics\/} {\bf 106}~(7), 2949--2960.

\bibitem[Rosti {\em et~al.\/}(2018)Rosti, Banaei, Brandt \&
  Mazzino]{rosti_flexible_2018}
{\sc Rosti, M.~E., Banaei, A.~A., Brandt, L. \& Mazzino, A.} 2018 Flexible
  fiber reveals the two-point statistical properties of turbulence. {\em
  Physical Review Letters\/} {\bf 121}~(4), 044501.

\bibitem[Saad \& Schultz(1986)]{saad_gmres_1986}
{\sc Saad, Y. \& Schultz, M.~H.} 1986 {{GMRES}}: {{A}} generalized minimal
  residual algorithm for solving nonsymmetric linear systems. {\em SIAM Journal
  on Scientific and Statistical Computing\/} {\bf 7}~(3), 856--869.

\bibitem[Saggiorato {\em et~al.\/}(2015)Saggiorato, Elgeti, Winkler \&
  Gompper]{saggiorato_conformations_2015}
{\sc Saggiorato, G., Elgeti, J., Winkler, R.~G. \& Gompper, G.} 2015
  Conformations, hydrodynamic interactions, and instabilities of sedimenting
  semiflexible filaments. {\em Soft Matter\/} {\bf 11}~(37), 7337--7344.

\bibitem[Saintillan {\em et~al.\/}(2005)Saintillan, Darve \&
  Shaqfeh]{saintillan_smooth_2005}
{\sc Saintillan, D., Darve, E. \& Shaqfeh, E. S.~G.} 2005 A smooth
  particle-mesh {{Ewald}} algorithm for {{Stokes}} suspension simulations:
  {{The}} sedimentation of fibers. {\em Physics of Fluids\/} {\bf 17}~(3),
  033301.

\bibitem[Sanderson \& Curtin(2016)]{sanderson_armadillo_2016}
{\sc Sanderson, C. \& Curtin, R.} 2016 Armadillo: A template-based {{C}}++
  library for linear algebra. {\em The Journal of Open Source Software\/} {\bf
  1}, 26.

\bibitem[Sanderson \& Curtin(2018)]{sanderson_user-friendly_2018}
{\sc Sanderson, C. \& Curtin, R.} 2018 A {{User}}-{{Friendly Hybrid Sparse
  Matrix Class}} in {{C}}++. {\em Lecture Notes in Computer Science\/} {\bf
  10931}, 422--430.

\bibitem[Schoeller \& Keaveny(2018)]{schoeller_flagellar_2018}
{\sc Schoeller, S.~F. \& Keaveny, E.~E.} 2018 From flagellar undulations to
  collective motion: Predicting the dynamics of sperm suspensions. {\em Journal
  of The Royal Society Interface\/} {\bf 15}~(140), 20170834.

\bibitem[Schoeller {\em et~al.\/}(2019)Schoeller, Townsend, Westwood \&
  Keaveny]{schoeller_github_2019}
{\sc Schoeller, S.~F., Townsend, A.~K., Westwood, T.~A. \& Keaveny, E.~E.} 2019
  {{GitHub}} repository. \url{https://github.com/ekeaveny/filaments/}.

\bibitem[Sheehan \& Carlstedt(1984)]{sheehan_hydrodynamic_1984}
{\sc Sheehan, J.~K. \& Carlstedt, I.} 1984 Hydrodynamic properties of human
  cervical-mucus glycoproteins in {{6M}}-guanidinium chloride. {\em The
  Biochemical Journal\/} {\bf 217}~(1), 93--101.

\bibitem[Shelley(2016)]{shelley_dynamics_2016}
{\sc Shelley, M.~J.} 2016 The dynamics of microtubule/motor-protein assemblies
  in biology and physics. {\em Annual Review of Fluid Mechanics\/} {\bf
  48}~(1), 487--506.

\bibitem[Sierou \& Brady(2001)]{sierou_accelerated_2001}
{\sc Sierou, A. \& Brady, J.~F.} 2001 Accelerated {{Stokesian Dynamics}}
  simulations. {\em Journal of Fluid Mechanics\/} {\bf 448}, 115--146.

\bibitem[Simons {\em et~al.\/}(2015)Simons, Fauci \& Cortez]{simons_fully_2015}
{\sc Simons, J., Fauci, L.~J. \& Cortez, R.} 2015 A fully three-dimensional
  model of the interaction of driven elastic filaments in a {{Stokes}} flow
  with applications to sperm motility. {\em Journal of Biomechanics\/} {\bf
  48}~(9), 1639--1651.

\bibitem[Smith(2009)]{smith_boundary_2009}
{\sc Smith, D.~J.} 2009 A boundary element regularized {{Stokeslet}} method
  applied to cilia- and flagella-driven flow. {\em Proceedings of the Royal
  Society A: Mathematical, Physical and Engineering Sciences\/} {\bf
  465}~(2112), 3605--3626.

\bibitem[Smith {\em et~al.\/}(2007)Smith, Gaffney \&
  Blake]{smith_discrete_2007}
{\sc Smith, D.~J., Gaffney, E.~A. \& Blake, J.~R.} 2007 Discrete cilia
  modelling with singularity distributions: Application to the embryonic node
  and the airway surface liquid. {\em Bulletin of Mathematical Biology\/} {\bf
  69}~(5), 1477--1510.

\bibitem[Smith {\em et~al.\/}(2019)Smith, {Montenegro-Johnson} \&
  Lopes]{smith_symmetry-breaking_2019}
{\sc Smith, D.~J., {Montenegro-Johnson}, T.~D. \& Lopes, S.~S.} 2019
  Symmetry-breaking cilia-driven flow in embryogenesis. {\em Annual Review of
  Fluid Mechanics\/} {\bf 51}~(1), 105--128.

\bibitem[Stockie \& Green(1998)]{stockie_simulating_1998}
{\sc Stockie, J.~M. \& Green, S.~I.} 1998 Simulating the motion of flexible
  pulp fibres using the immersed boundary method. {\em Journal of Computational
  Physics\/} {\bf 147}~(1), 147--165.

\bibitem[Supatto \& Vermot(2011)]{supatto_cilia_2011}
{\sc Supatto, W. \& Vermot, J.} 2011 From cilia hydrodynamics to zebrafish
  embryonic development. {\em Current Topics in Developmental Biology\/} {\bf
  95}, 33--66.

\bibitem[Swan \& Brady(2007)]{swan_simulation_2007}
{\sc Swan, J.~W. \& Brady, J.~F.} 2007 Simulation of hydrodynamically
  interacting particles near a no-slip boundary. {\em Physics of Fluids\/} {\bf
  19}~(11), 113306.

\bibitem[Teran {\em et~al.\/}(2010)Teran, Fauci \&
  Shelley]{teran_viscoelastic_2010}
{\sc Teran, J., Fauci, L. \& Shelley, M.} 2010 Viscoelastic fluid response can
  increase the speed and efficiency of a free swimmer. {\em Physical Review
  Letters\/} {\bf 104}~(3), 038101.

\bibitem[Tornberg \& Gustavsson(2006)]{tornberg_numerical_2006}
{\sc Tornberg, A.-K. \& Gustavsson, K.} 2006 A numerical method for simulations
  of rigid fiber suspensions. {\em Journal of Computational Physics\/} {\bf
  215}~(1), 172--196.

\bibitem[Tornberg \& Shelley(2004)]{tornberg_simulating_2004}
{\sc Tornberg, A.-K. \& Shelley, M.~J.} 2004 Simulating the dynamics and
  interactions of flexible fibers in {{Stokes}} flows. {\em Journal of
  Computational Physics\/} {\bf 196}~(1), 8--40.

\bibitem[{van de Rotten}(2003)]{van_de_rotten_limited_2003}
{\sc {van de Rotten}, B.~A.} 2003 {\em A limited memory {{Broyden}} method to
  solve high-dimensional systems of nonlinear equations\/}. {{PhD}} Thesis,
  Leiden University, Netherlands.

\bibitem[Vince(2011)]{vince_quaternions_2011}
{\sc Vince, J.} 2011 {\em Quaternions for {{Computer Graphics}}\/}.
  {Springer-Verlag}, London.

\bibitem[Wajnryb {\em et~al.\/}(2013)Wajnryb, Mizerski, Zuk \&
  Szymczak]{wajnryb_generalization_2013}
{\sc Wajnryb, E., Mizerski, K.~A., Zuk, P.~J. \& Szymczak, P.} 2013
  Generalization of the {{Rotne}}--{{Prager}}--{{Yamakawa}} mobility and shear
  disturbance tensors. {\em Journal of Fluid Mechanics\/} {\bf 731}.

\bibitem[Wiens \& Stockie(2015)]{wiens_simulating_2015}
{\sc Wiens, J.~K. \& Stockie, J.~M.} 2015 Simulating flexible fiber suspensions
  using a scalable immersed boundary algorithm. {\em Computer Methods in
  Applied Mechanics and Engineering\/} {\bf 290}, 1--18.

\bibitem[Yamamoto \& Matsuoka(1995)]{yamamoto_dynamic_1995}
{\sc Yamamoto, S. \& Matsuoka, T.} 1995 Dynamic simulation of fiber suspensions
  in shear flow. {\em The Journal of Chemical Physics\/} {\bf 102}~(5),
  2254--2260.

\bibitem[Yang {\em et~al.\/}(2008)Yang, Elgeti \&
  Gompper]{yang_cooperation_2008}
{\sc Yang, Y., Elgeti, J. \& Gompper, G.} 2008 Cooperation of sperm in two
  dimensions: {{Synchronization}}, attraction, and aggregation through
  hydrodynamic interactions. {\em Physical Review E\/} {\bf 78}~(061903).

\bibitem[Yang {\em et~al.\/}(2010)Yang, Marceau \& Gompper]{yang_swarm_2010}
{\sc Yang, Y., Marceau, V. \& Gompper, G.} 2010 Swarm behavior of
  self-propelled rods and swimming flagella. {\em Physical Review E\/} {\bf
  82}~(031904).

\bibitem[Yeo \& Maxey(2010)]{yeo_simulation_2010}
{\sc Yeo, K. \& Maxey, M.~R.} 2010 Simulation of concentrated suspensions using
  the force-coupling method. {\em Journal of Computational Physics\/} {\bf
  229}~(6), 2401--2421.

\bibitem[Zuk {\em et~al.\/}(2014)Zuk, Wajnryb, Mizerski \&
  Szymczak]{zuk_rotneprageryamakawa_2014}
{\sc Zuk, P.~J., Wajnryb, E., Mizerski, K.~A. \& Szymczak, P.} 2014
  Rotne\textendash{{Prager}}\textendash{{Yamakawa}} approximation for
  different-sized particles in application to macromolecular bead models. {\em
  Journal of Fluid Mechanics\/} {\bf 741}.

\bibitem[Zupan {\em et~al.\/}(2009)Zupan, Saje \&
  Zupan]{zupan_quaternion-based_2009}
{\sc Zupan, E., Saje, M. \& Zupan, D.} 2009 The quaternion-based
  three-dimensional beam theory. {\em Computer Methods in Applied Mechanics and
  Engineering\/} {\bf 198}, 3944--3956.

\end{thebibliography}
\bibliographystyle{jfm2}

\end{document}